\documentclass[prd,twocolumn,showpacs,nofootinbib,floatfix]{revtex4}
\usepackage{bm}
\usepackage{graphicx}
\usepackage{amsmath}
\usepackage{wasysym}
\begin{document}
\newcommand{\mwimp}{$m_\chi$}
\newcommand{\sigmapsi}{$\sigma_p^{SI}$}
\newcommand{\sigmapsd}{$\sigma_p^{SD}$}
\newcommand{\sigmansd}{$\sigma_n^{SD}$}
\newcommand{\kms}{km~s$^{-1}$}
\newcommand{\vsun}{$v_\odot$}
\newcommand{\ve}{$v_\oplus$}
\newcommand{\vescsun}{$v^{\odot}_{esc}$}
\newcommand{\vesce}{$v^{\oplus}_{esc}$}

\title{Dark matter in the solar system I: The distribution function of WIMPs at the Earth from solar capture.}
\author{Annika H. G. Peter}
\email{apeter@astro.caltech.edu}
\affiliation{Department of Physics, Princeton University, Princeton, NJ 08544, USA}
\affiliation{California Institute of Technology, Mail Code 105-24, Pasadena, CA 91125, USA}
\date{\today}

\begin{abstract}
The next generation of dark matter (DM) direct detection experiments and neutrino telescopes will probe large swaths of dark matter parameter space.  In order to interpret the signals in these experiments, it is necessary to have good models of both the halo DM streaming through the solar system and the population of DM bound to the solar system.  In this paper, the first in a series of three on DM in the solar system, we present simulations of orbits of DM bound to the solar system by solar capture in a toy solar system consisting of only the Sun and Jupiter, assuming that DM consists of a single species of weakly interacting massive particle (WIMP).  We describe how the size of the bound WIMP population depends on the WIMP mass \mwimp, spin-independent cross section \sigmapsi, and spin-dependent cross section \sigmapsd.  Using a standard description of the Galactic DM halo, we find that the maximum enhancement to the direct detection event rate, consistent with current experimental constraints on the WIMP-nucleon cross section, is $< 1\%$ relative to the event rate from halo WIMPs, while the event rate from neutrinos from WIMP annihilation in the center of the Earth is unlikely to meet the threshold of next-generation, km$^3$-sized (IceCube, KM3NeT) neutrino telescopes.
\end{abstract}


\maketitle

\section{Introduction}\label{sec:intro}
\subsection{Dark Matter and Detection}
There is overwhelming evidence that non-baryonic DM must exist in large quantities in the universe, yet its nature is unknown.  A popular candidate for DM is one or more species of WIMP.  Particles of this type occur naturally in many theories of physics beyond the Standard Model (SM); examples include the neutralino $\chi$ in supersymmetry \cite{jungman1996}, the lightest Kaluza-Klein photon $B^{(1)}$ in universal extra-dimension (UED) theories \cite{cheng2002,servant2002,hooper2007}, or the heavy photon $A_H$ in Little Higgs models \cite{birkedal2004,hubisz2005,birkedal2006}.  These particular candidates are all stable, self-annihilating, behave as cold dark matter, and are thermally produced in the early universe in roughly the amount needed to explain the dark matter \cite{komatsu2008}.

We may expect rapid progress in constraining the nature of DM due to the maturity of a number of technologies targeting different but complementary WIMP signals.  The next generation of particle colliders, in particular the Large Hadron Collider, may see signatures of physics beyond the SM.  A new generation of satellites is searching for photons (e.g., the \emph{Fermi Gamma-Ray Space Telescope} \cite{wai2007,kuhlen2008}) and other particles (e.g., ATIC \cite{chang2008}, PAMELA \cite{pamela2008}) resulting from WIMP annihilations in the Milky Way's DM halo.

There are also experiments to probe the local WIMP population.  Since the flux of particles from WIMP annihilation scales as the square of the WIMP density, any region in the solar system that has an unusually high density of WIMPs is a good target.  WIMPs generically interact with baryons, which means that WIMPs passing through the solar system may be trapped and settle into dense cores in the potential wells of the Sun or the planets.  The previous generation of neutrino telescopes (e.g., BAKSAN \cite{boliev1996}, MACRO \cite{ambrosio1999}, Super-Kamiokande \cite{desai2004}, AMANDA \cite{achterberg2006,ackermann2006}) places the strongest constraints on the spin-dependent WIMP-proton cross section $\sigma_p^{SD}$ ($\lesssim 10^{-39} \hbox{ cm}^2$ at $m_\chi \sim 100$ GeV) based on flux limits of neutrinos from WIMP annihilation in the Sun and Earth.  Even if the next generation of neutrino telescopes with detector volumes approaching 1 km$^3$ (e.g., Antares \cite{lim2007}, IceCube \cite{delosheros2008}, the proposed KM3NeT \cite{dewolf2008}) do not identify a WIMP annihilation signature, they are projected to improve constraints on $\sigma_p^{SD}$ by almost two orders of magnitude.

The best limits on the spin-dependent WIMP-neutron $\sigma_n^{SD}$ and spin-independent WIMP-nucleon $\sigma_p^{SI}$ cross sections come from direct detection experiments.  The signature of WIMPs in these experiments is a small ($\sim 10\hbox{ keV} - 100\hbox{ kev}$) nuclear recoil.  The next generation of direct detection experiments is slated to have target masses approaching 1000 kg (e.g., DEAP/CLEAN \citep{hime2007}, LUX \citep{gaitskell2007}, SuperCDMS \citep{brink2005,schnee2005,akerib2006c}, WARP \citep{brunetti2005}, XENON1T \citep{aprile2002}, XMASS \citep{xmass2005}) and to be sensitive to cross sections down to $\sigma_p^{SI} \sim 10^{-46} \hbox{ cm}^2$, $\sigma_p^{SD} \sim 10^{-40}\hbox{ cm}^2$, and $\sigma_n^{SD} \sim 10^{-42}\hbox{ cm}^2$ \cite{zacek2007,baudis2008}.  

\subsection{WIMPs in the Solar System}

For a given WIMP model, event rates in direct detection experiments and neutrino telescopes are determined by the phase space distribution function
(DF) of WIMPs in the solar system.  The fiducial assumption is that the 
direct detection event rate and WIMP capture rate in the Earth are
dominated by DM particles from the Galactic halo, passing through the
solar system on unbound orbits \cite{jungman1996,green2007b}.  There are potentially observable consequences if even a tiny fraction of WIMPs may become captured to the solar system, since bound WIMPs have lower speeds than halo WIMPs.   The push for many direct detection experiments is toward ever-lower nuclear recoil energy thresholds ($\sim 0.1\hbox{ keV}-1 \hbox{ keV}$), both in order to gain sensitivity to low mass WIMPs and because the event rate is much higher there than at lower energies \cite{aalseth2008,angle2008}.  At such low energies, low speed WIMPs contribute disproportionately to the event rate for kinematic reasons.

Low speed WIMPs have an even greater impact on the event rate of neutrinos from annihilation in the Earth.  The shallowness of the Earth's potential well means that only low speed WIMPs may be captured in the Earth.  In particular, if the WIMP mass is above 400 GeV, \emph{only} WIMPs bound to the solar system may be trapped in the Earth.

Two processes have been identified by which WIMPs may become captured to the solar system at rates large enough to be important for terrestrial dark matter experiments.  \emph{Gravitational Capture:} Gould \cite{gould1988,gould1991} pointed out that WIMPs may be captured from the halo by gravitationally scattering on the planets.  By treating WIMP orbits in the solar system as a diffusion problem, Gould \cite{gould1991} and \citet{lundberg2004} estimated that bound WIMPs dominate the annihilation rate of WIMPs in the Earth for WIMP masses $\gtrsim 100$ GeV. \emph{Solar Capture:} WIMPs captured in the Sun will reach thermal equilibrium with solar nuclei on timescales $t \sim \tau^{-1}P$, where $\tau$ is the optical depth of the Sun for WIMPs and $P$ is the orbital period of a bound WIMP.  However, \citet{damour1999} identified a population of solar-captured WIMPs that could survive for much longer periods of time due to a type of secular resonance that pulls their perihelia outside the Sun.  Using secular perturbation theory, they found that this population could produce a low-recoil direct detection rate comparable to that of halo WIMPs for $\sigma_p^{SI} \sim 10^{-42} - 10^{-40}\hbox{ cm}^2$, and could yield an annihilation rate in the Earth a factor of $\sim 100$ greater than the rate expected from unbound halo WIMPs for WIMP masses $\sim 100-150\hbox{ GeV}$ \cite{bergstrom1999}.

While these results are intriguing, the semi-analytic treatments used in these papers cannot fully capture the rich range of behavior in small-N systems such as the solar system.  It is important to check these results with numerical experiments.  Moreover, the annihilation rate of WIMPs in the Sun depends critically on whether WIMPs captured in the Sun thermalize rapidly with solar nuclei.  If the planets can pull the WIMPs out of the Sun for extended periods of time, or even eject the particles from the system, the annihilation rate will be depressed with respect to current estimates.  

In a set of three papers \cite{peter2009b,peter2009c}, we present simulations of WIMP orbits in the solar system, including both the gravitational effects of the dominant planet, Jupiter, and an accurate Monte Carlo description of WIMP-nuucleon elastic scattering in the solar interior, as well as a discussion of the likely contribution of bound WIMPs to direct detection experiments and neutrino telescopes.  In this paper, Paper I, we focus on WIMP capture in the Sun.  In order to put our results in context, we first summarize the treatment of \citet{damour1999}, and describe the mechanism they found that extended the lifetimes of solar captured WIMPs in the solar system and built up the DF of WIMPs at the Earth: the Kozai mechanism.

\subsection{Damour and Krauss (1999) and the Kozai Mechanism}\label{sec:intro_dk}

In the absence of gravitational torques from the planets, WIMPs captured onto Earth-crossing orbits by elastic scattering in the Sun will have a small number density at the Earth relative to the halo number density for two reasons.  (i) Unless the WIMP is massive ($m \gtrsim 1$ TeV), the characteristic energy a WIMP loses to a solar nucleus is large enough such that most captured WIMPs have aphelia that lie inside the Earth's orbit.  (ii) The characteristic time to the next scatter, which will almost certainly remove the WIMP from an Earth-crossing orbit unless $m \gtrsim 1$ TeV, is of order $t\propto P / \tau$.  For a WIMP with semi-major axis $a = 1\hbox{ AU}$, $P_\chi = 1$ year.  If, for example, $\sigma_p^{SI} = 10^{-41} \hbox{ cm}^2$ (or $\sigma_p^{SD} = 10^{-39}\hbox{ cm}^2$), $\tau \sim 10^{-3}$, so the WIMP lifetime in the solar system is only of order a thousand years, short compared to the age of the solar system.  

\citet{damour1999} recognized that the lifetimes of bound WIMPs in Earth-crossing orbits could be extended by orders of magnitude if gravitational torques from the planets decreased the WIMP eccentricity (increased the perihelion distance) enough that the WIMP orbit no longer penetrated the Sun.  For WIMPs in planetary systems such as our own, such behavior is possible if the rate of perihelion precession $\dot{\omega}$ is small, since then the torques from the planets act in a constant direction over many WIMP orbits.  This process was first examined by Kozai \cite{kozai1962} in the context of asteroid orbits and is sometimes called the Kozai resonance.  The signature of this resonance is large fluctuations in both the inclination and eccentricity while the semi-major axis is fixed.  The Kozai resonance can lead to both libration and circulation in the argument of perihelion $\omega$, and we use the term ``Kozai cycles" to describe these oscillations.  Kozai cycles have been studied in the context of comets \cite{thomas1996}, asteroids \cite{kozai1979,michel1996}, triple star systems \cite{ford2000}, and exoplanets \cite{fabrycky2007}.  

Damour \& Krauss found approximate analytic Kozai solutions for a solar system containing the inner planets and Jupiter on circular, coplanar orbits.  The requirement that $\dot{\omega}$ is small means that Kozai cycles are only significant for WIMPs with perihelia that are not too far inside the solar radius, so the solar potential is not far from that of a point mass.  Damour \& Krauss use an analytic approximation to the solar potential in the outer $r > 0.55 R_\odot$ of the Sun, where $R_\odot$ is the radius of the Sun.  They expanded the potentials of the planets to quadrupole order in the small parameter $a_P/a$, where $a_P$ is the semi-major axis of a planet, and neglected short-period terms and mean-motion resonances.  The solutions have an additional feature---if the orbital plane of the planets in the solar system is the $x-y$ plane, the $z$-component of the specific angular momentum, $J_z = \sqrt{ GM_\odot a (1-e^2) } \cos I$ ($I$ is the inclination) is a conserved quantity.  

To estimate the size of the solar-captured WIMP population at the Earth, Damour \& Krauss made the following additional assumptions.  (i) Jupiter-crossing WIMPs (with aphelia greater than Jupiter's semi-major axis, $r_a > a_{\jupiter} \approx 5.2 \hbox{ AU}$) were ignored, since their lifetimes in the solar system were assumed to be short.  Similarly, all WIMPs with $r_a < a_{\jupiter}$ that were not on Kozai cycles were also ignored.  (ii) They assumed that all WIMPs on Kozai cycles would survive for the lifetime of the solar system without rescattering in the Sun, regardless of the optical depth in the Sun for WIMPs.  Since the typical lifetime of Earth-crossing WIMPs is $\sim 10^3$ yr for $\sigma_p^{SI} \sim 10^{-41} \hbox{ cm}^2$, the extension of the lifetimes of even a few Earth-crossing particles to the age of the solar system results in a significant boost to the DF of bound WIMPs at the Earth.  

\subsection{This Work}
To investigate the validity of these assumptions and to provide a more accurate assessment of the contribution of bound WIMPs to direct detection experiments and neutrino telescopes, we perform a set of numerical simulations of WIMP orbits in the solar system.  In this paper, Paper I, we present a suite of simulations of solar-captured WIMP orbits in a toy solar system consisting only of the Sun and Jupiter.  Jupiter is the only planet included in the simulations for two reasons. (i) As the largest planet in the solar system, it dominates the dynamics of minor bodies in the system.  We address the issue of other planets in Section \ref{sec:discussion} of this paper as well as in later papers.  (ii) Since some of our numerical methods (described in Section \ref{sec:num}) are new, and since it is important to have a physical understanding of the principal mechanisms that determine the key features of the bound WIMP population, it is useful to simulate a simple system first.  In particular, particle orbits in our toy solar system enjoy a constant of motion (Eq. \ref{eq:ch3_cj}), which provides a check on the numerical accuracy of the integrations.

We describe the simulations in Section \ref{sec:simulations}, and the DFs derived from the simulations in Section \ref{sec:df}.  Also in Section \ref{sec:df}, we show how the DFs depend on the WIMP mass and cross section $m_\chi$, $\sigma_p^{SI}$, and $\sigma_p^{SD}$.  Predictions for the event rates in direct detection experiments and neutrino telescopes are made in Sections \ref{sec:dd} and \ref{sec:id}.  We discuss our results in the context of the previous work on solar captured-WIMPs by \citet{damour1999} and \citet{bergstrom1999}, the presence of other planets, and the assumptions concerning the halo DF in Section \ref{sec:discussion}, and summarize the main results of this work in Section \ref{sec:conclusion}.  

We defer the topic of annihilation of WIMPs in the Sun to Paper II \cite{peter2009b}, and the simulations of gravitationally captured WIMPs to Paper III \cite{peter2009c}.

\section{Orbit Integration}\label{sec:num}
The problem of determining the long-term trajectories of bound dark matter particles imposes a set of difficult challenges to the integration algorithm.  The algorithm must (i) be stable and accurate over 4.5 Gyr; (ii) accurately follow highly eccentric ($e > 0.995$) orbits with no numerical dissipation; (iii) accurately integrate trajectories that are influenced by perturbing forces that may be comparable to the force from the Sun for short intervals (including close encounters with and passages through planets); and (iv) be fast, in order to obtain an adequate statistical sample of orbits. 

Much progress has been made in the past fifteen years to address the first and last criteria. This progress has largely been motivated by interest in the long-term stability of planetary systems.  The most significant development has been the advent of geometric integrators (symplectic  and/or time-reversible integrators), which have the desirable property that errors in conserved quantities (such as the Hamiltonian) are oscillatory rather than growing.  However, the most commonly used algorithms \cite{wisdom1991,saha1994,chambers1999} are not immediately applicable to the present problem, for two main reasons.  First, one would like to use an adaptive time step to quickly but accurately integrate a highly eccentric orbit (using very small time steps near perihelion and larger ones otherwise), or to resolve close encounters with the planets.  It is difficult to introduce an adaptive time step in a symplectic or time-reversible way since varying the time step by criteria that depend on phase space position destroys symplecticity.  Secondly, since for practical purposes the integrations of planetary or comet orbits end when two bodies collide, there has been little attention to integrating systems for which the potential can deviate significantly from the Keplerian point-mass potential, as it does in the solar interior.

In the following sections, we describe an algorithm to efficiently carry out the long-term integration of dark matter particles in the solar system.  In Section \ref{sec:int}, we outline an adaptive time step symplectic integrator (simultaneously formulated by Preto and Tremaine \cite{preto1999} and Mikkola and Tanikawa \cite{mikkola1999}) that is used for most of the orbital integrations.  We explain the error properties of the integrator in Section \ref{sec:err}.  In Section \ref{sec:special}, we discuss procedures to handle special cases, such as close planetary encounters.  We discuss the merits of various coordinate systems in Section \ref{sec:coords}.

\subsection{The Adaptive Time Step Integrator}\label{sec:int}
We closely follow the arguments of \citet{mikkola1999} and \citet{preto1999} in the description of the adaptive time step symplectic integrator.  

A separable Hamiltonian $H(\mathbf{q}, \mathbf{p}, t) = T(\mathbf{p}) + U(\mathbf{q},t)$ ($T$ is the kinetic energy and $U$ is the potential energy), a function of the canonical position $\mathbf{q}$ and momentum $\mathbf{p}$, can be implemented as a symplectic integrator with fixed time step $\Delta t$.  The key to finding a symplectic integrator with a variable time step is to promote the time $t$ to a canonical variable, and make it a function of a new ``time'' coordinate $s$,
\begin{eqnarray}\label{t}
	\mathrm{d} t = g(\mathbf{q}, \mathbf{p}, t) \mathrm{d} s.
\end{eqnarray}
 find a separable Hamiltonian $\Gamma$ in the extended phase space that describes the motion, and take fixed time steps $\Delta s$ when integrating the new equations of motion.  The new canonical coordinates are $q_0 = t$ and $p_0 = - H$, so the new set of canonical variables $\mathbf{Q}=(q_0, \mathbf{q})$ and $\mathbf{P} = (p_0,\mathbf{p})$.  \citeauthor{preto1999} and \citeauthor{mikkola1999} find such an extended phase space Hamiltonian, 
\begin{eqnarray}
	\Gamma(\mathbf{Q}, \mathbf{P}) = g(\mathbf{Q}, \mathbf{P}) \left[ H(\mathbf{q},\mathbf{p},t) - p_0 \right],
\end{eqnarray}
which can be made separable with the choice
\begin{eqnarray}\label{gcomp}
	g(\mathbf{Q}, \mathbf{P}) = \frac{ f(T(\mathbf{p})+p_0) - f(-U(\mathbf{Q}))}{T(\mathbf{p}) + U(\mathbf{Q})+p_0},
\end{eqnarray}
so that the extended Hamiltonian is
\begin{eqnarray}
	\Gamma(\mathbf{Q},\mathbf{P}) = f(T(\mathbf{p})+p_0) - f( - U(\mathbf{Q}) ). \label{eq:gamma_f}
\end{eqnarray}
The equations of motion for this Hamiltonian are:
\begin{eqnarray}
	\frac{\mathrm{d} q_0}{\mathrm{d} s} &=& f^\prime(T(\mathbf{p}) + p_0) \label{eq:eom_first}\\
	\frac{\mathrm{d} \mathbf{q}}{\mathrm{d} s} &=& f^\prime(T(\mathbf{p}) + p_0) \frac{\partial T}{\partial \mathbf{p}} \\
	\frac{\mathrm{d} p_0}{\mathrm{d} s} &=& - f^\prime(-U(\mathbf{Q}) \frac{\partial U(\mathbf{Q})}{\partial q_0}\\
	\frac{\mathrm{d} \mathbf{p}}{\mathrm{d} s} &=& - f^\prime(-U(\mathbf{Q}) \frac{\partial U(\mathbf{Q})}{\partial \mathbf{q}} \label{eq:eom_last}.
\end{eqnarray}
\newline\indent
To determine a useful choice for $f(x)$, Preto and Tremaine expand Eq. (\ref{gcomp}) in a Taylor series about the small parameter $T + p_0 + U$ ($=0$ if the Hamiltonian is exactly conserved) to show that
\begin{eqnarray}
	g(\mathbf{Q}, \mathbf{P}) \approx f^\prime( - U(\mathbf{Q}) ).
\end{eqnarray}
Outside the Sun, the gravitational potential of the solar system is
\begin{eqnarray}
	U(\mathbf{q},t) = -\frac{GM_\odot}{|\mathbf{q} - \mathbf{q}_\odot|} + \sum_i \Phi_i(\mathbf{q}, \mathbf{q}_i),
\end{eqnarray}
where the first term in the potential denotes the Keplerian potential of the Sun and $\Phi_i$ is the potential from planet $i$, and the potential from the Sun dominates most of the time.  Preto and Tremaine show that for a choice of
\begin{eqnarray}
	g(\mathbf{Q},\mathbf{P}) &&= |\mathbf{q} - \mathbf{q}_\odot| \label{g}\\
				 &&\approx - \frac{ GM_\odot}{U(\mathbf{q},t)}
\end{eqnarray}
the two-body problem can be solved exactly, with only a time (phase) error $\delta t / P \propto N^{-2}$, where $P$ is the orbital period and $N$ is the number of steps per orbit.  This is a good feature because phase errors are far less important for our purposes than, for example, systematic energy drifts or numerical precession.  Note that the time step is proportional to the particle's separation from the Sun, so that small time steps are taken near the perihelion of the orbit and large steps near the aphelion.  We use Eq. (\ref{g}) as our choice for $g(\mathbf{Q},\mathbf{P})$, for which the functional form of $f(x)$ is
\begin{eqnarray}
	f(x) = GM_\odot \log (x).
\end{eqnarray}
\indent
The adaptive time step equations of motion are implemented via a second-order leapfrog integrator (also called a Verlet integrator) with $\Delta s \simeq \Delta t / g = h$, where $h$ is determined by the number of steps desired per orbit.  Since the goal is to understand the behavior of a large ensemble of orbits, we are more interested in maintaining small energy errors over long times rather than precisely integrating orbits over short times, and so a second-order symplectic integrator is sufficient. For our choice of $f(x)$, and given $T = v^2 / 2$ and $U = U(\mathbf{r},t)$, the change over a single fictitious time step $h$ can be written as the mapping
\begin{eqnarray}
	\mathbf{r}_{1/2} &=& \mathbf{r}_0 + \frac{1}{2} h \frac{ GM_\odot \mathbf{v}_0}{ \frac{1}{2} v_0^2 + p_{0,0}} \label{eq:verlet_first}\\
	t_{1/2} &=& t_{0} + \frac{1}{2} h \frac{GM_\odot}{\frac{1}{2} v_0^2 + p_{0,0}} \\
	\mathbf{v}_{1} &=& \mathbf{v}_0 + h \frac{GM_\odot}{U(\mathbf{r}_{1/2},t_{1/2})}\frac{\partial U(\mathbf{r}_{1/2},t_{1/2})}{\partial \mathbf{r}} \label{thalf}\\
	p_{0,1} &=& p_{0,0} + h \frac{GM_\odot}{U(\mathbf{r}_{1/2},t_{1/2})}\frac{\partial U(\mathbf{r}_{1/2},t_{1/2})}{\partial t} \label{p1}\\
	 \mathbf{r}_{1} &=& \mathbf{r}_{1/2} + \frac{1}{2} h \frac{ GM_\odot \mathbf{v}_1}{ \frac{1}{2} v_1^2 + p_{0,1}} \\
	t_{1} &=& t_{1/2} + \frac{1}{2} h \frac{GM_\odot}{\frac{1}{2} v_1^2 + p_{0,1}}, \label{eq:verlet_last}
\end{eqnarray}
where the subscript $i = 0,1/2,1$ labels what fraction of the total time step $h$ has been taken.

\subsection{Errors Along the Path}\label{sec:err}
We explore the behavior of the energy errors in the adaptive time step integrator as a function of energy, eccentricity, distance from the Sun, and number of steps per orbit.  This study allows us, in conjunction with the results of Section \ref{sec:planets}, to determine which the fictitious time step $h$ to use to meet accuracy requirements.  The choice of $h$ for the simulations is described in Section \ref{sec:start_end}.  For the current study, we use short integrations in order to focus on the errors of the adaptive time step integrator alone.  We will discuss the long-term behavior of the whole integration scheme after we discuss the other pieces of our algorithm.

Since our toy solar system (Sun + Jupiter + WIMP) is a restricted three-body problem, there is one constant of motion, the Jacobi constant 
\begin{eqnarray}
C_J = -2( E - n_{\jupiter} J_z), \label{eq:ch3_cj}
\end{eqnarray}
where $E$ is the particle energy in an inertial frame, $n_{\jupiter}$ is the mean motion of Jupiter, and $J_z$ is the $z$-component of the particle's angular momentum, assuming that the motions of the Sun and Jupiter are confined to the $x-y$ plane.  Therefore, we can parameterize errors in terms of the Jacobi constant.  There are no analogous conserved quantities for particles orbiting in planetary systems with more than one planet or if the planetary orbit is eccentric.  In those systems, one can quantify errors for integrators of the type described in Section \ref{sec:int} in terms of the relative energy error $\Delta E / E = (E+p_0)/E$, where $E$ is determined by the instantaneous position and velocity of the particle and $p_0$ is the $0-$component of the momentum in the extended phase space.  If the equations of motion (\ref{eq:eom_first})--(\ref{eq:eom_last}) were integrated with no error, then $p_0 = - E$ and $\Delta E/E = 0$.
\newline\indent
In this section, we treat the Sun as a point mass, and consider trajectories with aphelia well inside Jupiter's orbit.  We consider two different initial semi-major axes, $a = a_{\jupiter}/3$ and $a= a_{\jupiter}/6$ respectively, where $a_{\jupiter}$ is the semi-major axis of Jupiter.  To determine the size of the errors in $C_J$ as a function of eccentricity, we integrate orbits with initial eccentricity $e=0.9,0.99,0.999$ and $0.9999$.  We perform integrations for each combination of $a$ and $e$ for 10 different initial, random orientations and an ensemble of step sizes.  We run each integration for a total of $2\times 10^4$ Kepler periods.  The integrations are started at perihelion (to mimic the initial conditions of scattering in the Sun) with a very small $h=10^{-8} R^{-1}_\odot$ year.  We use such a small time step because the magnitude of the errors in the integrator are largest if the integration is started at pericenter, and smallest when started at apocenter.  Once the particle reaches its first aphelion, $h$ is adjusted so that it will provide the desired number of steps per orbit. The fictitious time step is related to the number of steps per orbit by the step in the eccentric anomaly $\Delta u$ and semi-major axis $a$ by
\begin{eqnarray}
	h = 2 \frac{ 1 - \cos \Delta u}{(GM_\odot / a)^{1/2} \sin \Delta u}, \label{eq:nstep}
\end{eqnarray}
for the symplectic mapping of Eqs. (\ref{eq:verlet_first})--(\ref{eq:verlet_last}) in the case of the Kepler two-body problem.  The number of steps per orbit is given by
\begin{eqnarray}
	N = \frac{2\pi}{\Delta u}. \label{eq:nstep2}
\end{eqnarray}

\begin{figure}
	\includegraphics[width=3.3in]{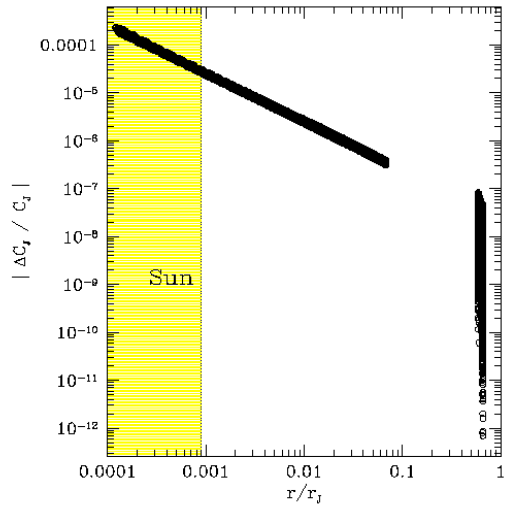}
	\caption{\label{fig:rad}Jacobi constant errors as a function of distance from the primary for a trajectory with $a = 1.73$ AU, followed for $2\times 10^4$ Kepler periods.  This trajectory was integrated with $500$ steps/orbit.  Errors are calculated at perihelion and aphelion.   Points to the left of the vertical line lie within the volume of the Sun; however, we used a point-mass Sun for this integration.}
\end{figure}

We show the dependence of the error on the distance from the Sun in Fig. \ref{fig:rad}.  In this figure, we plot the perihelion and aphelion Jacobi constant errors for a trajectory with initial $a = a_{\jupiter}/3$ and $e=0.999$, integrated with 500 steps/orbit, representative of all the simulations.  We plot only errors at perihelion and aphelion for clarity; a plot showing errors at each time step would be similar but with more scatter.  The interior of the Sun is in the shaded region (though the integrations were done for a point-mass Sun).  From Figure \ref{fig:rad}, it appears that 
\begin{eqnarray}\label{delta_cj}
\left| \Delta C_J / C_J\right| \propto r^{-1}
\end{eqnarray}
This is a generic feature of the integrator, and implies that the maximum Jacobi constant or energy error occurs at perihelion.  The errors are oscillatory,  i.e., there is no secular growth in the error envelope with time.

\begin{figure}
	\includegraphics[width=3.3in]{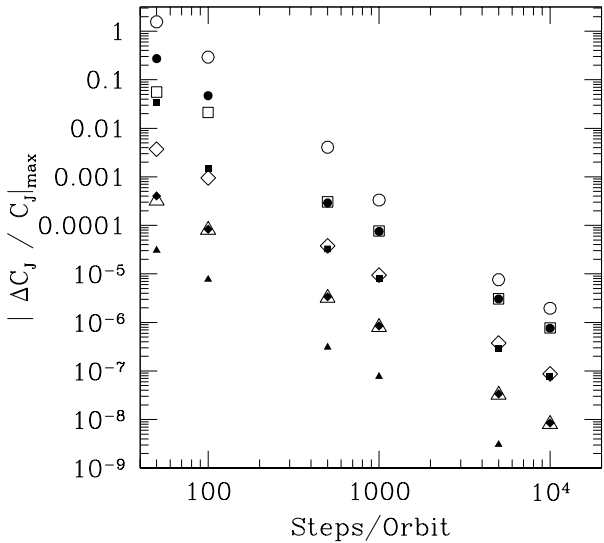}
	\caption{Errors in the Jacobi constant as a function of eccentricity and semi-major axis.  Each point shows the maximum error for 10 trajectories initialized with the same eccentricity but with random initial orientation, and followed for $2\times 10^4$ Kepler orbits.  Open points denote those trajectories for which the semi-major axis $a = a_{\jupiter} / 3 = 1.73$ AU; closed points refer to trajectories with $a = a_{\jupiter}/6 = 0.87$ AU.  Circles mark trajectories with initial eccentricity $e_i = 0.9999$, squares denote those with $e_i = 0.999$, diamonds indicate those with $e_i = 0.99$, and triangles those with $e_i = 0.9$.}\label{fig:step}
\end{figure}

In Fig. \ref{fig:step}, we show the maximum Jacobi constant error as a function of initial semi-major axis $a_i$ and eccentricity $e_i$.  To find the maximum error, we calculate the error in Jacobi constant every time $e$ is in the range $e_i \pm 0.1(1-e_i)$.  The restriction on $e$ isolates the effect of eccentricity on $|\Delta C_J/C_J |$, since Fig. \ref{fig:rad} demonstrates that the maximum error in a simulation depends on the largest eccentricity in the orbit.  We then plot the maximum error found among all simulations for the same initial $a_i$ and $e_i$.  For each type of simulation, the maximum error occurs at perihelion.  Fig. \ref{fig:step} indicates that the maximum Jacobi constant error is a decreasing function of the number of steps per orbit, and an increasing function of semi-major axis and eccentricity.  Furthermore, the maximum error for $e \in e_i \pm 0.1(1 -e_i)$ within each simulation is a function of the initial conditions.  In the simulations with fixed eccentricity and $a = a_{\jupiter}/6$, the spread in these central values is less than a factor of two, while the spread is about a factor of ten in the $a = a_{\jupiter}/3$ simulations.  This is described more in \cite{peter2008}.

To set the fictitious time step $h$ for the simulations detailed in Section \ref{sec:start_end}, it is preferable to consider errors at a fixed, small distance from the Sun rather than exclusively at perihelion.  This is because we use a mapping technique to follow perihelion passages where $r_p \le 2 R_\odot$.  Therefore, we want to impose a maximum Jacobi constant (or energy) error for the simulations at $r=2 R_\odot$.  However, we also want to optimize $h$ such that passages near planets can be integrated accurately with the least overall CPU time.  A full discussion which values $h$ are used for the main set of simulations in this work will be deferred to Section \ref{sec:start_end}, after we discuss  close encounters with Jupiter in Section \ref{sec:planets}.

\subsection{Special Cases}\label{sec:special}

While we would like to use this adaptive time step integrator as much as possible, keeping the fictitious step $h$ fixed, there are two situations which must be handled separately.

\subsubsection{The Sun}\label{sec:sun}

The interior of the Sun has a potential that deviates strongly from Keplerian.  The integrator described in Section \ref{sec:int} works badly inside the Sun because it is designed for nearly Keplerian potentials. Thus, we replace the integration through the Sun by a map.  We exploit the fact that tidal forces from the planets are small near the Sun.  Since the two-body problem can be solved exactly, we can define a region about the Sun (called a ``bubble,'' with a typical radius of 0.1 AU) for which we use the exact solutions to the two-body problem.  In reality, we create a map for the bubble but only use it if the orbital perihelion lies within $r = 2 R_\odot$.  The bubble wall is larger than $2R_\odot$ so that a particle does not accidentally step into the Sun when stepping into the bubble.  In the WIMP orbital plane, we map the incoming position and velocity to the outgoing position and velocity using look-up tables for
\begin{multline}
	\Delta t( a, e) = \frac{2}{\sqrt{GM_\odot}} \\
	\times \, \int_{r_p (a,e)}^{r_b}\frac{\mathrm{d}r}{\sqrt{2[\pm \frac{1}{2a} - \tilde{\Phi}_\odot(r)] \pm a(1-e^2)/r^2}} \label{delta_t}
\end{multline}
and
\begin{multline}
	\Delta \phi (a, e) =  2 \sqrt{\pm a(e^2-1)} \\
	\times \, \int_{r_p (a,e)}^{r_b} \frac{ \mathrm{d}r}{r^2 \sqrt{ 2[\pm \frac{1}{2a} - \tilde{\Phi}_\odot(r) ] \pm a(1-e^2)/r^2}}, \label{delta_omega}
\end{multline}
which are the time $\Delta t$ and phase $\Delta \phi$ through which the particle passes in the bubble region.  By convention, $a$ is always positive, such that $E = GM_\odot/2a$ for hyperbolic orbits and $E = -GM_\odot / 2a$ for eccentric orbits.  The $+/-$ signs in Eqs. (\ref{delta_t}) and (\ref{delta_omega}) correspond to hyperbolic $( e>1)$ and elliptical orbits ($e < 1$) respectively.  We have normalized the solar potential $\tilde{\Phi}_\odot = \Phi_\odot/GM_\odot$.  Note that $r_b$ is the bubble radius and $r_p$ is the true perihelion, defined by
\begin{eqnarray}
	\frac{\mathrm{d}r}{\mathrm{d}t}\Bigg|_{r_p}  &=& 0 \\
						&=& \sqrt{2\left( \pm \frac{1}{2a} - \tilde{\Phi}_\odot(r_p) \right) \pm a(1- e^2) / r_p^2}.
\end{eqnarray}
We parameterize the look-up tables in terms of the semi-major axis and Keplerian perihelion $r_K = | a(1-e)|$.
\newline\indent
There is one subtlety in matching the  map through the bubble to the integrator outside of bubble.  In the Keplerian two-body problem, one solves the equations of motion $\mathrm{d} \mathbf{p}/ \mathrm{d} t$ and $\mathrm{d} \mathbf{q}/ \mathrm{d} t$ instead of $\mathrm{d} \mathbf{p}/ \mathrm{d} s$ and $\mathrm{d} \mathbf{q}/ \mathrm{d} s$.  If one divides $\mathrm{d}\mathbf{q}/\mathrm{d}s$ and $\text{d}\mathbf{p}/\text{d}\mathbf{s}$ by the differential equation for the time coordinate, the time-transformed equations of motion are
\begin{eqnarray}
	\frac{\mathrm{d}\mathbf{q}}{\mathrm{d}t} \Bigg|_{\Gamma} &=& \frac{ \mathrm{d} \mathbf{q}/ \mathrm{d} s}{ \mathrm{d}t / \mathrm{d} s} \\
	&=& \frac{f^\prime(T + p_0) \mathrm{d}T / \mathrm{d} \mathbf{p} }{f^\prime(T+p_0)} \\
	&=& \mathbf{p}
\end{eqnarray}
\begin{eqnarray}
	\frac{\mathrm{d}\mathbf{p}}{\mathrm{d}t} \Bigg|_{\Gamma} &=& \frac{ \mathrm{d} \mathbf{p}/ \mathrm{d} s}{ \mathrm{d}t / \mathrm{d} s} \\
		&=& - \frac{ f^\prime( -U) \partial U / \partial \mathbf{q}}{ f^\prime(T+p_0) } \\
		&=& -\frac{f^\prime(-U)}{f^\prime(T+p_0)} \frac{ \partial U}{\partial \mathbf{q}} \label{dpdt_mod}.
\end{eqnarray}
The second of these differs from the equations of motion of the original Hamiltonian $H$ by a multiplicative factor 
\begin{eqnarray}\label{mu}
\mu = f^\prime(-U)/f^\prime(T+p_0),  
\end{eqnarray}
in other words, the particle follows a Kepler orbit about a Sun of mass $\mu M_\odot$.  Therefore, we calculate the orbital elements using
\begin{eqnarray}
	a &=& \left| \frac{p_0}{2 \mu GM_\odot} \right| \\
	e &=& \sqrt{ 1 \pm J^2/(\mu GM_\odot a)},
\end{eqnarray} 
where the upper (lower) sign should be used for hyperbolic (elliptical) orbits.  We use a look-up table for $\Delta t$ and $\Delta \phi$ with the modification that $\Delta t$, as calculated for $a$ and $e$ with $\mu = 1$, must be multiplied by a factor of $\mu^{-1/2}$.  The change in phase is unaffected by the choice of central mass since it is a purely geometric quantity.
\newline\indent 
Similar lookup tables are also used to determine the perihelion $r_p$ and the optical depth as a function of semi-major axis and eccentricity.  We discuss additional scattering in the Sun in Appendix \ref{sec:tau}.

\begin{figure*}
\includegraphics[width=6.in]{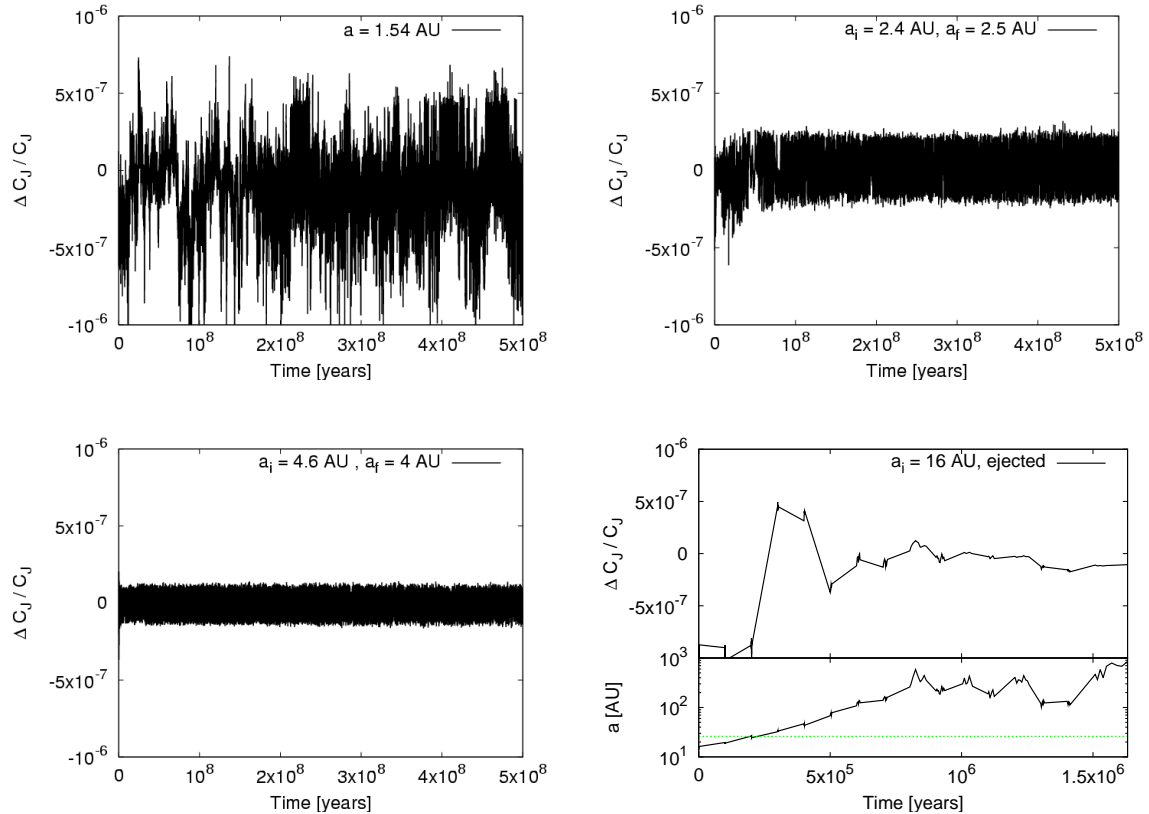}
\caption{\label{fig:cj}Error in the Jacobi constant as a function of time for several particles.  The Jacobi constant is recorded at aphelion at $10^5$ yr intervals.  \emph{Top left:} A particle with $a = 1.54$ AU.  This particle repeatedly goes through the Sun (about $10^7$ times), but never goes through the bubble around Jupiter.  It is integrated with $h=6\times 10^{-5} R^{-1}_\odot$ yr, which corresponds to $\approx 100$ steps/orbit.  \emph{Top right:}  A particle that gets stuck near a Sun-skimming 2:1 resonance with Jupiter.  This particle repeatedly goes through the Jupiter bubble.  It is integrated with $h = 2\times 10^{-5} R^{-1}_\odot$ yr, or $\approx 350$ steps/orbit.  \emph{Bottom left:} A particle gets stuck near a 3:2 resonance with Jupiter.  This orbit was integrated with $h = 1.5\times 10^{-5} R^{-1}_\odot$ yr, or $\approx 650$ steps/orbit.  \emph{Bottom right:} This particle repeatedly crosses $r_{c}$, the transition radius between barycentric and heliocentric coordinates (dashed line marks $r_c/2$, the crossing semi-major axis for an orbit with $e\sim 1$) and has its last aphelion before ejection from the solar system at $t=1.6 \times 10^6$ years.  It is integrated with $h = 2 \times 10^{-6} R^{-1}_\odot$ yr, or $9\times 10^3$ steps/orbit.}
\end{figure*}

We demonstrate the robustness of the map in the upper left panel of Fig. \ref{fig:cj}, where we show errors in the Jacobi constant over a 500 Myr time span for an orbit with $a\approx 1.54$ AU.  The orbit enters the Sun $\sim 10^7$ times in this time span.  We sample the orbit at the first aphelion after a $10^5$ yr interval from the previous sample, and there are approximately 100 steps/orbit.  This figure shows that there are only oscillatory errors throughout this long-term integration, and these fractional errors never exceed $10^{-6}$ at aphelion.  Long-term integrations of the two-body problem using the map demonstrate energy errors only at the level of machine precision.  

\subsubsection{The Planets}\label{sec:planets}

While the adaptive time step integrator works well in a near-Keplerian potential, one must treat close encounters with planets more carefully.  If the time step is too large near a planet, the particle fails to resolve the force from the planet.  This can cause growing errors in the particle's trajectory.  Since we use an $f(x)$ that is optimized to the potential of the Sun, the only way to achieve a small time step near each planet is to either make the fictitious time step $h$ small or to switch to a different integration method near each planet while using the method of Section \ref{sec:int} with a reasonably large $h$ for the rest of the orbit.  The advantage of the former approach is that it does not break the symplectic nature of the integrator.  However, it is also prohibitively computationally expensive.  Therefore, we use the latter approach.

We define a spherical region (``bubble'') about each planet for which we allow a different integration scheme, while continuing to use the adaptive time step symplectic integrator (Section \ref{sec:int}) outside the spheres.  The transition between the integration schemes is not symplectic, but reduce errors in the integration by enforcing an accuracy requirement on $|\Delta E / E| = |(p_0 + E)/E| = |(-H+E)/E|$ in the bubble of each planet. 

In the bubble of each planet, we continue to use the adaptive time step integrator, but the value of $h^\prime$ (the prime denotes the fact that this fictitious time step is \emph{only} used within a planet bubble) used in the bubble is selected to keep the quantity $|\Delta E/E|$ as small as possible while also keeping the total integration time short.  To find the optimal value of $h^\prime$, we use the following algorithm. When a particle first enters a bubble, we record the particle's energy error at the first step, $|\Delta E_i / E_i|$.  Then, we integrate the particle's trajectory through the bubble using the default value of $h$.  As the particle is about to exit the bubble, we calculate the energy error $\left| \Delta E_f / E_f \right|$.  If the energy error meets the accuracy criterion, or if it is less than $\left| \Delta E_i/E_i \right|$, then the integration is allowed to continue normally.  If, however, $\left| \Delta E_f/E_f \right| $ does not satisfy the accuracy criterion, we restart the integration in the bubble from the point at which the particle first entered with a smaller fictitious time step $h^\prime$.  This process iterates until either the energy accuracy condition is satisfied or the energy error plateaus in value.  If the energy error plateaus in value, whichever trajectory (corresponding to a particular choice of $h^\prime$) has the minimum $\left| \Delta E_f / E_f \right| $ is chosen.

The choice of the bubble size $l_{\text{\jupiter}}$ is related to the choice of fiducial value of $h$ and to the mass of the planet.  A larger value of $h$ means that the bubble needs to be larger to ensure that the planet's gravitational potential is properly resolved.  Planets with larger masses will require larger bubbles than smaller planets.  We choose to keep the bubble size fixed for all orbits.  In general, we tune $h$ so that the typical energy errors for all energies are similar near each planet, and to keep the error in the Jacobi constant small $|\Delta C_J/C_J| < 10^{-4}$ at $r=2R_\odot$.  The optimum sizes of the Jupiter bubble is $l_{\text{\jupiter}} \sim 1-3$ AU if we require that $\left| \Delta E_f / E_f \right| \sim 10^{-7} - 10^{-6}$.  

A complication arises when particles experience large changes in energy in their passage through the planetary bubble.  In this case, the value of $h$ that guaranteed a certain precision in $\left| \Delta E/E \right|$ in the pre-encounter orbit  may be either too large (for adequate precision) or too small (it will slow down the orbital integration).  Therefore, we change the value of $h$ at the next aphelion.  Again, this procedure breaks the symplectic nature of the integrator, but by changing $h$ at aphelion, our experiments show that we minimize errors.  In Section \ref{sec:start_end}, we outline how $h$ is chosen for the initial orbits, and how $h$ is changed if the particle experiences significant changes in energy from planetary encounters.

We demonstrate the performance of the bubble for the case of the 3-body problem in Fig. \ref{fig:cj}.  In this figure, the fractional error of the Jacobi constant is plotted against the time since the initial scatter in the Sun that produced a bound orbit, and we show the first $500$ Myr of the integrations.  The Jacobi constant is measured at the aphelion of the orbit at $10^5$ year intervals.  The trajectories of the particles in the upper right and bottom panels repeatedly pass through the bubble around Jupiter.  For these integrations, $l_{\text{\jupiter}} = 2.3$ AU, and the energy criterion was $ \left| \Delta E_f / E_f \right| < 2 \times 10^{-7}$.  There are no secular changes of the Jacobi constant with time.  Therefore, even though the planet bubble disrupts the symplecticity of the integrator, the integrator tracks the Hamiltonian well.

\subsection{Coordinate Choice}\label{sec:coords}

For most of the integration, we use a heliocentric coordinate system for both the particles and the planets.  There are two main reasons why we choose a heliocentric system.  First, it is much simpler to use heliocentric coordinates for passages through the Sun.  Secondly, consider the gravitational potential of the planets in the heliocentric frame,
\begin{eqnarray}
	\Phi(\mathbf{r})_{P} &=& \Phi_{d}(\mathbf{r}) + \Phi_{i}(\mathbf{r}) \\
		&=& - \sum_P \frac{GM_P}{|\mathbf{r} - \mathbf{r}_P|} + \sum_P \frac{GM_P \mathbf{r} \cdot \mathbf{r}_P}{x_P^3}, \label{planet_potential}
\end{eqnarray}
where the indirect term ($i$) arises from the fact that this coordinate system is not the center-of-mass coordinate system, and $d$ denotes the direct term.  For orbits that are well within a planet's orbit, the direct term can be expanded into spherical harmonics
\begin{eqnarray}
	\Phi_{d}(\mathbf{r}) &=& \sum_P \frac{GM_P}{|\mathbf{r} - \mathbf{r}_P|} \\
		&=& \sum_P \Bigg[ - \frac{GM_P}{r_P} - \frac{GM_P}{r_P^3}\mathbf{r}\cdot \mathbf{r}_P \\
	& &- \frac{GM_P}{r_P}\sum_{l=2}^\infty \left( \frac{r}{r_P} \right)^l P_{l} \left( \frac{ \mathbf{r} \cdot \mathbf{r}_P}{rr_P} \right) \Bigg] \label{eq:legendre},
\end{eqnarray} 
where the $P_{l}$ are Legendre polynomials.  The dipole term of the direct potential is canceled by the indirect potential.  Therefore, the primary contributor to the force on the particle by the planet comes from the $l = 2$ tidal term of the potential, whereas the $l=1$ term is dominant in the barycentric frame.  
\newline\indent
While there are distinct advantages to using the heliocentric frame, the indirect term of the potential dominates the potential at large distances from the Sun.  This poses a problem for the adaptive time step integrator, since the choice of $g = - GM_\odot / U = |\mathbf{r} - \mathbf{r}_\odot|$ is only optimal if the Keplerian solar potential is dominant.   Therefore, we choose to work in the barycentric frame at large distances.

In practice, this means switching between heliocentric and barycentric coordinate systems for long-period orbits.  We choose the crossover radius such that
\begin{eqnarray}
	\max \left| \Phi_{i,P} (r_c, \theta_P = 0) \right| = \epsilon \frac{GM_\odot}{r_c}, \label{eq:crossover_epsilon}
\end{eqnarray}
where $\theta_P$ is the angle between $\mathbf{r}$ and $\mathbf{r}_P$, $r_c$ is the crossover radius, the ``max'' signifies the planet for the planet $P$ for which the indirect potential is strongest, and $\epsilon$ is a factor $\lesssim 1$.  In our solar system, the planet for which the indirect potential is strongest is Jupiter.  The choice of $\epsilon \approx 0.1$ works well.  The crossover radius is thus
\begin{eqnarray}
	\frac{ M_{\jupiter} r_c}{a_{\jupiter}^2} = \epsilon \frac{M_\odot}{r_{c}},
\end{eqnarray}
or
\begin{eqnarray}
	r_{c} \approx \sqrt{ \epsilon M_\odot/M_{\jupiter}} a_{\jupiter}.
\end{eqnarray}
In changing coordinates, one breaks the symplectic flow of the integrator.  Therefore, one must treat the Hamiltonian, and therefore $p_0$, carefully at the crossover.  We choose to treat the transition the same way we treat the transition into the bubble about the Sun.  Namely, we calculate $\mu$ (Eq. \ref{mu}), the factor by which the gravitational potential is modified in the integrator (see Eq. \ref{dpdt_mod}), in the initial coordinate frame $i$.  Then we set
\begin{eqnarray}
	p_{0} \left|_{f} = - \mu_{i} E(\mathrm{r},t) \right|_f,
\end{eqnarray}
where quantities calculated in the final frame are denoted by $f$, and $E$ is the energy derived from the position and velocity coordinates of the particle.  While this transition is not symplectic, in practice it conserves the Jacobi constant to adequate precision.  This is demonstrated in the lower left panel in Figure \ref{fig:cj}, an orbit for which the initial semi-major axis is 50 AU.  In this integration, $\epsilon = 0.1$, which translates to $r_c = 53$ AU.

\section{Simulations}\label{sec:simulations}

\subsection{Dark Matter Model}\label{sec:dm_model}
In order to perform the orbit simulations, it is necessary to specify some dark matter properties.  The particle mass and elastic scattering cross sections completely determine scattering properties in the Sun, and hence, these are the only WIMP-dependent parameters necessary to run the simulations and find the WIMP distribution function at the Earth.  The particle physics model and parameter space within each model do not need to be specified for the simulations, although we assume that the dark matter particle is a neutralino when we estimate event rates in neutrino telescopes in Section \ref{sec:id}.  Thus, we use $m_\chi$ to denote the WIMP mass. 

The relative strengths of the spin-dependent and spin-independent elastic scattering cross sections are important in the context of scattering in both the Sun and the Earth.  For simplicity in interpreting the simulations, we would like to use either a spin-independent \emph{or} spin-dependent cross section, but not a mixture of the two.  We choose to focus on the spin-independent cross section for the simulations, but in Section \ref{sec:results_sd}, we show how to extend our results to the case of non-zero spin-dependent interactions.  In Section \ref{sec:start_end}, we discuss the specific choices for the WIMP mass and $\sigma_p^{SI}$ used in the simulations.

We adopt the Maxwellian distribution function (DF)
\begin{eqnarray}
	f_{h}(\mathbf{x},\mathbf{v}) = \frac{n_\chi}{(2\pi \sigma^2)} e^{ - \mathbf{v}^2 / 2\sigma^2} \label{eq:local_maxwell}
\end{eqnarray}
to describe the dark matter distribution function in the solar neighborhood in Galactocentric coordinates and far outside the gravitational sphere of influence of the Sun.  Here, $\sigma$ is the one-dimensional dark matter velocity dispersion, set to $\sigma = v_\odot / \sqrt{2}$. We set the speed of the Sun around the Galactic center to be $v_\odot = 220$ km/s, for which the observational uncertainty is about 10\% \cite{gunn1979,kerr1986}.  The WIMP number density is $n_\chi = \rho_\chi/ m_\chi$.  We assume that the dark matter density is smooth and time-independent in the neighborhood of the Sun, and that $\rho_\chi = 0.3$ GeV cm$^{-3}$.  Even if the dark matter were somewhat lumpy, the results of the simulations will still be valid if $\rho_\chi$ is interpreted as the average density in the solar neighborhood \citep{kamionkowski2008}.
\newline\indent
Transforming to the heliocentric frame via a velocity transformation $\mathbf{v}_s = \mathbf{v} - \mathbf{v}_\odot$,
\begin{eqnarray}
	f_{s}(\mathbf{x},\mathbf{v}_s)\mathrm{d}^3\mathbf{x}\mathrm{d}^3\mathbf{v}_s &=& f_{h}(\mathbf{x},\mathbf{v}_s + \mathbf{v}_\odot) \mathrm{d}^3\mathbf{x} \mathrm{d}^3 \mathbf{v}_s, \label{eq:local_maxwell_sun}
\end{eqnarray}
where the subscript $s$ refers to quantities measured in the heliocentric frame.  This distribution is anisotropic with respect to the plane of the solar system (the ecliptic).  The direction of the anisotropy with respect to the ecliptic depends on the phase of the Sun's orbit about the Galactic center.  In order to avoid choosing a specific direction for the anisotropy (i.e., to avoid choosing to start our simulations at a particular \emph{phase} of the Sun's motion about the Galactic center), we angle-average this anisotropic distribution function to obtain an isotropic DF of the form
\begin{eqnarray}
	\bar{f}_{s}(x,v_s) &=& \frac{1}{4 \pi}\int f_{s}(\mathbf{x},\mathbf{v}_s) d\Omega\\
	&=& \frac{1}{2 (2\pi)^{3/2}} \frac{n_\chi}{\sigma v_\odot v_s} \Big[ e^{-(v_s-v_\odot)^2/2\sigma^2} \nonumber \\
	& & \hskip2cm -e^{-(v_s+v_\odot)^2/2\sigma^2} \Big]. \label{eq:dfhaloaveraged}
\end{eqnarray}
Using the angle-averaged DF is approximately valid for two reasons:  (i) Scattering in the Sun is isotropic, so any bound WIMPs produced by elastic scattering will initially be isotropically distributed.  (ii) The time-averaged distribution function (averaged over the Sun's $\approx$ 200 Myr orbit about the Galactic center) only has a small anisotropic component \citep{gould1988}, a consequence of the large (60$^\text{o}$) inclination of the ecliptic pole with respect to the rotation axis of the Galaxy.  If the flux at the Earth is dominated by particles whose lifetime in the solar system is greater than the period of the Sun's motion about the Galactic center, the use of time-averaged distribution function is appropriate.  
\newline\indent
We use Liouville's theorem to find the halo DF for an arbitrary point in the Sun's potential well.  Neglecting the gravitational potential of the planets and the extremely rare interactions between dark matter particles, each particle's energy $E$ is conserved:
\begin{eqnarray}
	E &=& \frac{1}{2} v_s^2 \\
	  &=& \frac{1}{2} v^2 + \Phi_\odot (r),
\end{eqnarray}
where $v$ is the speed of particle with respect to and in the gravitational sphere of influence of the Sun, and $\Phi_\odot (r)$ is the gravitational potential of the Sun ($\Phi_\odot = - GM_\odot/r$ for $r > R_\odot$, where $R_\odot$ represents the surface of the Sun).  Thus, the DF within the Sun's potential well is
\begin{eqnarray}
	f(r,v) &=& \bar{f}_s(r,v_s(r,v)), \label{df}\\
	v_s(r,v) &=& \sqrt{ 2\Phi_\odot (r) + v^2 }.
\end{eqnarray}
An important consequence of this result is that the distribution function is identically zero for local velocities $v < \sqrt{-2\Phi_\odot (r)} = v_{esc} (r)$ below the escape velocity at that distance.

\subsection{Astrophysics Assumptions}

\emph{The Sun: }The Sun is modeled as spherical and non-rotating.  The gravitational potential and chemical composition are described by the BS(OP) solar model \citep{bahcall2005}.  We include the elements $^1$H, $^4$He, $^{12}$C, $^{14}$Ni, $^{16}$O, $^{20}$Ne, $^{24}$Mg, $^{28}$Si, and $^{56}$Fe in computing the elastic scattering rate.

We treat the Sun with the ``zero-temperature'' approximation (i.e., the solar nuclei are at rest) since the kinetic energy of nuclei in the Sun is much less than the kinetic energy of dark matter particles.  At the center of the Sun, the temperature is $T_c \sim 10^7$ K, so the average kinetic energy of a nucleus is of order
\begin{eqnarray}
	K_A &= \frac{3}{2} kT_c \\
	     &\sim 1 \hbox{ keV}.
\end{eqnarray}
In the cooler outer layers of the Sun, the nuclei have even less kinetic energy.  In contrast, the kinetic energy of dark matter particles in the Sun is of order 
\begin{eqnarray}
	K_\chi &\sim & m_\chi v^2_{esc} \\
		&\sim &10^3 \left( \frac{m_\chi}{100 \mathrm{~GeV}} \right) \mathrm{ keV}.
\end{eqnarray}
The rms velocity of the nuclear species $A$ is $\langle v^2_A \rangle ^{1/2} = \sqrt{ 2 K_A / m_A}\approx 500 (m_A/\mathrm{GeV})^{-1/2}$ km s$^{-1}$, lower than the $\sim 10^3$ km/s speed of dark matter particles.  Therefore, to good approximation, one can treat the baryonic species in the Sun as being at rest (i.e., having $T = 0$)

\emph{The Solar System: }The solar system is modeled as having only one planet, Jupiter, since Jupiter has the largest mass of any planet by a factor of 3.3 and therefore dominates gravitational scattering.  We place Jupiter on a circular orbit about the Sun, with a semi-major axis $a_{\jupiter} = 5.203$ AU, its current value, for the entire simulation, since its eccentricity is only $e \approx 0.05$ \citep{murray2000}, and the fractional variation in its semi-major axis is $\lesssim 10^{-9}$ over the lifetime of the solar system \citep{ito2002}.  Jupiter is modeled as having constant mass density to simplify calculations of particle trajectories.  This is not a realistic representation of Jupiter's actual mass density but only a small fraction of particles scattered by Jupiter actually penetrate the planet.  WIMP-baryon interactions in Jupiter are neglected since the optical depth of Jupiter is small enough that the probability of even one scatter occurring in the simulation is significantly less than unity.
\newline\indent
The orbit of Jupiter defines the reference plane for the simulation.  The Earth's orbit is assumed to be coplanar with the reference plane, since the actual relative inclination of the two orbits is currently only 1.3$^\circ$.

\subsection{Initial Conditions}\label{sec:ic_weak}

The goal of this section is to compute the rate of elastic scattering of halo WIMPs by baryons in the Sun onto bound orbits, as a function of the energy and angular momentum of particles after the scatter.  We also show how we use this to choose the initial starting positions and velocities of the particles.  There are two natural approaches to this problem: (i) Sample the dark matter flux through a shell a distance $R > R_\odot$ from the center of the Sun, treating scatter in the Sun probabilistically, and keeping only those particles which scatter onto Earth-crossing bound orbits.  (ii) Calculate the scattering probability in the Sun directly.  The second approach is more efficient, and this is the one described below.

The initial energy of a dark matter particle is
\begin{eqnarray}
	E &=& \frac{1}{2} m_\chi v^2 + \Phi_\odot(r)  \label{eq:E_ic}\\ 
	  &=& \frac{1}{2} m_\chi \left[v^2 - v_{esc}^2(r) \right], \label{E}
\end{eqnarray}
where we have expressed the gravitational potential in terms of the local escape velocity from the Sun.  The final energy of the dark matter particle is
\begin{eqnarray}
	E^{\prime} &=& E - Q \label{E_Q} \\
		   &=& \frac{1}{2} m_\chi \left[ v^{\prime 2} - v_{esc}^2 (r) \right],
\end{eqnarray}
where $Q$ is the energy transfer between the dark matter particle and the nucleus during the collision. The energy transfer can be expressed in terms of the center-of-mass scattering angle $\theta$ as (cf. Eq. \ref{eq:Q})
\begin{eqnarray}\label{Q}
Q(v,\cos \theta) = 2 \frac{\mu_A^2}{m_A} v^2 \left( \frac{1-\cos\theta}{2} \right),
\end{eqnarray}
where 
\begin{eqnarray}
\mu_A = \frac{m_A m_\chi}{ m_A + m_\chi}
\end{eqnarray}
is the reduced mass for a nuclear species of nucleon number $A$.  The maximum energy transfer $Q_{max} = 2 \mu_A^2 v^2/ m_A$ occurs if the dark matter particle is back-scattered, i.e., $\theta = \pi$.  Since we are interested in particles that scatter onto bound, Earth-crossing orbits,\footnote{In principle, particles scattered to bound orbits with $a < a_\oplus / 2$ could later evolve onto Earth-crossing orbits.  However, the torque from Jupiter is never high enough to pull a particle with $a < a_\oplus / 2$ onto an Earth-crossing orbit unless $((a_\oplus/2)-a)/a \lesssim 10^{-3}$.  Moreover, each additional scatter in the Sun reduces the energy of the orbit in the limit of a cold Sun, so the semi-major axis may only shrink.} the interesting range of outgoing energy is
\begin{eqnarray}
	-\frac{GM_\odot m_\chi}{ 2 ( 0.5 a_\oplus) } \le E^\prime \le 0,
\end{eqnarray}
where $a_\oplus$ is the semi-major axis of the Earth's orbit, with the lower bound determined by the fact that the aphelion of a highly eccentric orbit is $2a$.  
\newline\indent
For a given incoming energy $E$, it may not be kinematically possible to scatter into the full range of bound, Earth-crossing orbits.  In particular, if $E - Q_{max} = E_{min}^\prime > 0$, the particle cannot scatter onto a bound orbit at all.  Therefore, the lower bound on allowed outgoing energy is
\begin{eqnarray}
	E^{\prime}_{min} = \max \left( -\frac{GM_\odot m_\chi}{ 2 ( 0.5 a_\oplus) }, \min(E - Q_{max}(v),0) \right),
\end{eqnarray}
while the upper bound remains
\begin{eqnarray}
	E_{max}^\prime = 0.
\end{eqnarray}
Thus scattering rate of particles onto bound, Earth-crossing orbits is
\begin{multline}\label{dotN}
	\frac{\text{d}\dot{N}_\oplus}{\mathrm{d}r\mathrm{d}\Omega_r\mathrm{d}v\mathrm{d}Q} = 4\pi \sum_A r^2 n_A(r) v^3 \frac{\mathrm{d}\sigma_A}{\mathrm{d}Q} f(r, v) \\
	\times \, \Theta( R_\odot - r) \Theta[ - E^\prime] \\
	\times \, \Theta[ E^\prime - E^{\prime}_{min}],
\end{multline}
where we have imposed spherical symmetry on the Sun, $\Omega_r$ is the solid angle in the Sun, $f(x,v)$ is the distribution function in Eq. (\ref{df}), $\text{d}\sigma_A / \text{d}Q$ is the WIMP-nucleus cross section (Eq. \ref{dsdQ}), and $\Theta (x)$ is the step function.  Since
\begin{eqnarray}
	\mathrm{d} E^\prime = \mathrm{d} Q,
\end{eqnarray}
we can write Eq. (\ref{dotN}) as
\begin{multline}\label{dotN2}
	\frac{\text{d}\dot{N}_\oplus}{\mathrm{d}r\mathrm{d}\Omega_r\mathrm{d}v\mathrm{d}E^\prime} = 4\pi \sum_A r^2 n_A(r) v^3 \frac{\mathrm{d}\sigma_A}{\mathrm{d}Q} f(r, v) \\
	\times \, \Theta( R_\odot - r) \Theta[ - E^\prime] \\
	\times \, \Theta[ E^\prime - E^{\prime}_{min}],
\end{multline}
By sampling this distribution, we find the initial energy, speed, and scattering position vector of the WIMPs.

The outgoing energy is distributed uniformly unless there is kinematic suppression.  The kinematic suppression is most pronounced for large WIMP masses and very negative energies because, in order for a particle to scatter onto a bound orbit,
\begin{eqnarray}\label{vlim}
	v_s \leq 2 \frac{\sqrt{m_\chi m_A}}{m_\chi - m_A}v_{esc}(r)	
\end{eqnarray}
where $v_s$ is the particle velocity at infinity.  Heavy dark matter particles can only scatter onto bound orbits if their velocities at infinity are only a small fraction of the escape velocity from the Sun a distance $r$ from the Sun.  
For energies for which the kinematic suppression is minimal, we express the uniformity of $\mathrm{d}\dot{N}_\oplus/\mathrm{d}E^\prime$ in terms of the semi-major axis.  Since $E^\prime = -GM_\odot / 2a$ for particles on elliptical orbits, $\mathrm{d}\dot{N}_\oplus/\mathrm{d}a \propto a^{-2}$, or
\begin{eqnarray}\label{dlogNdotdloga}
	\frac{\mathrm{d} \log \dot{N}_\oplus}{\mathrm{d} \log a} = -1.
\end{eqnarray}
Therefore, most particles scatter onto relatively small orbits.
\newline\indent
The angular momentum of each scattered particle is in the range $J\in [0,rv^\prime]$, where $r$ is the radius from the center of the Sun at which the particle scatters.  To determine the distribution of magnitudes and directions for the angular momentum, we assume that the direction of the final velocity $\mathbf{v}^\prime$ is isotropically distributed with respect to the position vector $\mathbf{r}$.  If we specify $\theta_v$ to be the colatitude of the velocity vector with respect to the position vector, and the magnitude of the angular momentum is $J = rv^\prime \sin \theta_v$, then the distribution in angular momentum at fixed $\mathbf{r},\mbox{ }\mathbf{v}^\prime$ has the form
\begin{eqnarray}
	\mathrm{d}\dot{N}_\oplus \propto \mathrm{d}\cos\theta_v = \frac{\mathrm{d}{J^2}}{2r^2v^{\prime 2} \sqrt{1-J^2/(r^2 v^{\prime 2})}}. \label{eq:dNdJ}
\end{eqnarray}
The effect of kinematic suppression due to a large WIMP mass is that the particles that do scatter onto bound orbits can only do so close to the center of the Sun where $v_{esc}$ is highest.  This reduces the maximum angular momentum of the outgoing particle, and so eccentricity is an increasing function of WIMP mass.

By sampling Eq. (\ref{dotN2}), Eq. (\ref{eq:dNdJ}), and the azimuth of $\mathbf{v}^\prime$ with respect to the position vector, we fully specify the 6-dimensional initial conditions of the WIMPs, sampled to the same density as they would actually scatter in the Sun.

\subsection{Simulation Specifics}\label{sec:start_end}

The goals of the simulations are to predict the direct and indirect detection signals from particles bound to the solar system (relative to the signal from unbound particles) as a function of $m_\chi$ and the elastic scattering cross section.  The simulate orbits for a variety of WIMP parameters and then interpolate the results for other values of those parameters.

\begin{figure}
		\includegraphics[width=3.3in]{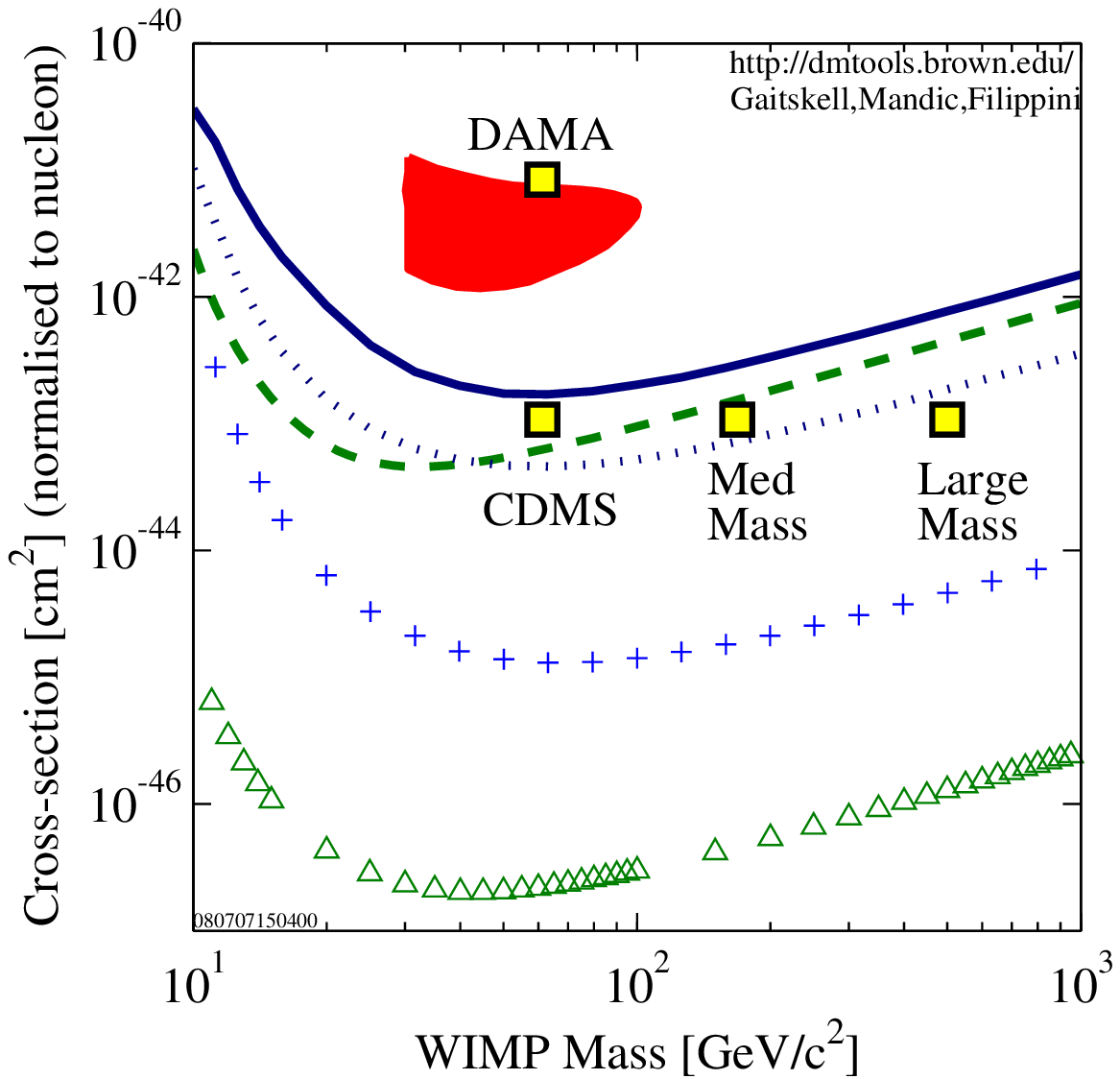} \\
		\includegraphics[width=3.3in]{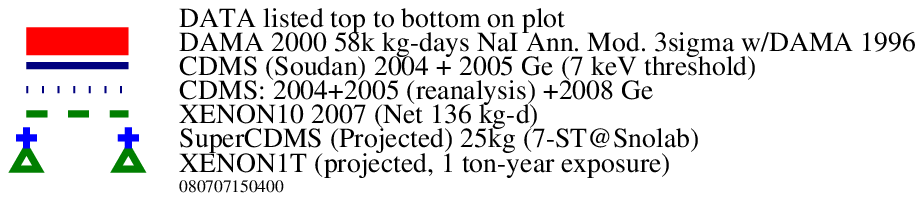}
		\caption{Points in the $\sigma_{p}^{SI} - m_\chi$ parameter space used for the solar capture simulations, plotted along with exclusion curves from recent experiments.  This plot was generated with\newline
http://dendera.berkeley.edu/plotter/entryform.html .}\label{fig:simpoints}
\end{figure}

We ran four sets of simulations with different choices of $m_\chi$ and $\sigma_p^{SI}$.  The first simulation, called ``DAMA'', used $m_\chi = 60$ atomic mass units (AMUs) and $\sigma_p^{SI} = 10^{-41}$ cm$^2$.  These parameters lie in the DAMA annual modulation region \citep{belli2000,bernabei2000a}.  A second simulation, called ``CDMS'', used the same WIMP mass as in the DAMA simulation but a cross section two orders of magnitude lower, $\sigma_p^{SI} = 10^{-43}$ cm$^2$, below the minimum of the CDMS 2006 exclusion curve (Fig. \ref{fig:simpoints}).  Two more simulations were chosen to have $\sigma_p^{SI} = 10^{-43}$ cm$^2$ but with larger WIMP masses.  The ``Medium Mass'' simulation uses $m_\chi = 150$ AMU, and the ``Large Mass'' simulation uses $m_\chi = 500$ AMU, selected to explore the dependence of the simulations on WIMP mass.  The choices for $m_\chi$ and $\sigma_p^{SI}$ are plotted in Fig. \ref{fig:simpoints} in addition to some recent direct detection results.  The details on the initial conditions of the simulations are summarized in Table \ref{tab:weakic}.

Since integrating the orbits of particles in the solar system is computationally expensive, it is more important to integrate just enough orbits to determine the approximate size of the bound DF relative to the unbound distribution, and to get a sense of which effects matter the most, than it is to get small error bars on the DF.  This principle guides our choices in the sizes of the ensembles of particles.

The number of particles simulated $N_p$ in each of the solar capture simulations is given in Table \ref{tab:weakic}.  We follow particles with semi-major axes slightly below the Earth-crossing threshold so that if the semi-major axis increases modestly during the simulation, the contribution to the Earth-crossing flux is included.  Fewer particles were simulated in the runs with lower cross section because the lifetimes were far longer than in the DAMA run.  

\begin{table}
	\caption{Solar capture simulations}\label{tab:weakic}
	\begin{center}
	\begin{ruledtabular}
	\begin{tabular}{l|c|c|c}

		Name &$m_\chi$ [AMU] &$\sigma_p^{\text{SI}}$ [cm$^2$] &$N_p$ [$a > 0.48$ AU] \\
		\hline

		DAMA &60 &$10^{-41}$ &1078586 \\
		CDMS &60 &$10^{-43}$ &145223 \\
		Medium Mass &150 &$10^{-43}$ &144145 \\
		Large Mass &500 &$10^{-43}$ &148173
	\end{tabular}
	\end{ruledtabular}
	\end{center}
	
\end{table}

\begin{figure*}[t]
	\includegraphics[width=6in]{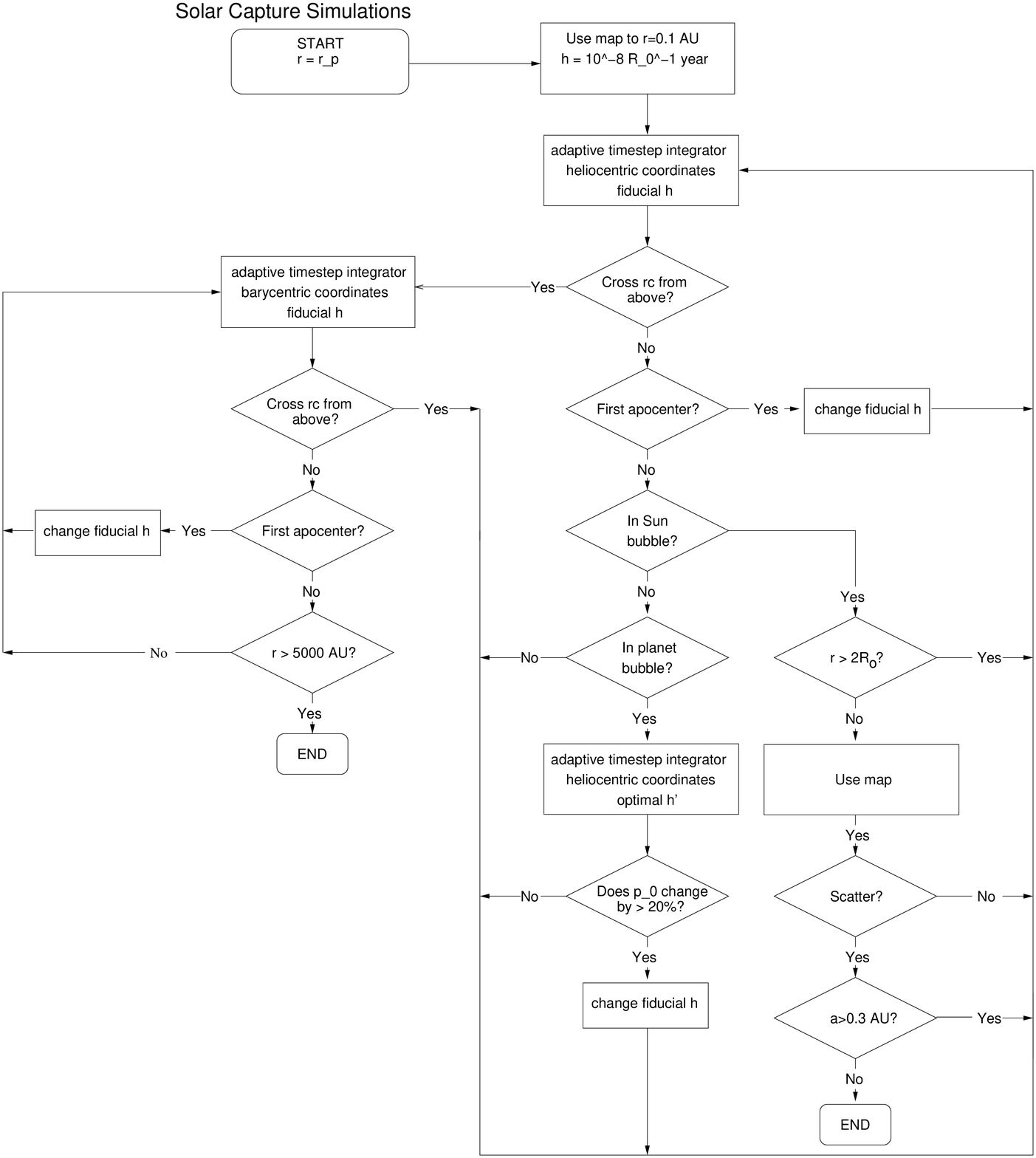}
	\caption{\label{fig:weak_flowchart}Flowchart for the simulation algorithm for the solar capture experiments.}
\end{figure*}

We use the flowchart in Fig. \ref{fig:weak_flowchart} to sketch the simulation algorithm.  There are six things that need to be set in order to run the simulations: starting conditions; the radius $r_c$ at which the heliocentric-barycentric coordinate change needs to be made; methods for initializing $h$ and potentially changing $h$ at later times; conditions for passing through and scattering in the Sun; the size of the bubble about Jupiter, $l_{\jupiter}$, and the accuracy criterion $|\Delta E/ E|$;  and conditions for terminating the simulation.  Following the flowchart, we discuss each point in turn.

\emph{Starting Conditions}
We sample the distribution of WIMPs initially scattered into the solar system according to Eqs. (\ref{dotN2}) and (\ref{eq:dNdJ}).  Once we have determined the initial position and velocity of each WIMP, we trace the WIMPs back to perihelion and start the integration there.  We follow all particles after the initial scatter, using the map to evolve the particles to the Sun bubble wall (0.1 AU) using map described in Section \ref{sec:sun}.  In order to account for the fact that particles may experience a second scatter before leaving the Sun for the first time, we perform a rescattering Monte Carlo when we construct the DFs.
\newline\indent
Once the particles have reached the bubble boundary, we initialize the adaptive time step symplectic integrator (Section \ref{sec:int}), setting $h = 10^{-8} R^{-1}_\odot$ yr and integrating the equations of motion in heliocentric coordinates.  With this choice of initial $h$, a particle with initial semi-major axis $a = 1$ AU will be integrated with $4.7\times 10^5$ steps/orbit, while a particle with $a = 100$ AU will be integrated with $4.7\times 10^6$ steps/orbit.  We choose such a small $h$ to minimize errors near perihelion, which is the point in the orbit at which errors are largest (Section \ref{sec:err}).  If the semi-major axis exceeds $r_c/2$, it may be necessary to change to barycentric coordinates before the particle reaches aphelion for the first time.

\emph{Coordinate Change}
For the weak scattering simulations, we set $\epsilon = 0.1$ (Eq. \ref{eq:crossover_epsilon}), thus setting the transition radius between the heliocentric and barycentric coordinated regimes to $r_c = 53$ AU.  This is large enough that only a small percentage of particles routinely cross the transition radius, but small enough that the heliocentric potential does not have too large a contribution from the indirect potential.

\emph{Setting $h$:}
After the particles reach their first aphelion, $h$ is reset according the values listed in Table \ref{tab:weakh}.  These values of $h$ are chosen such that both errors at perihelion ($|\Delta E / E| < 10^{-4}$) and near Jupiter ($|\Delta E / E| < 10^{-6}$) are small.  The combination of the values of $h$ and the Jupiter bubble radius $l_{\jupiter}$ (see below) were determined empirically.  We used slightly smaller values of $h$ for some semi-major axes in the CDMS, Medium Mass, and Large Mass runs compared to the DAMA run in order to check that the behavior of long-lived WIMPs was not an artifact of the choice of parameters.

\begin{table}
	\caption{The fictitious time step $h$ as a function of semi-major axis $a$ for the DAMA simulation and the simulations with $\sigma_p^{SI} = 10^{-43}\hbox{ cm}^2$ (CDMS, Medium Mass, Large Mass).}\label{tab:weakh}
	\begin{center}
	\begin{ruledtabular}
	\begin{tabular}{l|c|r}
		&DAMA	&$\sigma_p^{SI} = 10^{-43}$ cm$^2$ \\
	$a$ [AU]	&$h$ $[R_\odot^{-1}\hbox{ yr}]$	&$h$ $[R_\odot^{-1}\hbox{ yr}]$ \\
	\hline

	$a<0.75$	&$1\times 10^{-4}$	&$1\times 10^{-4}$ \\
	$0.75 \leq a < 1.1$	&$7\times 10^{-5}$ &$7\times 10^{-5}$ \\
	$1.1 \leq a < 1.6$	&$6\times 10^{-5}$	 &$6\times 10^{-5}$ \\
	$1.6 \leq a < 2.2$	&$5\times 10^{-5}$ 	& $2\times 10^{-5}$ \\
	$2.2 \leq a < 3.5$	&$4\times 10^{-5}$ 	&$2\times 10^{-5}$ \\
	$3.5 \leq a < 6.2$	&$3\times 10^{-5}$ 	&$1.5\times 10^{-5}$ \\
	$6.2 \leq a < 11$	&$2\times 10^{-5}$ 	&$1\times 10^{-5}$ \\
	$11 \leq a < 13$	&$9\times 10^{-6}$ 	&$2 \times 10^{-6}$ \\
	$13 \leq a < 22$	&$2\times 10^{-6}$ 	&$2 \times 10^{-6}$ \\
	$22 \leq a < 30$	&$2\times 10^{-6}$ 	&$2 \times 10^{-6}$ \\
	$30 \leq a < 45$	&$1\times 10^{-6}$ 	&$1\times 10^{-6}$ \\
	$45 \leq a < 120$	&$6\times 10^{-7}$ 	&$6\times 10^{-7}$ \\
	$120 \leq a < 200$	&$4\times 10^{-7}$ 	&$4\times 10^{-7}$ \\
	$200 \leq a < 500$	&$3\times 10^{-7}$ 	&$3\times 10^{-7}$ \\
	$a> 500$ or unbound	&$2\times 10^{-7}$	&$2\times 10^{-7}$ 

	\end{tabular}
	\end{ruledtabular}
	\end{center}
\end{table}

A particle's energy (and hence, semi-major axis) may change throughout the simulation.  If the energy changes by 20\% or more from when the particle enters the Jupiter bubble to when it exits, the particle is flagged to have $h$ adjusted at the next aphelion.  We do not readjust $h$ after every aphelion, or after each time the particle passes through the bubble, because very frequent changes in $h$ can induce growing numerical errors in the Jacobi constant.  We impose any changes in $h$ at aphelion instead of the bubble boundary, since we have determined experimentally that aphelion is the optimal point at which to change $h$. 

\emph{The Sun Bubble}
When a particle first crosses into the bubble about the Sun, we calculate its perihelion.  If the perihelion is smaller than $2R_\odot$, we proceed to map its current position and velocity to its position and velocity as it exits the bubble according to the prescription of Section \ref{sec:sun}.  If the perihelion lies within the Sun, we employ a Monte Carlo simulation of scattering in the Sun (Appendix \ref{sec:tau}).  The vast majority of the time, the particle does not scatter, and we simply use the map to move the particle to the edge of the bubble.  If the particle rescatters onto an orbit with semi-major axis $a < 0.3$ AU, we terminate the integration.  If the particle rescatters onto an orbit with $a > 0.3$ AU, we evolve the new orbit to the edge of the bubble, and then resume the adaptive time step symplectic integration.

\emph{The Jupiter Bubble}
For the DAMA, CDMS, and Medium Mass simulations, we set the Jupiter bubble boundary to be $l_{\jupiter} = 1.7$ AU, and the accuracy criterion to be $|\Delta E_f/E_f| < 10^{-6}$.  We adjusted this value for some particles in order to speed up the integration in cases where particles had generically slightly smaller initial $|\Delta E_i / E_i|$ than $|\Delta E_f / E_f|$, and took a longer time with $|\Delta E_f / E_f| = 10^{-6}$ to reach a sufficiently flat plateau in $|\Delta E_f / E_f|$ than with a slightly larger accuracy criterion.  We set $l_{\jupiter} = 3.4$ AU past 500 Myr for the Medium Mass simulation to determine if there were any effects of a larger $l_{\jupiter}$ on the orbits.  There were no effects on the DF resulting from this change.  For the Large Mass simulation, we experimented with a lower value of the accuracy criterion ( $|\Delta E_f/E_f| < 2\times 10^{-7}$ for the first 500 Myr, $|\Delta E_f/E_f| < 3\times 10^{-7}$ later) and a larger bubble, $l_{\jupiter} = 2.1$ AU.  The only effect the larger accuracy criterion had was to raise slightly the maximum energy error per orbit.

\emph{Stopping Conditions}
There are three reasons for terminating an integration.  If the particle crosses outward through the shell $r = 5000$ AU, the integration stops.  Particles crossing this shell will rarely cross the Earth's orbit again.  Secondly, if the particle rescatters in the Sun onto an orbit with $a < 0.3$ AU, we halt the integration since the particle will soon thermalize in the Sun and will not cross the Earth's orbit.  Thirdly, we stop the integration if the particle survives for a time $t_\odot = 4.5$ Gyr, approximately the lifetime of the solar system.

\subsection{Computing}\label{sec:sim_comp}
Simulations were performed on three Linux beowulf computing clusters at Princeton University.  Each set of simulations required $10^5$ CPU cycles on 3 GHz dual core processors.

\section{Distribution Functions}\label{sec:df}
Before presenting the results of the simulations, we define some terms that will be used frequently in this section.  The  ``heliocentric frame'' describes an inertial frame moving with the Sun.  ``Heliocentric speeds'' will refer to speeds relative to the Sun, measured at the Earth assuming the Earth has zero mass.  The ``geocentric frame'' refers to an inertial frame moving with the Earth.  Unless otherwise noted, geocentric WIMP speeds are those outside the potential well of the Earth.  The ``free space'' distribution function, Eq. (\ref{eq:dfhaloaveraged}), is the angle-averaged halo distribution function in an inertial frame moving with the Sun, outside of the gravitational sphere of influence of the Sun.  The ``unbound'' distribution function refers to the Liouville transformation of the free space distribution function to the position of the Earth (Eq. \ref{df}), including the effects of the gravitational field of the Sun but not the Earth.  

\begin{figure*}
	\begin{tabular}{ll}
	(a) &(b) \\
	\includegraphics[width=3in,angle=270]{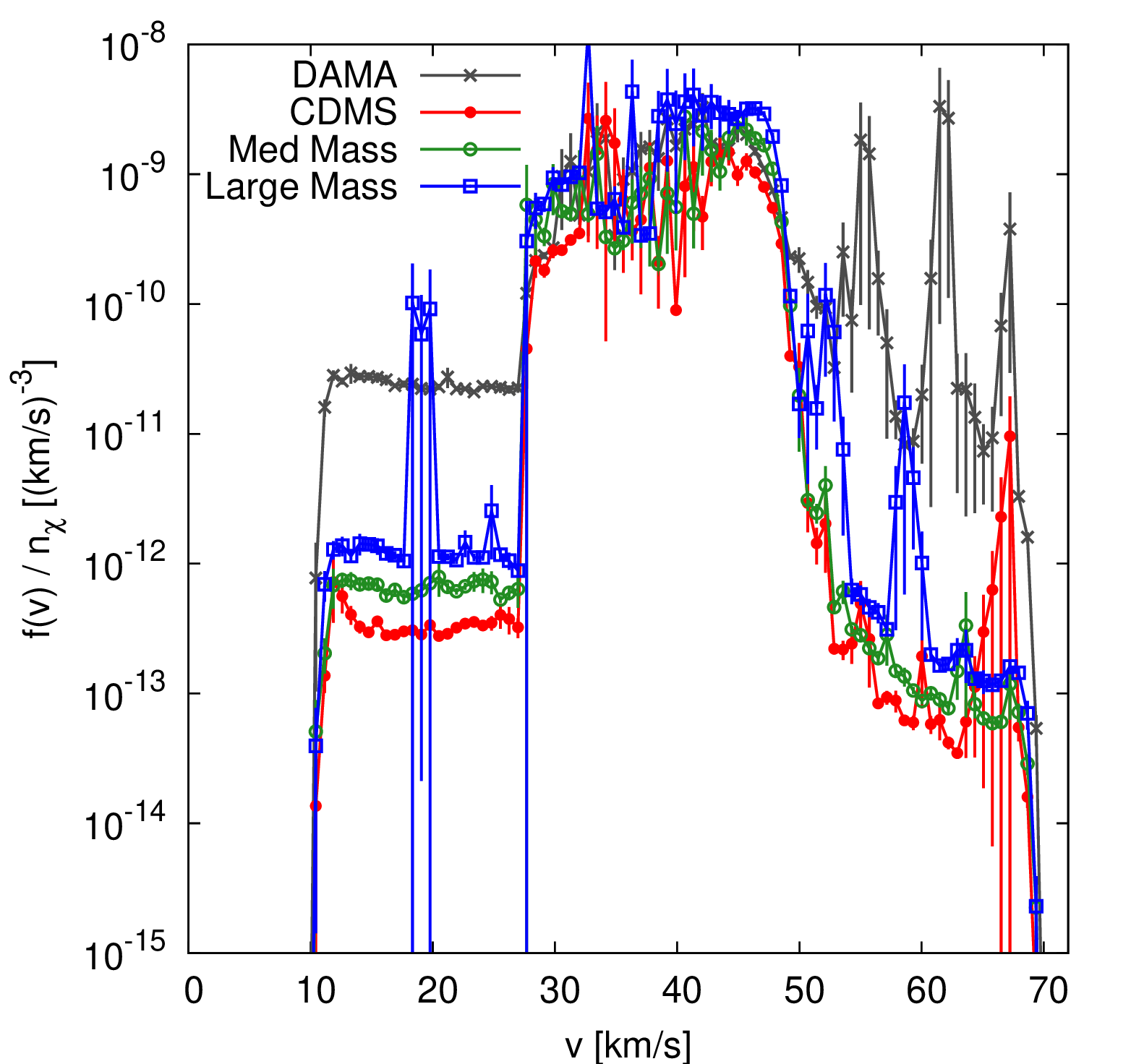} &\includegraphics[width=3in,angle=270]{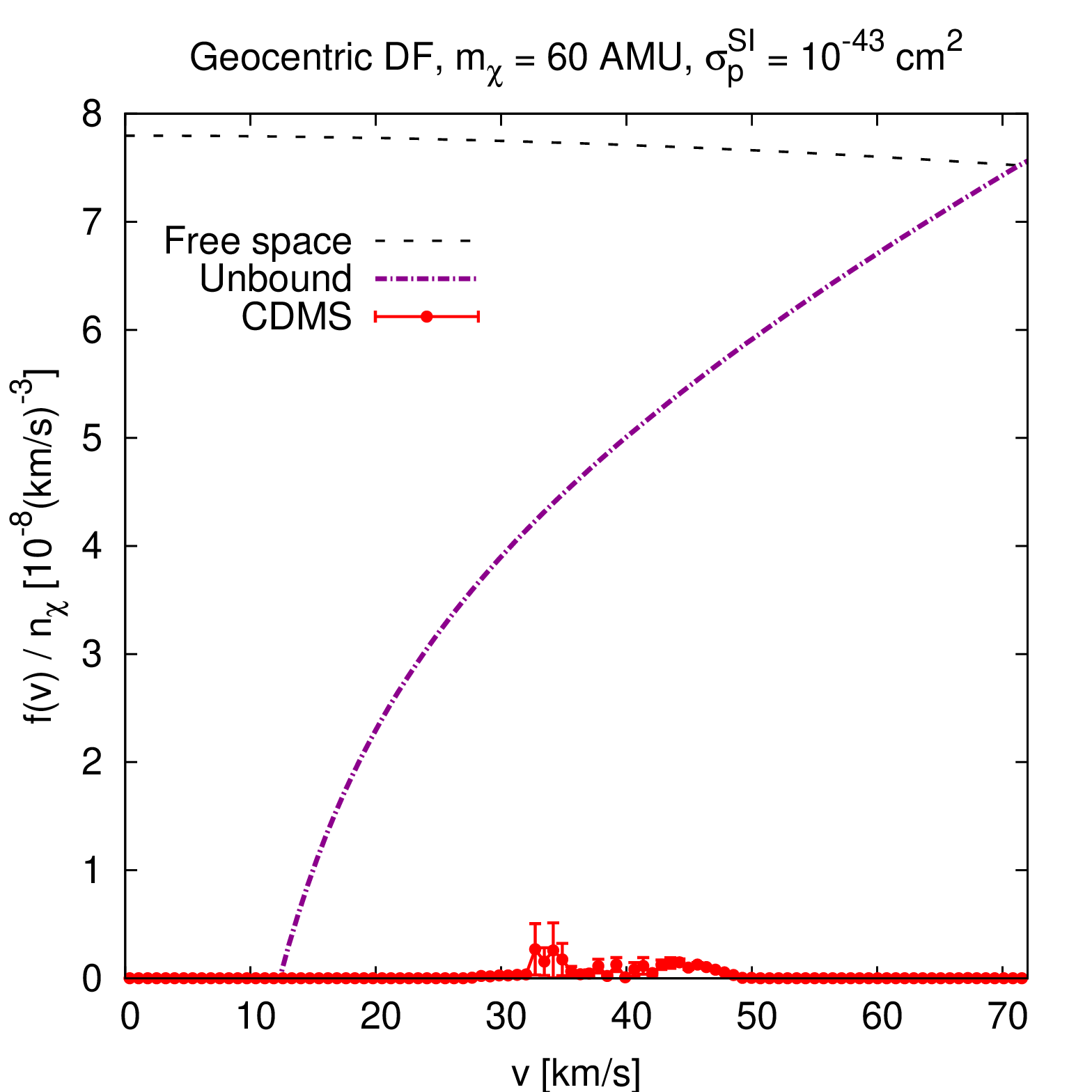}
	\end{tabular}
	\caption{\label{fig:weak_sim_df}Geocentric distribution functions from the simulations.  (a) Results from all simulations.  (b) The CDMS distribution function relative to theoretical distribution functions for unbound WIMPs.}
\end{figure*}

In Fig. \ref{fig:weak_sim_df}, we present the one-dimensional geocentric DFs divided by the halo WIMP number density $n_\chi$ (defined in Section \ref{sec:dm_model}) for each simulation.  These DFs have already been integrated over angles, and are normalized such that the bound dark matter density $n_{\chi, bound} = \int \text{d} v v^2 f(v)$, where $f(v) = \int \text{d} \Omega f(\mathbf{v})$.  We plot the DFs in Fig. \ref{fig:weak_sim_df}(a) on a logarithmic scale in order to highlight the structures, while we plot the CDMS simulation (Table \ref{tab:weakic}) DF on a linear scale in Fig. \ref{fig:weak_sim_df}(b) to compare the simulation results with theoretical DFs.  The DFs are based on $(1-4) \times 10^8$ passages of particles within a height $|z| < z_c = 10 R_\oplus$ of the Earth's orbit, and estimated using the technique described in Appendix \ref{sec:df_estimator}. The DFs are insensitive to $z_c$, at least in the range $1 \lesssim z_c \lesssim 25 R_\oplus$.  Error bars are based on 500 bootstrap resamplings of the initial scattered particle distributions for each simulation.  

The most striking feature of the DFs is the smallness with respect to the DF of halo WIMPs unbound to the solar system.  This is in stark contrast to the prediction of \citet{damour1999}.  In order to show why this is the case, we examine the simulations in detail.  In particular, we (i) identify the features in the DF with specific types of orbits, (ii) find the lifetime distribution of such orbits, and (iii) determine what sets the lifetime of those orbits (e.g., ejection vs. rescattering and thermalization in the Sun).  With these data, we may also determine how the DF varies with WIMP mass and cross section, and estimate the maximum DF consistent with limits on the spin-independent cross section.

\begin{figure*}
	\begin{tabular}{ll}
	(a) &(b) \\
	\includegraphics[width=2.5in,angle=270]{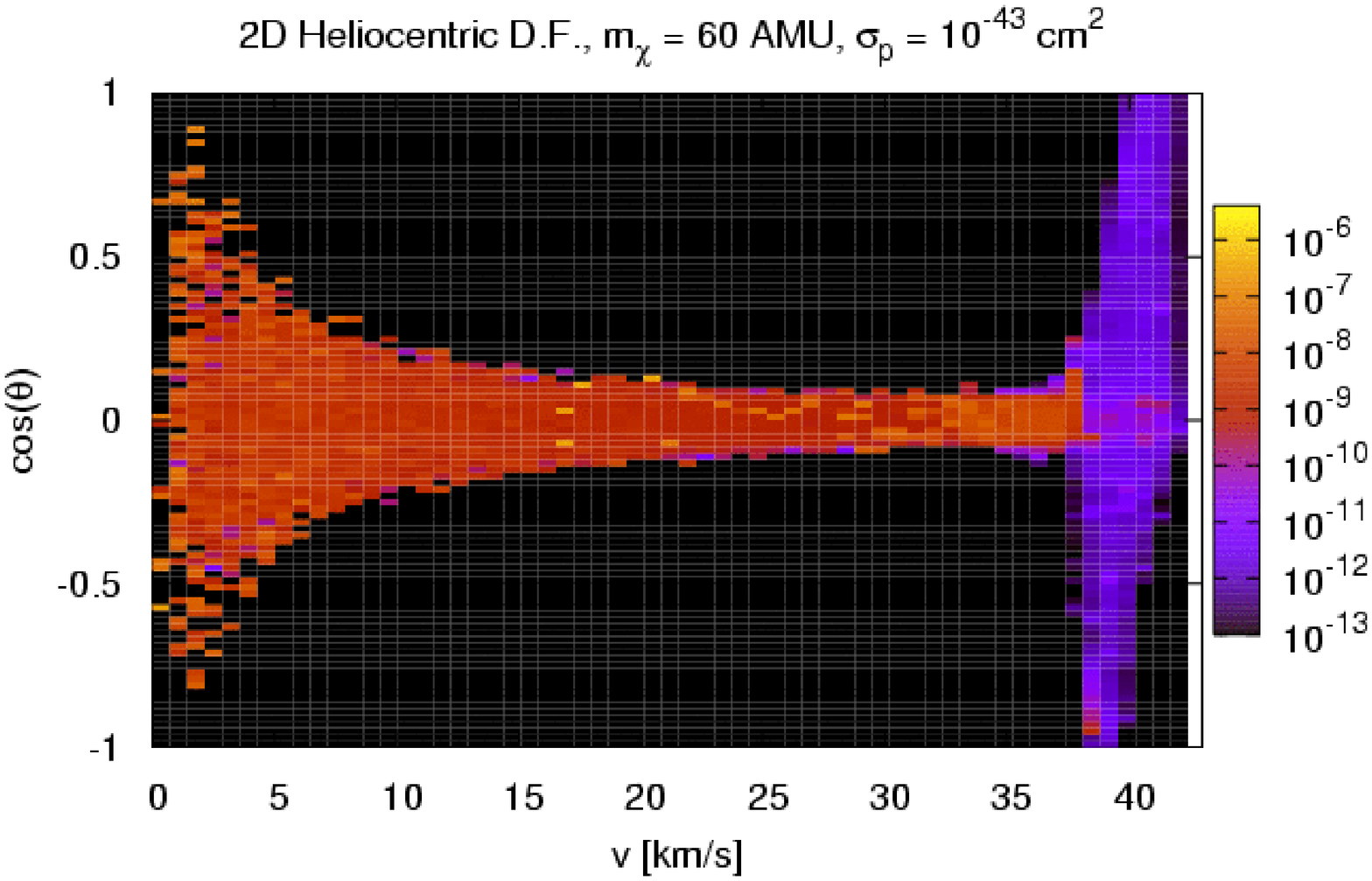} &\includegraphics[width=2.5in,angle=270]{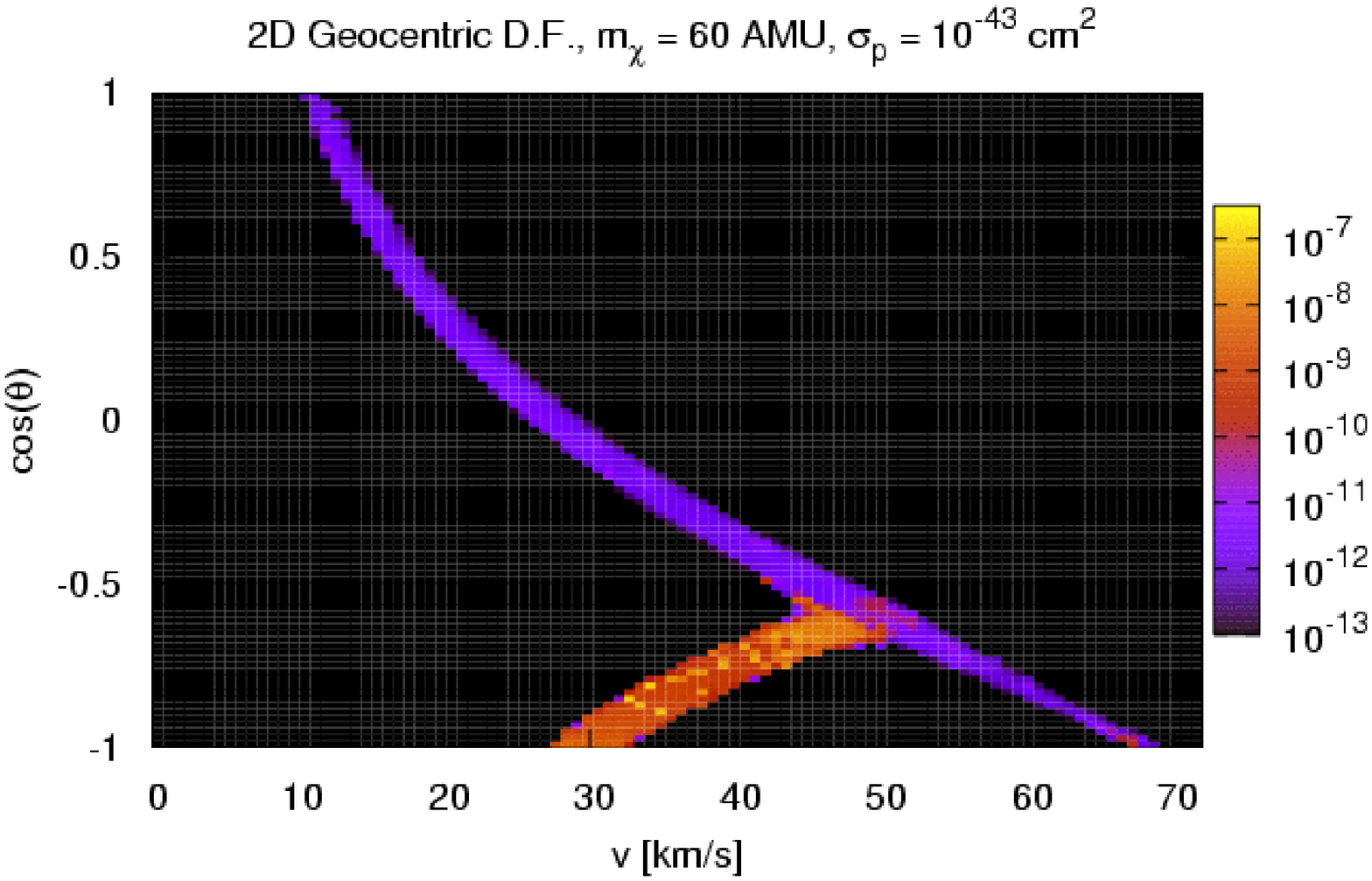}
	\end{tabular}
	\caption{\label{fig:2d}Distribution functions divided by $n_\chi$ in the $v-\cos\theta$ plane (integrated over $\phi$) for both (a) heliocentric and (b) geocentric frames.  These DFs come from the CDMS simulation, and the units are $( \text{km s}^{-1} )^{-3}$ }
\end{figure*}

The DFs from the four simulations show similar morphologies, although the normalization of the features differs.  The most prominent feature in all four DFs in Fig. \ref{fig:weak_sim_df} is the ``high plateau'' between $27 < v < 48 \hbox{ km s}^{-1}$.  In order to identify which orbits contribute to this plateau, it is useful to examine the two-dimensional DF.  In Fig. \ref{fig:2d}, we show the two-dimensional DF $f(v,\cos\theta) = \int \text{d} \phi f(v,\cos\theta,\phi)$ for the CDMS simulation (Table \ref{tab:weakic}) in both (a) heliocentric and (b) geocentric coordinates.  The angle between the velocity vector and the direction of the Earth's motion is $\theta$, while $\phi$ is an azimuthal angle, with $\phi = 0$ corresponding to the direction of the north ecliptic pole if $\theta = \pi / 2$.  The DFs are plotted on a logarithmic scale to highlight structure.  We only show the CDMS simulation results in this figure since the phase space structure of the DF is virtually the same in all simulations.

From Fig. \ref{fig:2d}(b), we identify the short arc between $27 < v < 50 \hbox{ km s}^{-1}$ below $\cos\theta < -0.5$ with the high plateau, although there is a small contribution from the other, longer arc.  We find that the short arc in the geocentric DF corresponds to the trumpet-shaped feature in the heliocentric DF below $v < 38 \hbox{ km s}^{-1}$.  For bound orbits, the heliocentric speed at $r = 1 \hbox{ AU}$ is
\begin{eqnarray}
	v(a) = \left[2-\left( \frac{a}{a_\oplus} \right)^{-1} \right]^{1/2}v_\oplus,
\end{eqnarray}
where $v_\oplus$ is the orbital speed of the Earth.  The heliocentric speed $v = 38 \hbox{ km s}^{-1}$ corresponds to the lowest Jupiter-crossing orbit, so the trumpet feature in the two-dimensional heliocentric DF corresponds to orbits that do not cross Jupiter's orbit.  

\begin{figure*}
	\begin{center}
	\begin{tabular}{ccc}
	(a) &\includegraphics[width=3.1in]{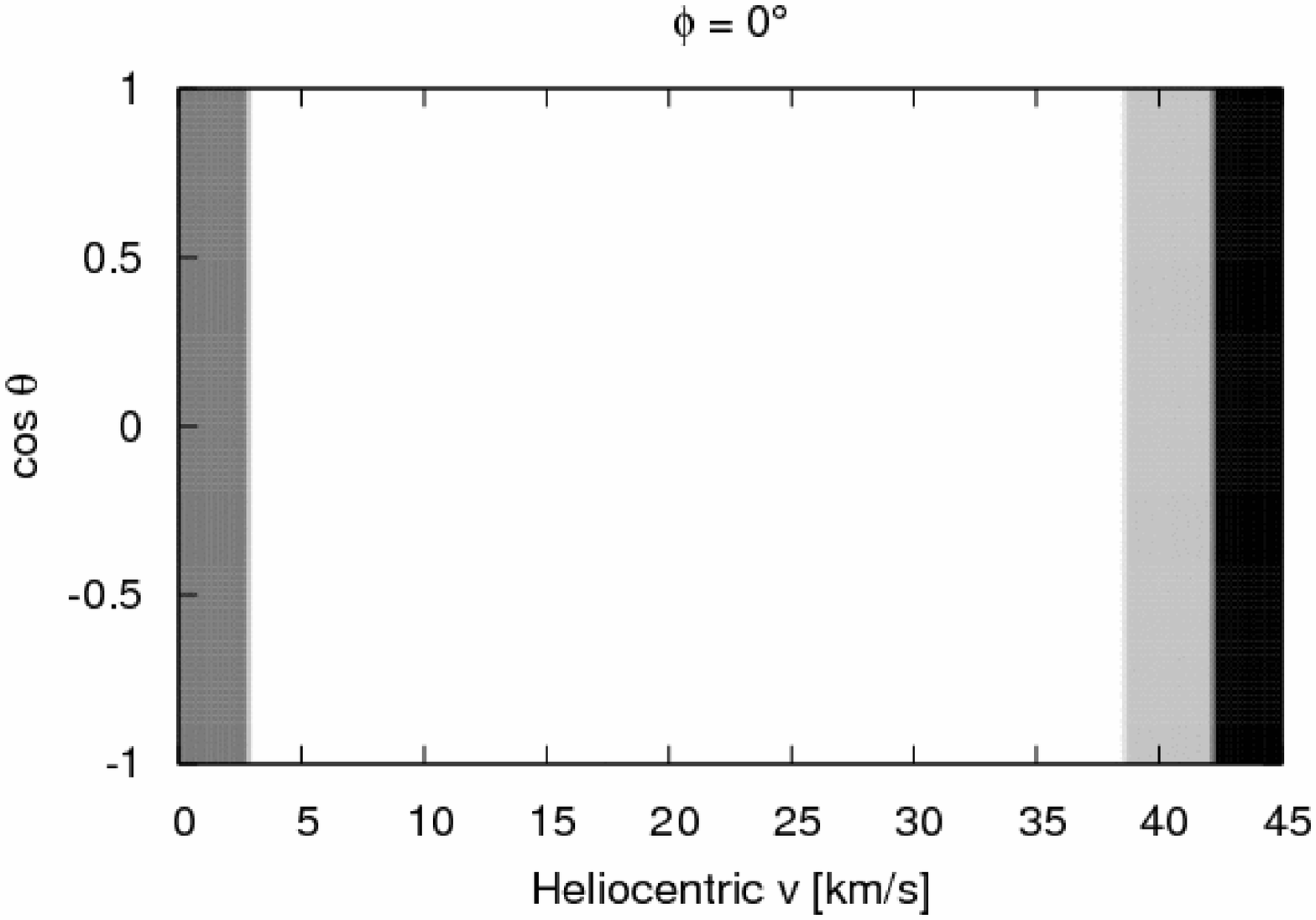}(b) &\includegraphics[width=3.1in]{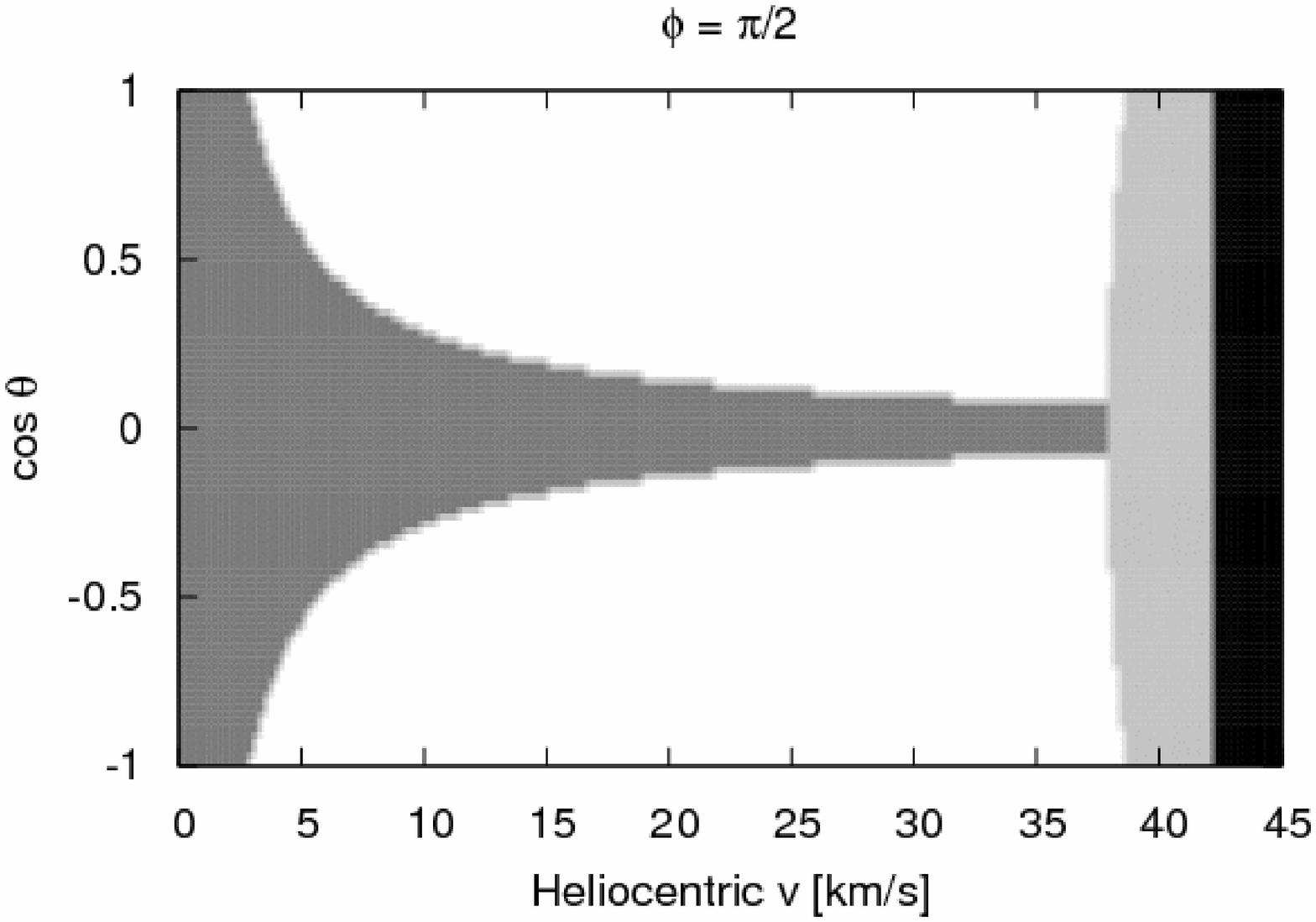} \\
	(c) &\includegraphics[width=3.1in]{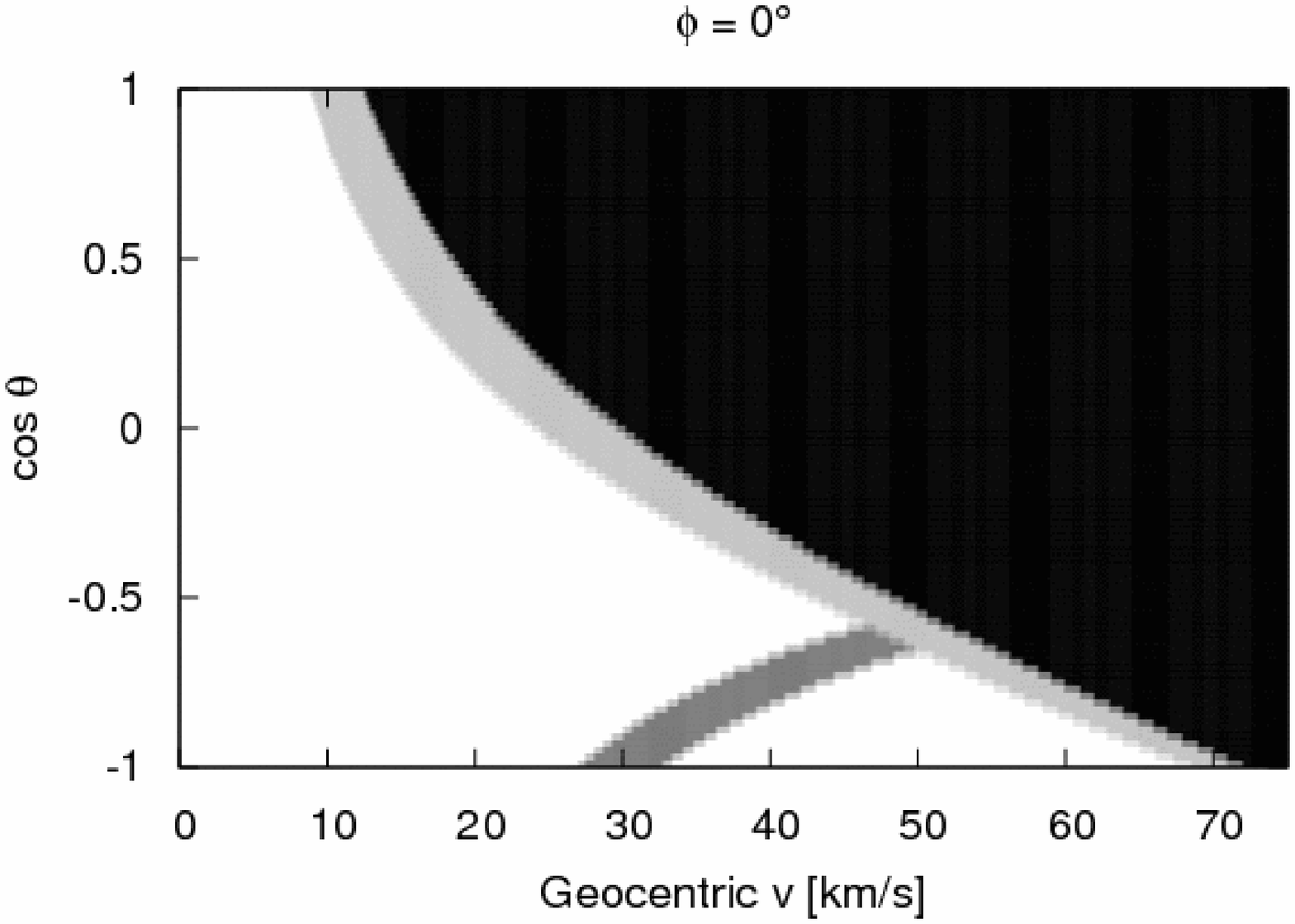}(d) &\includegraphics[width=3.1in]{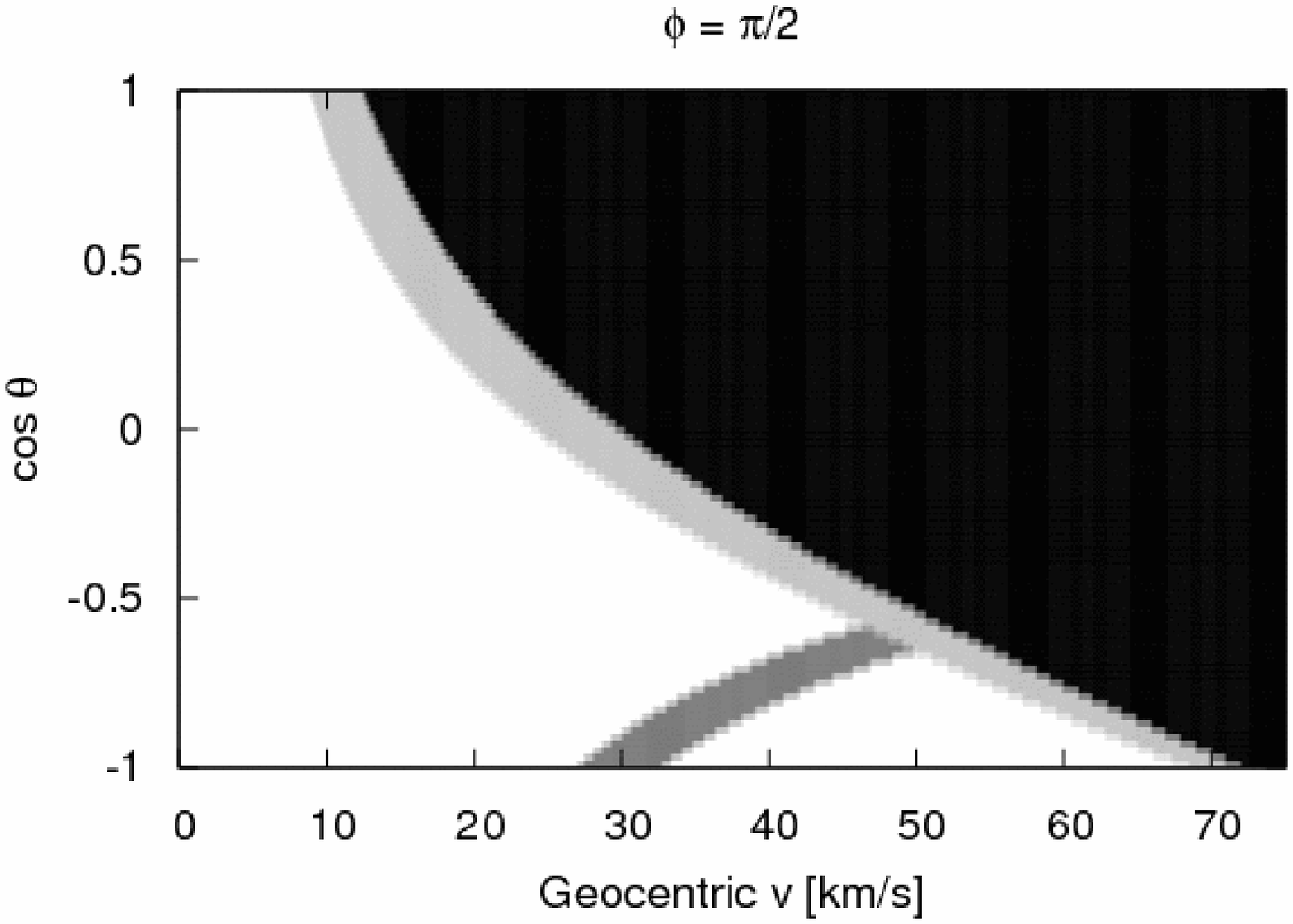}\\
	\end{tabular}
	\end{center}
	\caption{\label{fig:halfpi}Locations of various types of orbits in the (a) $\phi = 0$ and (b) $\phi = \pi / 2$ slices of heliocentric velocity space, and (c) $\phi = 0$ and (d) $\phi = \pi / 2$ slices of geocentric velocity space.}
\end{figure*}

The trumpet shape of the heliocentric DF (and the narrow band in the geocentric DF) in Fig. \ref{fig:2d} can be simply explained.  In Fig. \ref{fig:halfpi}, we calculate the energy and angular momentum for each point in velocity space.  The black region of velocity space represents unbound orbits.  All points for which orbits are bound and have perihelia inside the Sun are marked in dark grey, while orbits that are bound and cross Jupiter's orbit are marked in light grey.  The white regions correspond to bound orbits that neither enter the Sun nor cross Jupiter's path.  If we were to integrate the $\phi$-slices shown if Fig. \ref{fig:halfpi}, we would find that the region in Fig. \ref{fig:halfpi} corresponding to Sun-penetrating orbits that do not cross Jupiter exactly matches the parts of phase space we identified with the high plateau.

Fig. \ref{fig:halfpi} was computed for a system without planets.  Once Jupiter is added, another type of orbit that may exist is a bound orbit for which $J_z$ is fixed but $J$ is not.  An example of this type of orbit is a Kozai cycle.  In this case, $J_z = a_\oplus v \cos \theta$ in the heliocentric frame.  In the special case that $\phi = \pi / 2$, $J = J_z$.  Therefore, the parts of $(v,\cos\theta)$ phase space in the $\phi = \pi / 2$ plane corresponding to Sun-penetrating orbits also cover orbits with $J_z$ fixed by the initial scatter in the Sun for other values of $\phi$.  Thus, the high plateau in the 1-dimensional geocentric DF is built up by WIMPs with $a < a_{\jupiter}/2$ and small $J_z$ but not necessarily small $J$.

The second feature of the distribution functions in Fig. \ref{fig:weak_sim_df} is the ``low plateau.''  This is the relatively flat part of the distribution function that extends from $\approx 10$ km s$^{-1}$ to $\approx 70$ km s$^{-1}$.  This feature corresponds in the long arc in the two-dimension DF in Fig. \ref{fig:2d} and the stripe between $38 < v < 42 \hbox{ km s}^{-1}$ in the heliocentric DF.  From Fig. \ref{fig:halfpi}, we identify this feature with bound, Jupiter-crossing orbits.  Small gaps exist in the heliocentric DF with $v > 40 \hbox{ km s}^{-1}$ and $\cos\theta < 0$ and $38 < v < 40 \hbox{ km s}^{-1}$ and $\cos\theta > 0$ because these regions of phase space are inaccessible to WIMPs initially scattered in the Sun in the restricted three-body problem.  This translates to a truncation of the low plateau at the extrema in geocentric speed.

The third common set of features in the one-dimensional geocentric distribution function are spikes in the low plateau.  These spikes result from the long-lifetime tail of Jupiter-crossing or nearly Jupiter-crossing particles that spend significant time near mean-motion resonances or on Kozai cycles.  The minimum semi-major axis for these spikes corresponds to the 3:1 resonance, $a \approx 2.5 $ AU.  Long lifetime tails due to ``resonance-sticking'' orbits have also been observed in simulations of comets \cite{duncan1997,malyshkin1999}.  The error bars on the spikes are large due to the small numbers of long-lived resonance-sticking particles in each simulation.  There is also some Poisson noise in the height of the spikes between simulations due to the small number long-lived WIMPs in each simulation.

\begin{figure}
	\begin{center}
		\includegraphics[width=3.2in]{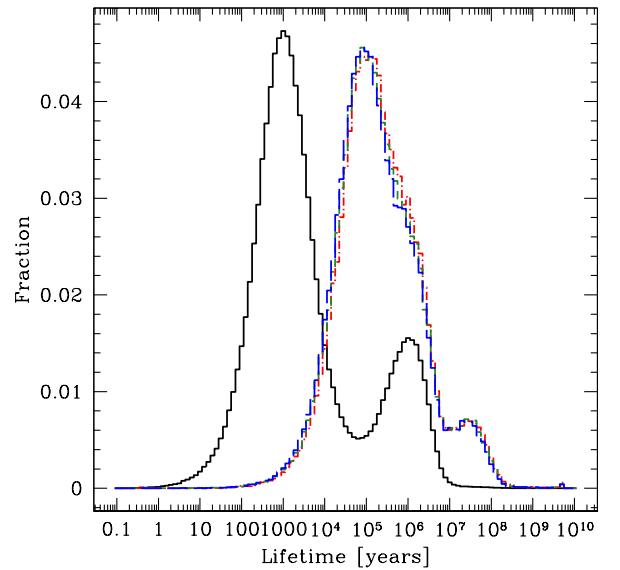}
	\end{center}
	\caption{\label{fig:weak_lifetimes}Particle lifetime distributions for the DAMA (\emph{solid}), CDMS (\emph{dot-dashes}), Medium Mass (\emph{short dashes}), and Large Mass (\emph{long dashes}) simulations.}
\end{figure}

Next, we show the lifetime distribution of bound WIMPs and demonstrate which mechanisms (thermalization or ejection) terminates the contribution of orbits to the DF.  In Fig. \ref{fig:weak_lifetimes}, we present the lifetime distributions for all WIMPs in the DAMA, CDMS, Medium Mass and Large Mass simulations.  There are several notable features in this plot.  First, and most striking, many of the bound particles survive for very long times---up to $10^6 - 10^8$ yr.  However, in none of the simulations is there a large population of particles that survive for times of order the age of the solar system, although there is a small population that does (the notch at $4.5$ Gyr in Fig. \ref{fig:weak_lifetimes}).  Secondly, the lifetime distribution functions of the CDMS, Medium Mass, and Large Mass runs are very similar.  However, these lifetime distributions are quite different from that of the DAMA simulation.  The differences in the lifetime distributions are due almost entirely to the elastic scattering cross section, at least for the range of WIMP masses we consider.

\begin{figure*}
	\begin{center}
		\begin{tabular}{cc}
		 \includegraphics[width=2.8in]{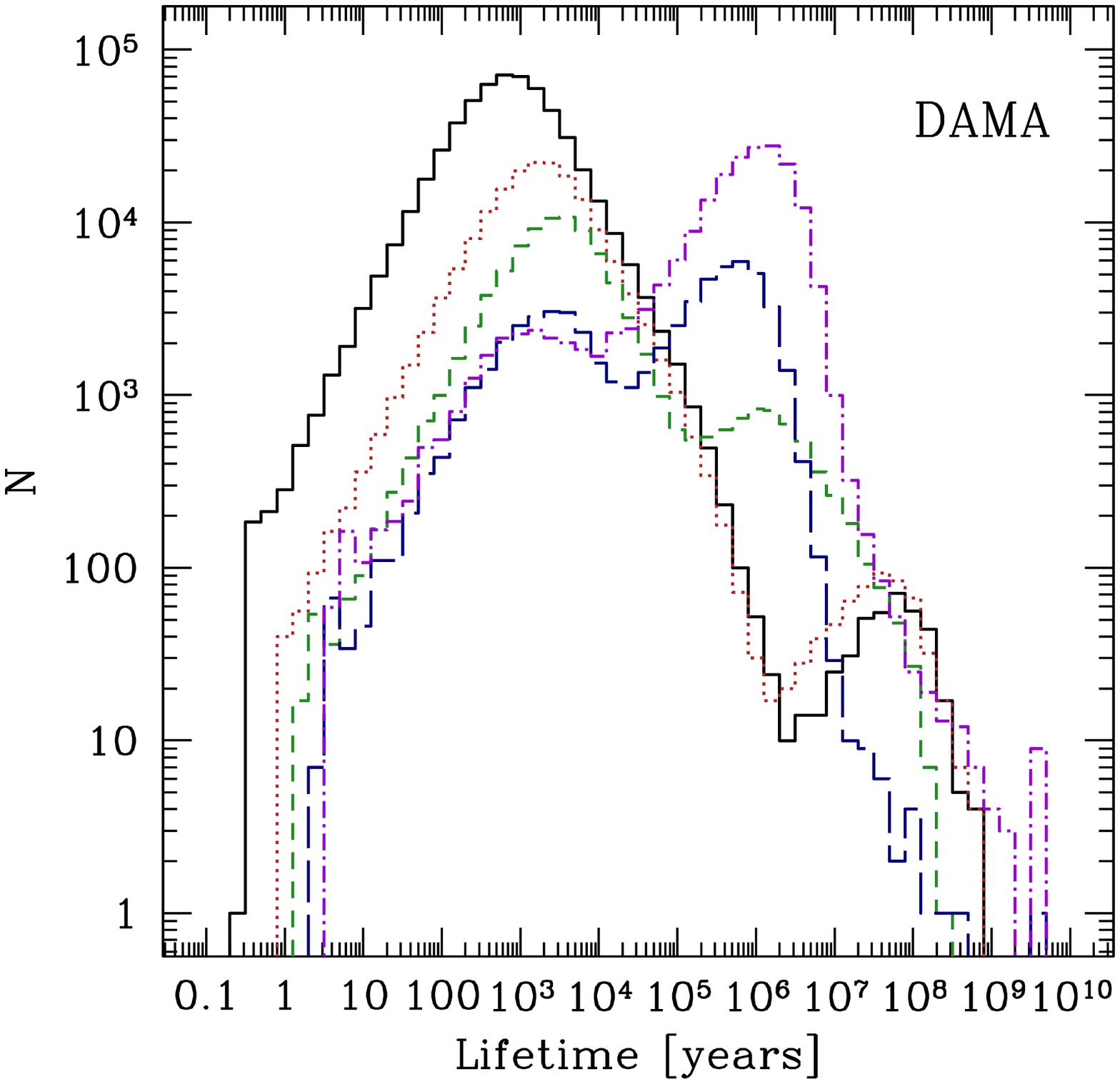} &\includegraphics[width=2.8in]{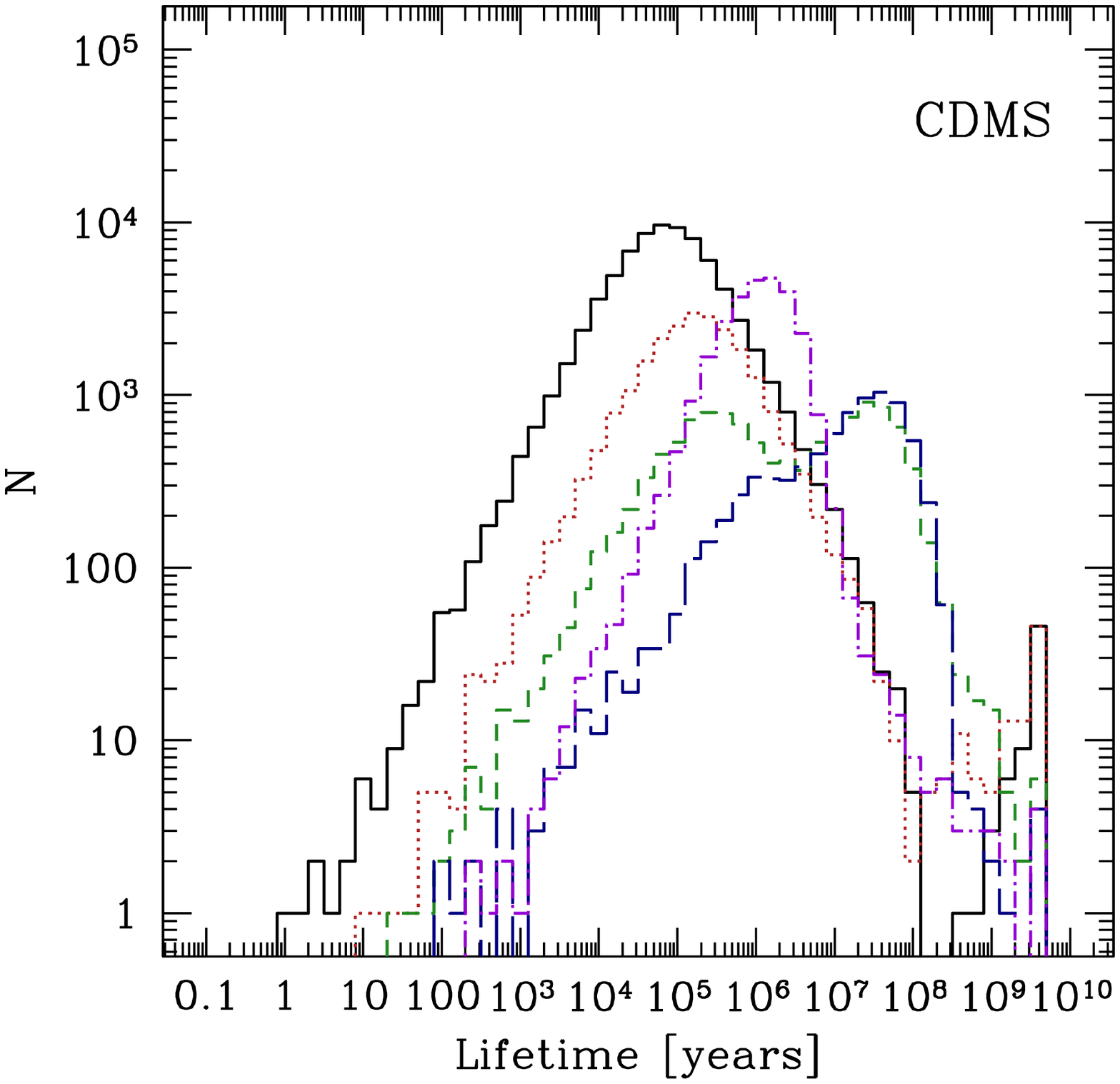} \\
		 \includegraphics[width=2.8in]{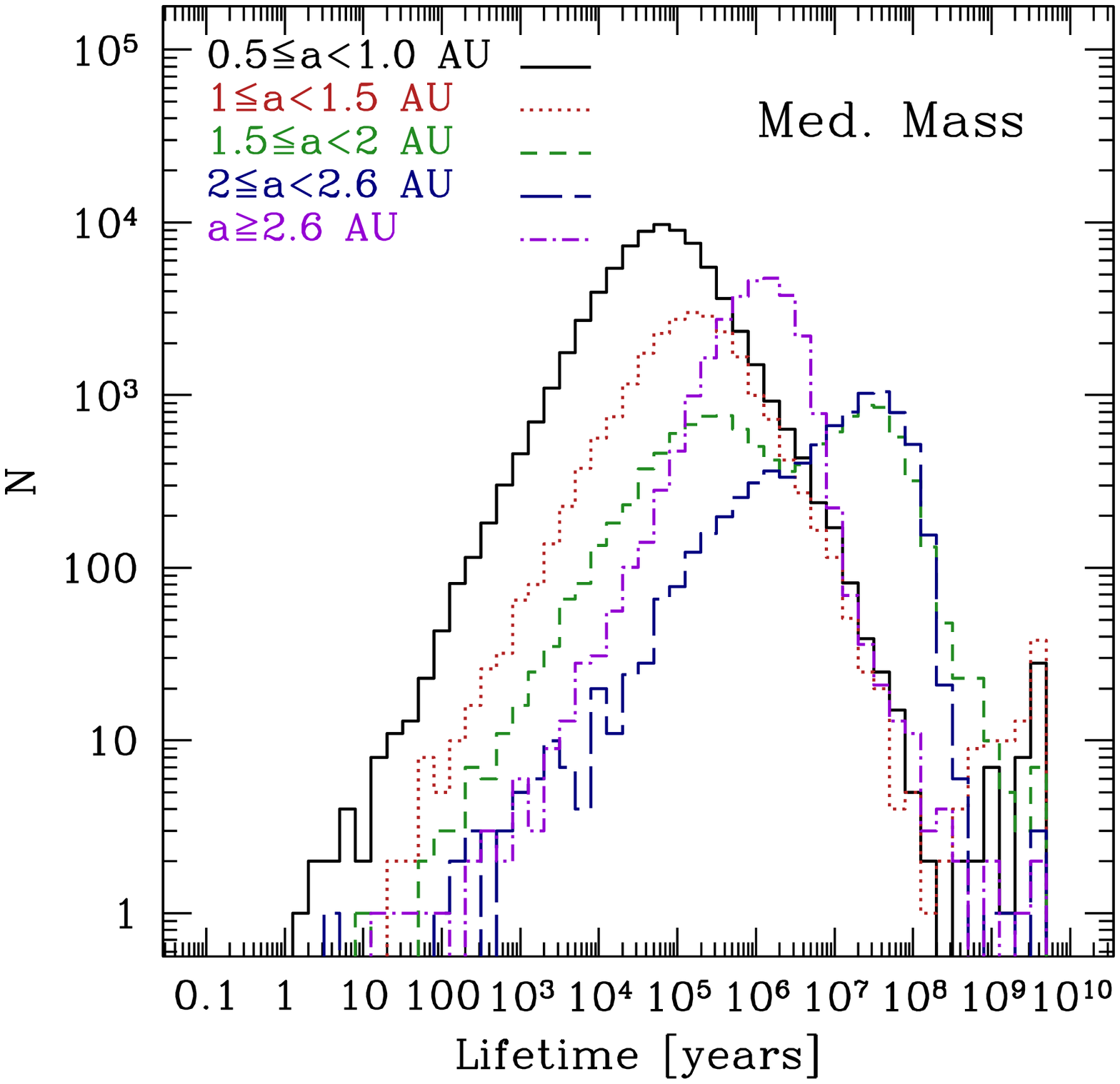} &\includegraphics[width=2.8in]{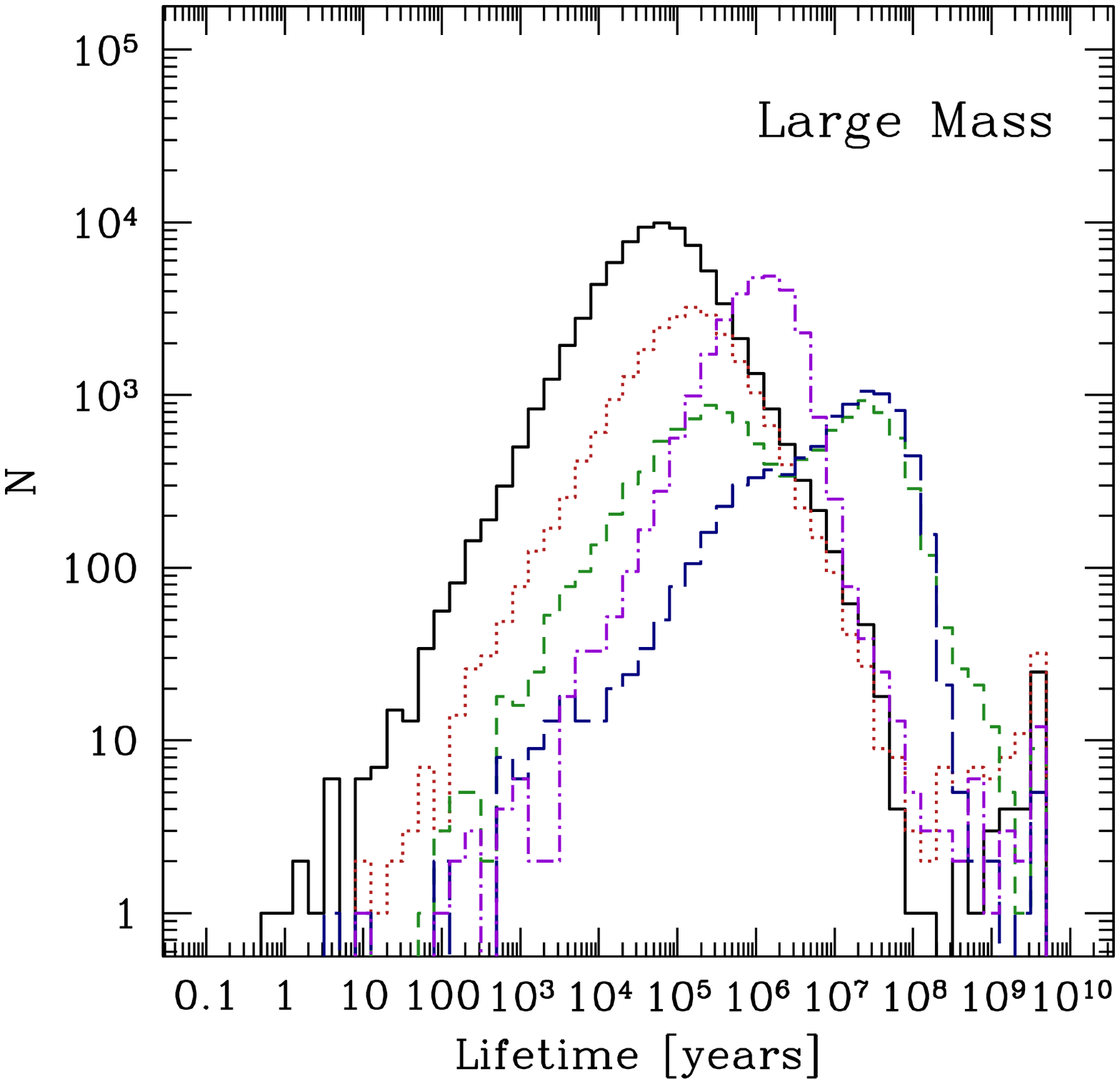}
		\end{tabular}
	\end{center}
	\caption{\label{fig:weak_tdist}Lifetime distributions as a function of initial semi-major axis.}
\end{figure*}

In order to both explain these features in the lifetime and phase space distribution functions, it is useful to examine the lifetime distributions as a function of the initial semi-major axis $a_i$, as shown in Fig. \ref{fig:weak_tdist}.  

The largest feature in Fig. \ref{fig:weak_lifetimes} is the strong peak at $t_{l} \sim 10^3$ yr for the DAMA simulation and $t_{l} \sim 10^5$ yr for the simulations with $\sigma_p^{SI} = 10^{-43}$ cm$^2$, which we call the ``rescattering peak.'' It encompasses the majority of particles in each simulation.  This feature results from WIMPs that rescatter in the Sun before they are ejected from the solar system by Jupiter or precess onto orbits that do not intersect the Sun.  This rescattering peak is offset between DAMA and the other simulations because the lifetime is inversely proportional to the scattering probability in the Sun, $t_{l} \propto \sigma_p^{-1}$.  

There is one important difference in the composition of the rescattering peak between the DAMA and other simulations.  In Fig. \ref{fig:weak_tdist}, we show that particles on Jupiter-crossing orbits exhibit a rescattering feature in the DAMA simulation but not in the other simulations.  Indeed, about 23\% of Jupiter-crossing particles in the DAMA simulation are rescattered in the Sun, while $<2\%$ are rescattered in the other simulations.  This is because the timescale on which Jupiter can perturb the perihelia of Jupiter-crossing orbits out of the Sun is significantly shorter than rescattering timescales for the $\sigma_p^{SI} = 10^{-43}$ cm$^2$ simulations, but the two timescales become closer at higher cross sections.

Another feature occurs at $t_{l} \sim 10^6$ yr, which we call the ``ejection peak.''  This feature occurs at the same time for each simulation, and from Fig. \ref{fig:weak_tdist} we see that this arises from Jupiter-crossing orbits.   The median time at which this feature occurs is independent of $\sigma_p^{SI}$ since ejection, not rescattering, sets the lifetime of these WIMPs.  The slope of the Jupiter-crossing lifetime distribution changes near $\sim 10$ Myr for all simulations.  WIMPs that have $t_l > 10$ Myr are resonance-sticking, and their lifetime distribution goes as $N(t) \propto t^{-\alpha}$, where $\alpha$ is slightly less than one.  

A third feature, called the ``quasi-Kozai peak,'' is centered at $t_{l} \sim 10^6$ yr in the DAMA simulation, and $t_{l} \sim 10^8$ yr in the other simulations.  The feature is seen in the $1.5 \text{ AU}\le a_i < 2$ AU and $2 \text{ AU} \le a_i < a_{\jupiter} / 2$ bins of Fig. \ref{fig:weak_tdist}.  The WIMPs in the quasi-Kozai peak are not on true Kozai cycles because of significant interactions with mean-motion resonances.  In the simulations, particles in the quasi-Kozai peak are observed to alternate between rescattering peak-type orbits with perihelion well inside the Sun, and orbits that look like Kozai cycles.  Both the semi-major axis and $J_z$ vary with time; neither is conserved although the combination giving the Jacobi constant $C_J$ is fixed (Eq. \ref{eq:ch3_cj}).  There are some orbits at the low end of the semi-major axis range $1.5 \text{ AU} < a \le a_{\jupiter}/2$ for which $a$ and $J_z$ are conserved and the Kozai description is accurate.

The median lifetime of WIMPs on quasi-Kozai cycles is well-described by $t_{l}/P_\chi \approx 300 / \tau$, where $P_\chi$ is the WIMP orbital period and $\tau$ is the optical depth through the center of the Sun.  This implies that particles are eventually removed from Earth-crossing orbits by rescattering in the Sun.  The height of the rescattering peak relative to the quasi-Kozai peak is greater in the DAMA simulation than the other simulations because the optical depth in the Sun is large enough that particles originating deep within the Sun rescatter before the torque from Jupiter can pull the perihelion towards the surface of the Sun.  

The fourth feature is not obvious in Fig. \ref{fig:weak_lifetimes}, but is once the lifetime distribution is displayed on logarithmic scales in Fig. \ref{fig:weak_tdist}.  This feature is the ``Kozai peak.''  This peak is located at about $t_{l} \sim 10^8$ yr for the DAMA simulation, and near $t_{l} \sim t_\odot = 4.5$ Gyr for the other simulations.  This feature results from particles whose orbital evolution can be described by Kozai cycles ($a$, $J_z$ conserved), of which we see both circulating and librating populations.  For the CDMS, Medium Mass, and Large Mass simulations, the peak is at $t_\odot$ because that is the time at which we terminate the simulations.  Particles on these orbits have $a_i < 1.5$ AU, and originate in the outer $r \gtrsim 0.5 R_\odot$ in the Sun.  They constitute only a small fraction ($\sim 0.1\%$) of all orbits with $ a_i < 1.5$ AU, but dominate the lifetime distribution of particles with lifetimes $\gtrsim 10^8$ yr.  The median lifetime of particles on Kozai cycles depends on the WIMP-nucleon cross section in the form $t_l / P_\chi \approx 10^5 / \tau$, where again $\tau$ is the optical depth through the very center of the Sun ($\tau \sim 10^{-3}$ for DAMA, $\tau \sim 10^{-5}$ for the other simulations).

Now that we have identified the types and lifetimes of bound WIMP orbits, we see how these come together to build up the phase space DF as a function of WIMP mass and cross section.

\subsection{The Distribution Function as a Function of $\sigma_p^{SI}$}

\begin{figure}
	\begin{center}
		\begin{tabular}{cc}
		(a) &\includegraphics[width=2.4in,angle=270]{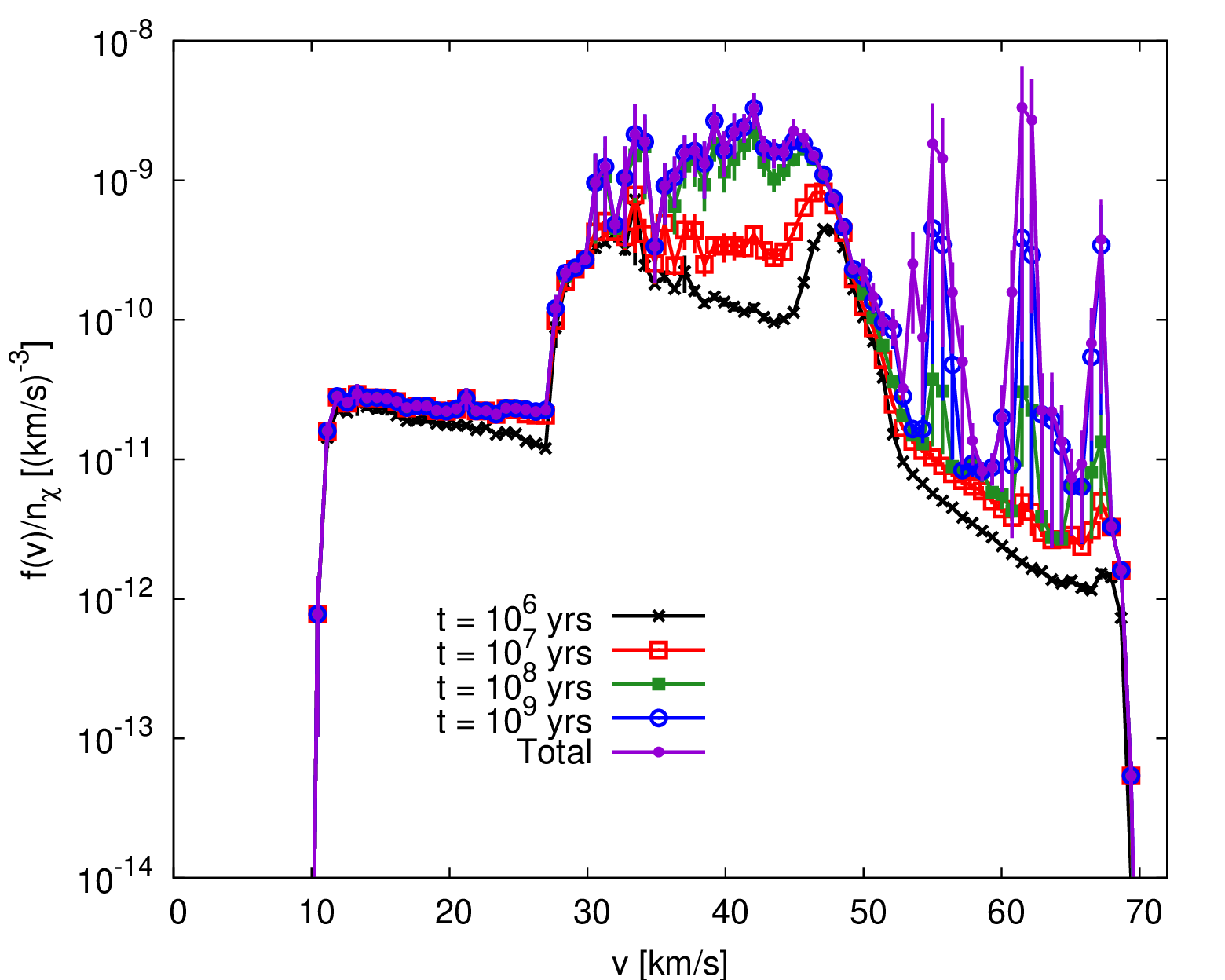} \\
		(b) &\includegraphics[width=2.4in,angle=270]{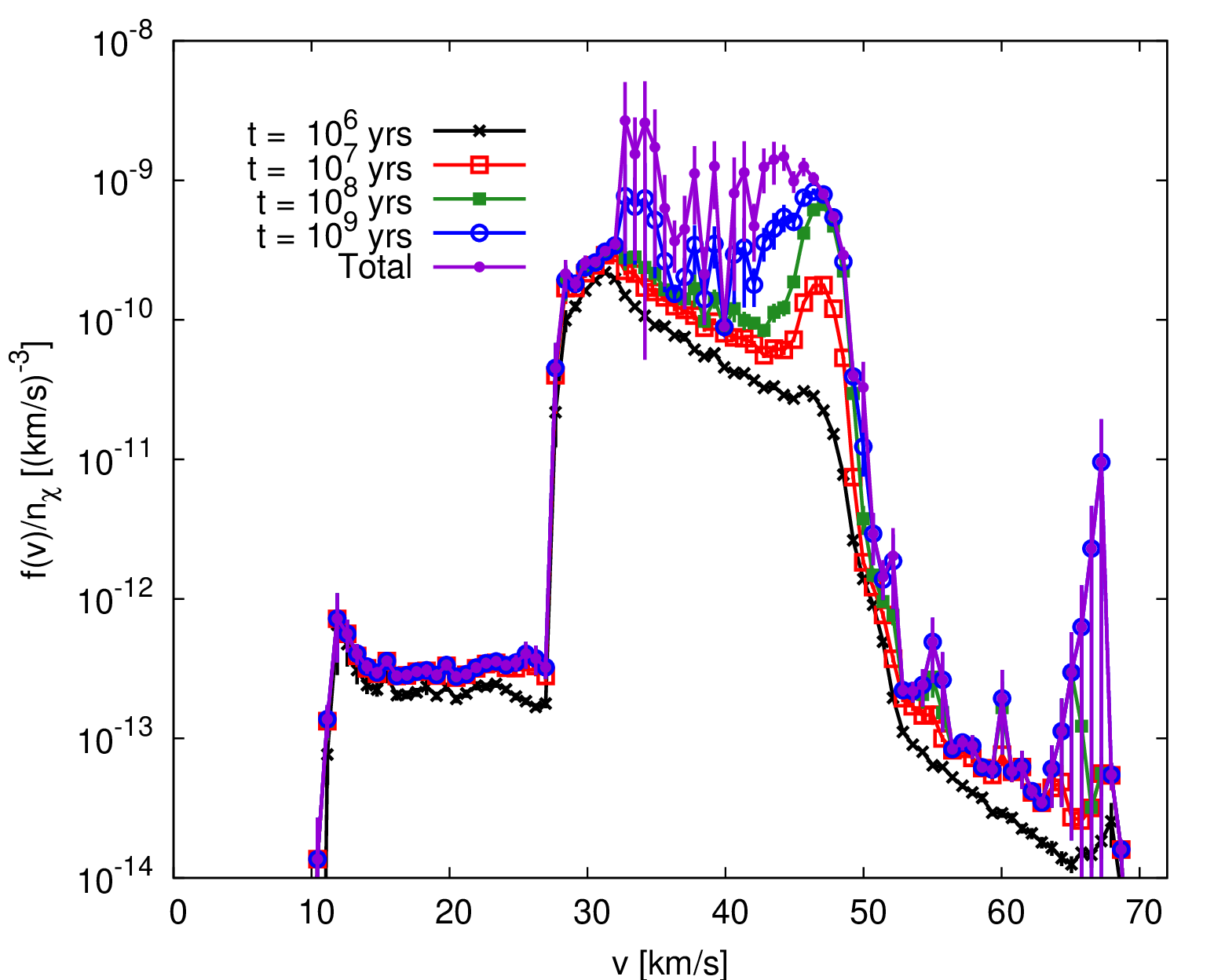}
	\end{tabular}
	\end{center}
	\caption{\label{fig:weak_time}Growth of the distribution functions as a function of time for the (a) DAMA and (b) CDMS simulations.}
\end{figure}

We have identified (i) which range of initial semi-major axis $a_i$ corresponds to the features in the geocentric DFs in Fig. \ref{fig:weak_sim_df}, and (ii) described the lifetime distributions of WIMPs.  Next, we describe the composition (not just in terms of the semi-major axis $a_i$ but by the type of orbit) and height of the DFs as a function of $\sigma_p^{SI}$.  This is most easily illustrated with the time-evolution of the DFs, which we show in Fig. \ref{fig:weak_time} for the (a) DAMA and (b) CDMS simulations.  We focus on these two simulations since the salient results of the Medium Mass and Large Mass simulations closely resemble those of the CDMS run.  In each plot, we show distribution functions as a function of time since the birth of the solar system, for $t= 10^6$ yr, $t = 10^{7}$ yr, $t = 10^{8}$ yr, $t =  10^9$ yr, and $t = t_\odot = 4.5$ Gyr. 

The low plateau, composed of Jupiter-crossing particles, has reached equilibrium in both $\sim 10^7$ year for both simulations.  The only growth in the low plateau after 10 Myr comes from particles on resonance-sticking orbits that pump up the spikes.  The time evolution of the low plateau (but not its final equilibrium height) is independent of cross section over two orders of magnitude in WIMP-baryon cross section because the equilibrium timescale is essentially the ejection timescale.  The height of the low plateau is proportional to the rate at which particles are initially scattered onto Jupiter-crossing orbits, $\dot{N}_{\jupiter}$.  Since the scattering rate is proportional to the cross section, the height of the low plateau is proportional to the spin-independent cross section, at least in these simulations.  One would expect that the plateau height would grow less rapidly with $\sigma_p^{SI}$ if the lifetimes of Jupiter-crossing WIMPs were dominated by rescattering in the Sun, not ejection from the solar system.

The absolute height of the spikes is similarly related to $\dot{N}_{\jupiter}$ and the relative ejection and rescattering timescales; the spikes in the DAMA simulation are more prominent than in the CDMS simulation because $\dot{N}_{\text{\jupiter}}$ is two orders of magnitude larger.  The time-evolution of the spikes can be explained by the following.  The lifetime distribution of spike WIMPs falls as $N(t) \propto t^{-\alpha}$, where $\alpha$ is slightly less than one.  The rate at which WIMPs cross the Earth's orbit is $\dot{N}_{c}(t) = \hbox{ const}$ if the long-lived WIMPs are resonance-sticking.  Therefore, the total contribution of the spike WIMPs to the DF beyond time $t$ goes as 
\begin{eqnarray}
\label{eq:growth} f_{spike} (>t) &\propto & \int_t^{t_\odot}  N(t^\prime) \dot{N}_c(t^\prime) \text{d} t^\prime \\
 & \propto & t_\odot^{1-\alpha} - t^{1-\alpha}.  
\end{eqnarray}
This argument is only strictly true if the types of spike orbits is independent of the lifetime distribution, which is uncertain due to the small number statistics of the spike WIMPs.  However, in Fig. \ref{fig:weak_time}, some of the spikes either grow linearly with time or do not grow at all for some stretches of time.  This phenomenon is due to the small numbers of long-lived resonance-sticking particles.  For an individual WIMP, $f(v) \propto t$ if $t < t_l$ and $f(v)$ is fixed for $t > t_l$.

The time evolution of the high plateau is different between the DAMA and CDMS simulations.  In the DAMA simulation, by $t = 1$ Myr, nearly all of the rescattering peak WIMPs have rescattered and thermalized in the Sun, as have a significant fraction of the quasi-Kozai WIMPs.  At this point, the high plateau in the range $27 \hbox{ km s}^{-1} < v < 45 \hbox{ km s}^{-1}$ is built up by roughly similar contributions from rescattering and Kozai peak WIMPs.  The contribution of Kozai WIMPs relative to rescattering peak WIMPs at a particular time can be estimated using 
\begin{eqnarray}
\frac{f_{Kozai}}{f_{rescatt}} \sim \beta \left( \frac{t}{t_{med}} \right),
\end{eqnarray}
where $f_{Kozai}$ is the DF due to Kozai cycling WIMPs, $f_{rescatt}$ that of rescattering peak WIMPs, $\beta$ is the fraction of WIMPs with $a < 1.5\hbox{ AU}$ on Kozai cycles, and $t_{med}$ is the median lifetime of rescattering peak WIMPs.  $\beta \approx 10^{-3}$ for WIMPs experiencing Kozai cycles, and $t_{med} \sim 10^{3}$ yr, so $f_{Kozai}/f_{rescatt} \sim 1$ at $t=1$ Myr.  This assumes that the increase in the DF for a Kozai cycling WIMP as a function of time is similar to that of a WIMP on a rescattering peak-type orbit.  The feature in the DF between $45 \hbox{ km s}^{-1} < v < 50 \hbox{ km s}^{-1}$ is due to quasi-Kozai WIMPs.

The high plateau grows substantially between $t = 1$ Myr and $t = 100$ Myr, although not strictly linearly because scatters in the Sun remove Kozai and quasi-Kozai WIMPs from Earth-crossing orbits.  The error bars on the DF increase with time as the ever-decreasing number of Earth-crossing WIMPs (Fig. \ref{fig:weak_tdist}) build up the DF.  After $100$ Myr, the high plateau grows very slowly  until it reaches equilibrium by $t = 1$ Gyr; in our simulation of $8\times 10^5$ WIMPs with orbits interior to Jupiter's orbit, only one WIMP has a lifetime of $1$ Gyr.  

Even though we simulate $\sim 10^3$ WIMPs on Kozai cycles, we are clearly undersampling those with $t_l > 10^9$ yr.  To estimate how much larger the DF could be, we note that the lifetime distribution of Kozai WIMPs with $t_l > 100$ Myr is well fit by $N(t) \propto t^{-2}$. If we assume that the rate at which the long-lived WIMPs contribute to the DF as a function of time is the same as for the Kozai WIMPs we simulate, then $\dot{N}_c(t) = \hbox{ const}$.  Therefore, according to Eq. (\ref{eq:growth}), the part of the DF built after time $t$ is $f_{Kozai}( v, >t ) \propto t^{-1}$.  For the high plateau, $f_{Kozai}(v, t > 10^8 \hbox{ yr}) \gg f_{Kozai}(v,t > 10^9\hbox{ yr})$, so we believe that we have not underestimated the high plateau.

A major consequence of this equilibrium distribution is that the high plateau of the DF $f(v)/n_\chi$ is fixed essentially fixed above a certain cross section.  Since the lifetime of Kozai orbits $t_l \propto (\sigma_p^{SI})^{-1}$, we find that the high plateau is is fixed for $\sigma_p^{SI} \gtrsim 10^{-42} \hbox{ cm}^2$.  

The time evolution of the distribution function is a bit different in the simulations for which $\sigma_p^{SI} = 10^{-43}$ cm$^2$.  At $t=1$ Myr, the high plateau is dominated by rescattering peak WIMPs, which have a median lifetime $t_{med} \sim 10^5$ yr.  Between $t = 1$ Myr and $t=100$ Myr, the growth in the high plateau is mostly due to the long-lifetime tail of the rescattering peak WIMPs and the quasi-Kozai WIMPs.  This is because $f_{Kozai}/f_{rescatt} \sim 1$ only for $t \sim 10^8$ yr.  For $t > 100$ Myr, the high plateau is dominated by Kozai WIMPs.  To determine if we sufficiently sampled the Kozai population, we compared the DF derived from the DAMA simulation when the integrated optical depths were equivalent to those in the CDMS simulation (if effect, comparing the DAMA simulation at $t=10^7$ yr with the CDMS simulation at $t=10^9$ yr).  We found the DFs to be consistent with each other.

Unlike in the DAMA simulation, a number of particles have lifetimes longer than the age of the solar system ($\sim 100$ out of $\sim 10^5$).  One consequence of this is that the DFs should be somewhat smaller than the DAMA distribution function, since the high plateau of the DAMA simulation has reached equilibrium by the present but the $\sigma_p^{SI} = 10^{-43}$ cm$^2$ distribution functions are still growing.  In fact, we find that the high plateau is about a factor of three smaller for the CDMS simulation than for the DAMA simulation.  As $\sigma_p^{SI}$ decreases, so should the height of the high plateau.  For $\sigma_p^{SI} \lesssim 10^{-45} \hbox{ cm}^2$, the high plateau should be dominated by rescattering peak orbits.

In summary, we find that while the DF for $\sigma_p^{SI} = 10^{-41}$ cm$^2$ is dominated by Kozai WIMPs, there is some contribution from long-lived Jupiter-crossing WIMPs (although the error bars are large due to small number statistics).  As the cross section decreases, the Jupiter-crossing component of the number density also decreases, and the Kozai and quasi-Kozai contributions dominate.  However, the Kozai WIMPs fail to reach equilibrium, so the overall DF goes down as a function of decreasing cross section.  Below $\sim 10^{-45}\hbox{ cm}^2$, we expect the DF to be dominated by the rescattering peak WIMPs.

\subsection{The Distribution Function as a Function of $m_\chi$}

There is little variation in the morphology of the lifetime distributions for the three simulations with $\sigma_p^{SI} = 10^{-43}$ cm$^2$.  The shape of the lifetime distribution appears to be determined almost solely by the elastic scattering cross section, not the particle mass, at least in the range of masses considered in these simulations.  It is possible that these distributions (in lifetime and density composition) for a very high or very low mass WIMP would perhaps look different from those in Fig. \ref{fig:weak_tdist}.

However, the DFs in Fig. \ref{fig:weak_sim_df} did show some variation with WIMP mass.  There are three effects that might induce a mass dependence on the DF.  
(i) The mass can affect the initial energy and angular momentum distribution of bound WIMPs.  As discussed in Section \ref{sec:ic_weak}, it is increasingly difficult to scatter halo WIMPs onto bound orbits as the WIMP mass increases.  The maximum energy transfer $Q_{max}$ approaches an asymptote for large WIMP masses, but the unbound WIMP energy increases since energy $E \propto m_\chi$.  Thus, the minimum scattered particle energy $E^\prime = E - Q_{min}$ increases for fixed initial speed but increasing WIMP mass.  However, this is not expected to be a major effect for the range of masses used in the simulations.
  
\begin{figure}
	\begin{center}
	\includegraphics[width=3.0in]{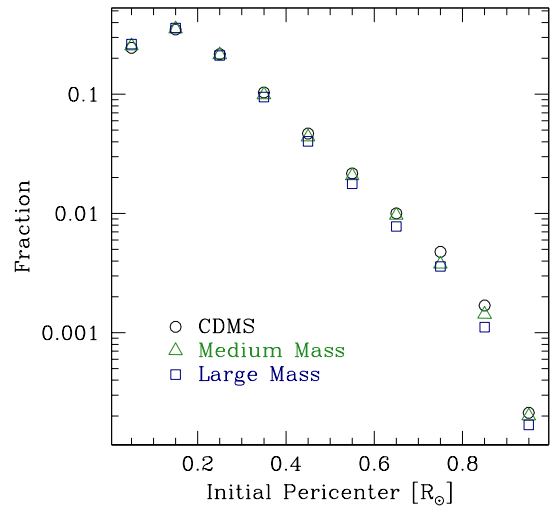}
	\end{center}
	\caption{Percentages of particles in each initial perihelion bin.  Poisson errors smaller than points.}\label{fig:weak_init}
\end{figure}

The angular momentum distribution is also affected by the WIMP mass, as parametrized by the initial particle perihelion in Fig. \ref{fig:weak_init}.  As discussed in Section \ref{sec:ic_weak}, the maximum angular momentum decreases with increasing $m_\chi$ since high mass particles scattering onto bound orbits must do so at smaller distances from the center of the Sun.  Thus, the Medium Mass and Large Mass simulations have a deficit of large perihelion particles relative to the CDMS simulation.  Since Kozai WIMPs originate in the outskirts of the Sun, this suggests that there will be fewer particles on Kozai cycles as the WIMP mass increases.

(ii) The particle mass affects the rescattering probability in the Sun.  In Eq. (\ref{eq:dtau_dl}), we show that the scattering probability along a path $l$ is proportional to $\text{d} \tau / \text{d} l \propto \left( 1 - e^{-Q_{max}/Q_A}\right)$, which is a mildly increasing function of WIMP mass $m_\chi$ (since $Q_{max}$ is mass-dependent, Eq. \ref{Q}).  The optical depth for the Large Mass simulation ($m_\chi = 500$ AMU) for a given path is about 15\% higher than for $m_\chi = 60$ AMU.  However, while high mass WIMPs have a higher scattering probability than low mass WIMPs, they also rescatter far more often onto Earth-crossing orbits.  Therefore, it is not clear from the outset whether high mass WIMPs will have longer or shorter lifetimes relative to low mass WIMPs.

\begin{figure*}
	\begin{center}
	\begin{tabular}{ll}
	(a) &(b) \\
	\includegraphics[width=2.8in]{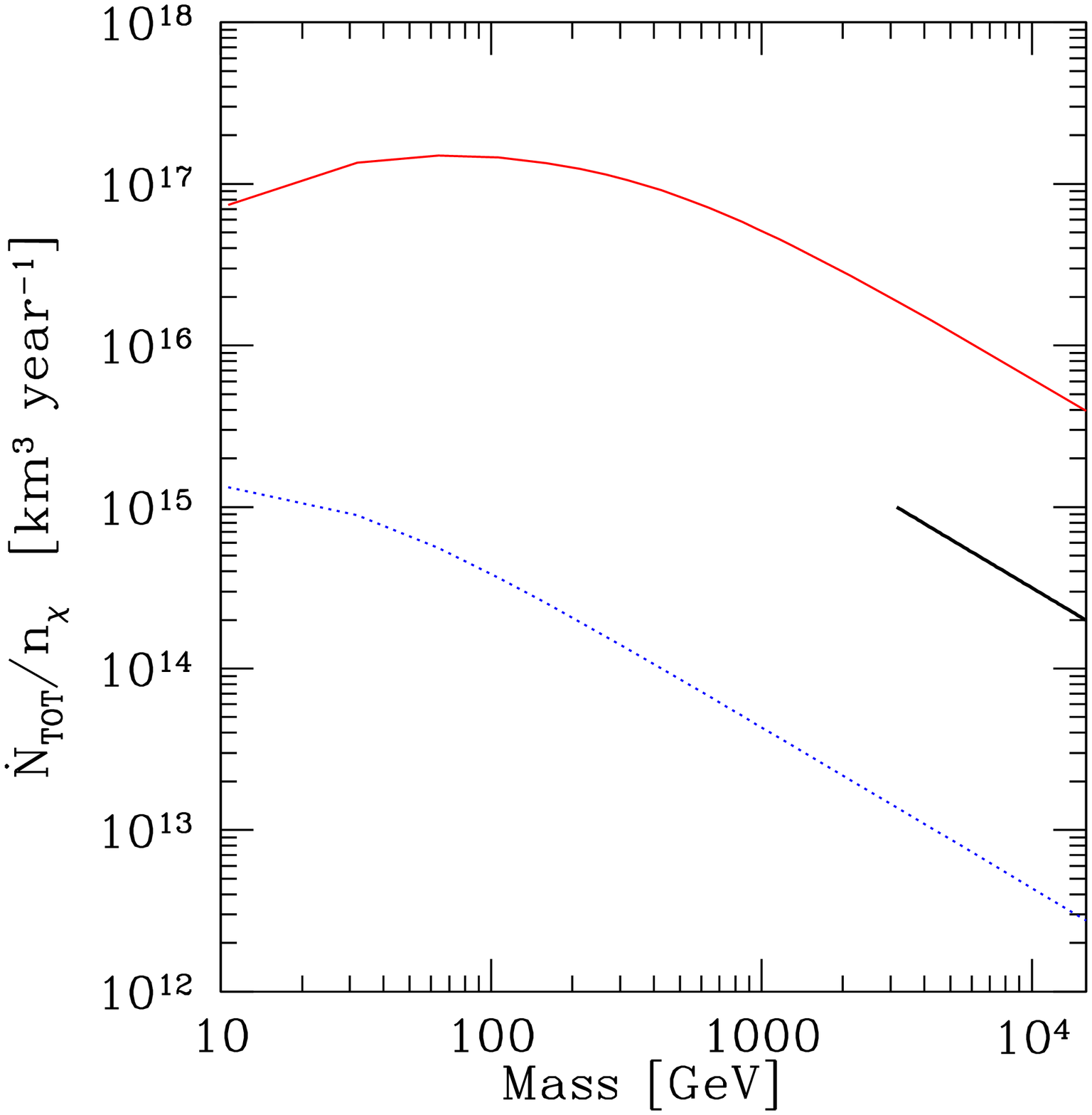} &\includegraphics[width=2.8in]{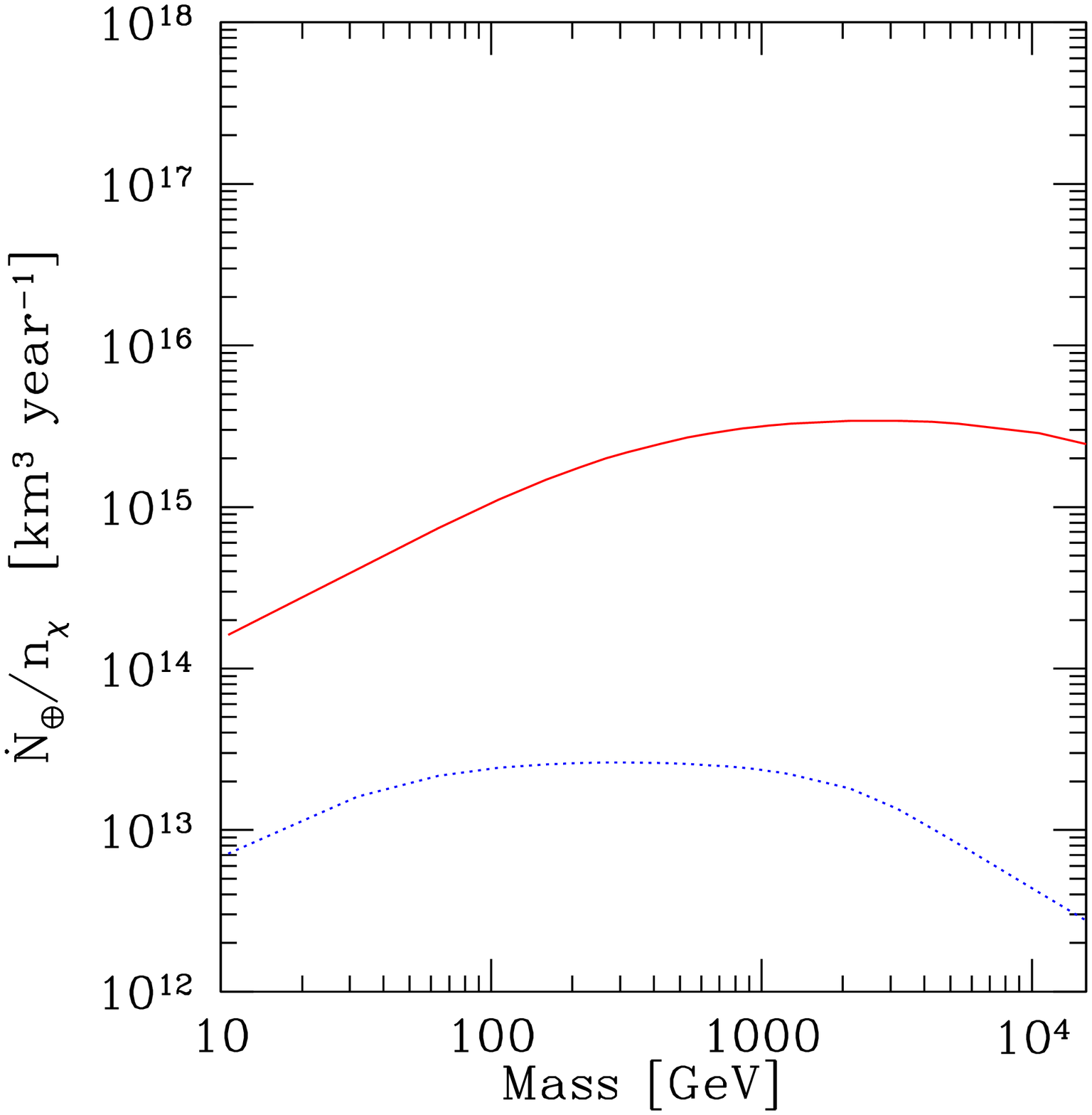} 
	\end{tabular}
	\end{center}
	\caption{In each plot, the red solid line denotes all species in the Sun, and the dotted blue line represents hydrogen.  (\emph{a}): The capture rate $\dot{N}$ of WIMPs by the Sun for $\sigma_{p}^{SI} = 10^{-43}$ cm$^2$, divided by the halo number density of WIMPs.  The short solid black line gives the slope $\dot{N}/n_\chi \propto m^{-1}_{\chi}$, the limiting slope for $m_\chi \gg m_A$ for a nuclear species $A$.  (\emph{b}): The capture rate $\dot{N}_{\oplus}$ to Earth-crossing orbits divided by the halo WIMP number density.}\label{fig:cap}

\end{figure*}

(iii) The WIMP mass also affects the overall amplitude of the final bound dark matter DF because the WIMP mass determines the scattering rate of halo particles onto bound, Earth-crossing orbits.  For high mass WIMPs, the \emph{total} capture rate of halo WIMPs in the Sun is \citep[e.g.,][]{gould1992}
\begin{eqnarray}
	\dot{N}_{tot} / n_\chi  \propto m_\chi^{-1}, &\text{ }m_\chi \gg m_A.
\end{eqnarray}
The function $\dot{N}_{tot} / n_\chi$ is plotted in Fig. \ref{fig:cap}(a) for the capture rate due to all species in the Sun (\emph{solid red}) and for scattering only on hydrogen (\emph{blue dots}; calculated in the limit of a cold Sun).  The capture rate of particles onto \emph{Earth-crossing} orbits is shown in Fig. \ref{fig:cap}(b).  Note that the capture rate $\dot{N}_{\oplus}/n_\chi$ onto Earth-crossing orbits is an increasing function of WIMP mass until about $m_\chi \approx $ TeV ($\approx 100$ GeV in the case of only hydrogen scattering).  This is because low mass WIMPs may be scattered onto very small orbits (whose aphelia may be within the Sun), which are kinematically suppressed for higher mass WIMPs.  Even though the total WIMP capture rate decreases for higher WIMP mass, those WIMPs that are captured are increasingly preferentially scattered onto Earth-crossing orbits.  The function $\dot{N}_\oplus / n_\chi$ turns over when most captured particles are on Earth-crossing orbits, and then the function follows the familiar $\dot{N}_\oplus / n_\chi \propto m_\chi^{-1}$.  

\begin{figure}
	\begin{center}
		\includegraphics[width=3.3in,angle=270]{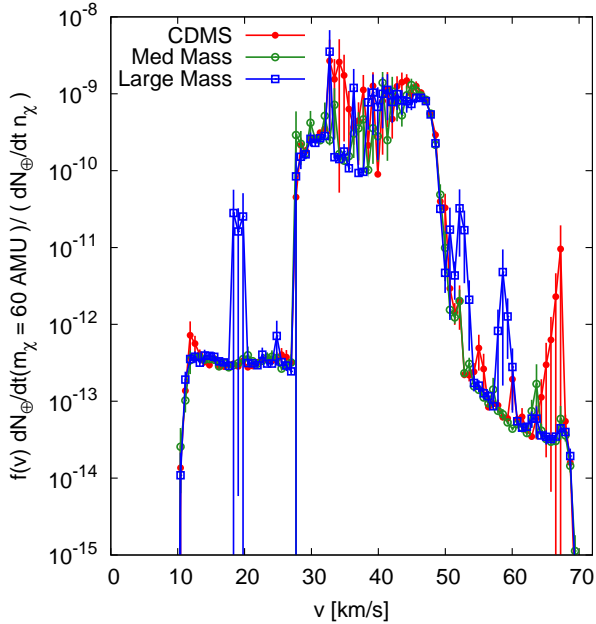}
	\end{center}
	\caption{DFs for the three simulations with $\sigma_p^{SI} = 10^{-43}$ cm$^2$ scaled by $\dot{N}_{\oplus}$.}\label{fig:weak_lowsigma}
\end{figure}

The consequence of these scattering rates of halo WIMPs in the Sun is that, if the DFs were otherwise independent of WIMP mass, the high mass DFs would be greater than the low mass DFs simply due to the prefactor $\dot{N}_\oplus$ in Eq. (\ref{eq:final_flux}). In order to isolate the effects of WIMP mass on the initial distribution of energy and angular momentum as well as subsequent rescattering, we divide the three DFs from the simulations with $\sigma^{SI}_p = 10^{-43}$ cm$^2$ in Fig. \ref{fig:weak_sim_df}(a) by $\dot{N}_\oplus$ and show these functions in Fig. \ref{fig:weak_lowsigma}.  The low plateaus do not appear to be significantly different.  There are some discrepancies in the spikes, which are due to the low numbers of long-lived resonance-sticking WIMPs in each simulation.  The high plateaus look relatively consistent with each other, given the large error bars.

\subsection{Maximum DF from Spin-Independent Solar Capture}

\begin{figure}
	\begin{center}
	\includegraphics[width=3.0in,angle=270]{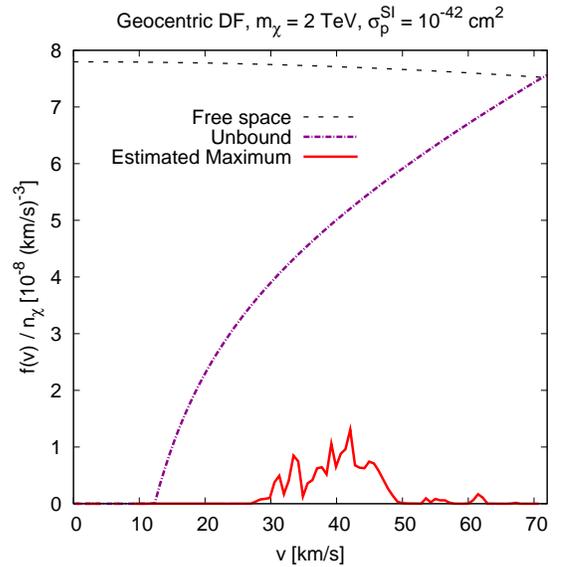}
	\caption{\label{fig:max_si}The estimated maximum DF consistent with current exclusion limits if spin-independent scattering dominates in the Sun, for a WIMP mass $m_\chi = 2$ TeV and WIMP-proton cross section $\sigma_p^{SI} = 10^{-42}$ cm$^2$.}
	\end{center}
\end{figure}

An important quantity to estimate is the maximum allowed DF consistent with experimental constraints on $\sigma_p^{SI}$.  We expect the point in $m_\chi - \sigma_p^{SI}$ yielding this maximum DF to lie on the exclusion curve, but the maximal point is determined by the shape of the curve for the following reason.  The best limits on $\sigma_p^{SI}$ are shown in Fig. \ref{fig:simpoints} and come from the XENON10 (below $m_\chi = 40$ GeV) and CDMS (above $m_\chi = 40$ GeV) experiments \cite{angle2008,cdms2008}.  The exclusion curves reach a minimum of $\sigma_p^{SI} \sim 4\times 10^{-44} \hbox{ cm}^2$ in the range $m_\chi = 20 - 70$ GeV.  Below these masses, the exclusion curve rises sharply due to kinematic reasons.  Above $m_\chi \sim 70$ GeV, the exclusion curves rise $\propto m_\chi$ because the flux of halo WIMPs at the detector goes as $\rho_\chi / m_\chi$.

Since extensions to the standard model generically predict $m_\chi \gtrsim 100$ GeV (with the notable exception of some gauge-mediated supersymmetry breaking models which predict $m_\chi \sim 1$ keV \cite{dine1993,dine1995,dine1996,baltz2003}), we focus on this part of the exclusion curve \cite{jungman1996,hubisz2005,hooper2007}.  In the previous sections, we found that (i) the high plateau dominates the DF at least up to $\sigma_p^{SI} = 10^{-41}$ cm$^2$, (ii) this plateau is a growing function of cross section until it reaches equilibrium for $\sigma_p^{SI} \gtrsim 10^{-42}$ cm$^2$, and (iii) for a fixed cross section, $f(v)/n_\chi \propto \dot{N}_\oplus / n_\chi$, which reaches its maximum for $m_\chi \sim 2$ TeV (Fig. \ref{fig:cap}(b)).  If the CDMS exclusion curve in Fig. \ref{fig:simpoints} were extended to higher mass, one would find that the exclusion curve hits $\sigma_p^{SI} = 10^{-42}$ cm$^2$ near $m_\chi = 2$ TeV, which is exactly the point at which both equilibrium in the high plateau is achieved and $\dot{N}_\oplus / n_\chi$ reaches its maximum.  

In Fig. \ref{fig:max_si}, we show the estimated DF for $m_\chi = 2$ TeV and $\sigma_p^{SI} = 10^{-42}$ cm$^2$, which we interpret as the maximum possible DF consistent with exclusion limits if spin-independent scattering dominates in the Sun.  This DF is based on the DAMA simulation DF, appropriately scaled by WIMP cross section and mass.  This DF yields a bound WIMP fraction (relative to the halo) which is a factor of $\sim 4$ greater than that of the DAMA simulation.    

In conclusion, we find that even for the maximal DF for bound WIMPs, the bound population is more than three orders of magnitude smaller than the total halo population at the Earth.

\subsection{Extension to Spin-Dependent Capture}\label{sec:results_sd}
So far, we have only explored the dark matter DF in the case where WIMP-nucleon scatters in the Sun are dominated by spin-independent, scalar interactions.  However, limits on the spin-dependent WIMP-proton cross section are $\mathcal{O}(10^7)$ times weaker than $\sigma_p^{SI}$, and spin-dependent cross sections are generally higher than spin-independent cross sections in large parts of parameter space for well-motivated WIMPs.  We showed that the low plateau of the solar capture DF, consisting of Jupiter-crossing WIMPs, grows as $\dot{N}_\oplus/n_\chi \propto \sigma_p^{SI}$ or $\sigma_p^{SD}$.  Since the constraints on $\sigma_p^{SD}$ are so much weaker than on $\sigma_p^{SI}$, the low plateau could become large, reaching equilibrium when rescattering in the Sun occurs on timescales shorter than the time to pull the Jupiter-crossing WIMP perihelia outside of the Sun.

\begin{figure}
	\begin{center}
	\includegraphics[width=2.9in,angle=270]{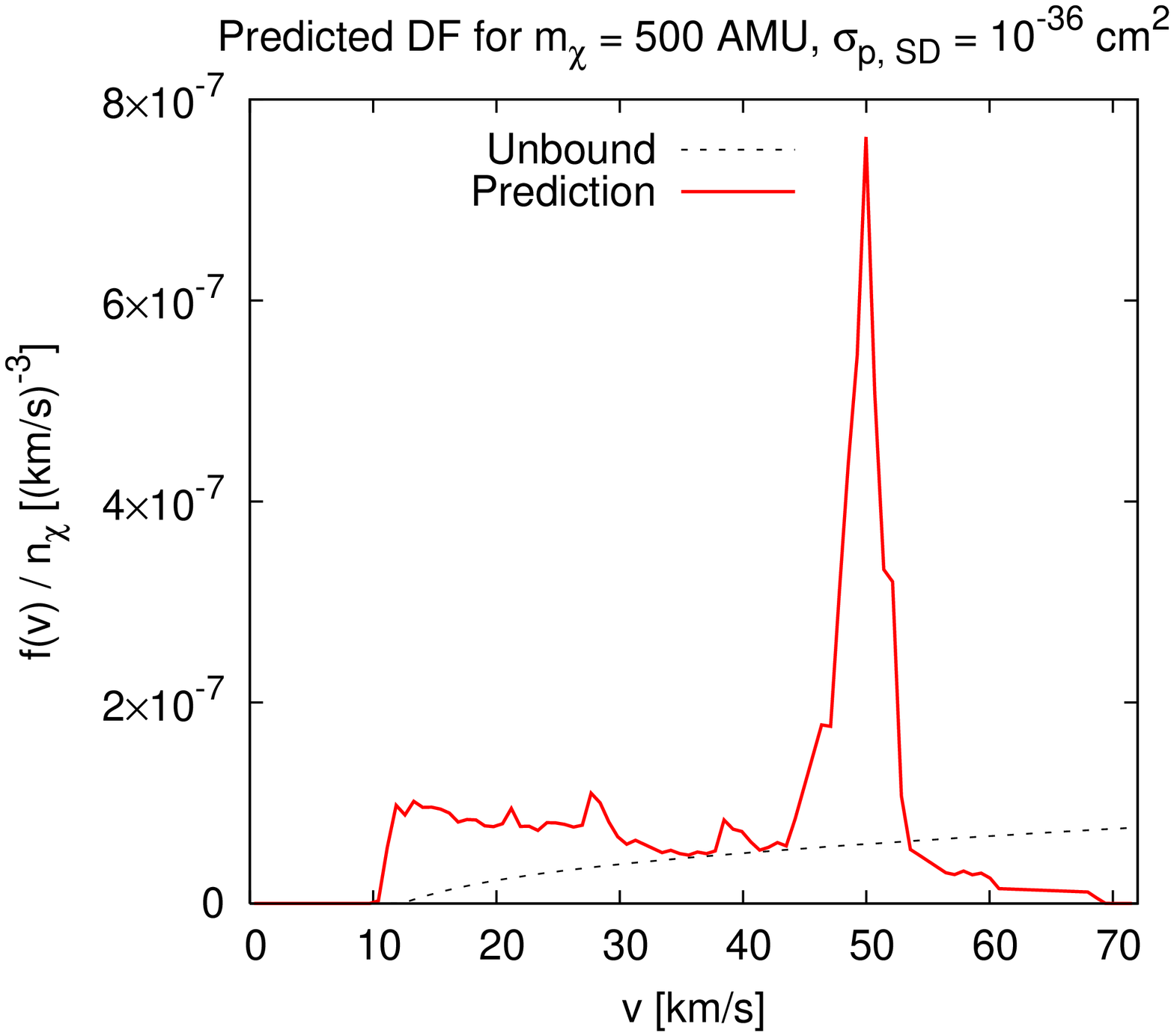}
	\end{center}
	\caption{Predicted geocentric DF if $\sigma_p^{SD} = 10^{-36}$ cm$^2$, assuming $f(v) \propto \sigma^{SI,SD}_p$ for Jupiter-crossing orbits.  This prediction is based on the output of the DAMA simulation.}\label{fig:predict}
\end{figure}

In Fig. \ref{fig:predict}, we show a prediction for the low plateau for $m_\chi = 500$ AMU and $\sigma_p^{SD} = 10^{-36}$ cm$^2$, if the only dependence of the DF on the WIMP cross section is $f(v) \propto \dot{N}_\oplus$.  The cross section is above the best $m_\chi - \sigma_p^{SD}$ unless $m_\chi > 1$ TeV \cite{lee2007,behnke2008}, but is chosen to demonstrate an approximate maximum possible bound DF.  At higher cross sections, the Sun becomes optically thick to WIMP scattering, at which point we expect the WIMP DF at the Earth to drop dramatically.  The large central peak in the predicted DF arises from the nearly radial orbits.  If the low plateau scales strictly with cross section until the Sun becomes optically thick, the Jupiter-crossing particles dominate the bound DF, and can swamp the unbound DF at low speeds ($v < 50 \hbox{ km s}^{-1}$).

However, there are some indications within the simulations that the low plateau will grow less rapidly with cross section than in this simple model.  Recall that $\approx 98\%$ of Jupiter-crossing WIMPs are ejected in the CDMS, Medium Mass, and Large Mass simulations, but a smaller fraction ($\approx 73\%$) of WIMPs are ejected in the DAMA simulation.  Therefore, a more careful estimate of the DF is required.

To find how large the WIMP DF can get, we estimated the bound WIMP DF for various large spin-dependent cross sections ($\sigma_p^{SD} > 10^{-40}$ cm$^2$) using the DAMA simulation as a starting point, since it has the highest $\sigma_p^{SI}$ and best statistics of all the spin-independent simulations.  We scaled the total optical depth of each particle in the DAMA simulation by an estimate of the optical depth for a particular spin-dependent cross section.  For particles that were not on Jupiter-crossing orbits, we scaled the lifetimes by the ratio of the optical depth for the particular spin-dependent cross section and the DAMA optical depth.  For the particles on Jupiter-crossing orbits, we used the optical depth data from the DAMA simulation to find the approximate time at which each particle hit a total optical depth $\tau = 1$ for the new cross section, which we interpreted as the new WIMP lifetime.  We calculated the DFs using the methods in Appendix \ref{sec:df_estimator}, with the inclusion of a Monte Carlo treatment of the initial conditions to determine if captured WIMPs scattered multiple times before they could leave the Sun.  

There are several assumptions in this approach.  First, we used the initial distribution of semi-major axis and eccentricity derived from the DAMA simulation without any kinematic corrections due to the extreme mass difference between hydrogen atoms and WIMPs.  Thus, we tend to overestimate the Kozai contribution to the DF since scattering in the outer part of the Sun is suppressed for high $m_\chi$.  This also underestimates the contribution of Jupiter-crossing particles since the semi-major axis distribution skews to higher $a$ for large imbalances between the WIMP a.  However, between $m_\chi = 60$ AMU and $m_\chi = 500$ AMU, the fraction of Earth-crossing particles that are also Jupiter-crossing only increases from $18.9\%$ to $21.5\%$ if the particles scatter only on hydrogen.  

Secondly, we did not recalculate optical depths for each passage through the Sun.  This would be too time-consuming.  Instead, we scaled the optical depths of each particle by the ratio of the scattering rate of $E = 0$ halo particles with the new cross section to the scattering rate of $E = 0$ halo particles in the DAMA simulation.  Since bound Earth-crossing particles do not have energies that vary significantly from $E=0$ relative to typical energies of unbound halo particles, using the ratio of the scattering rates to scale the DAMA optical depths should be a reasonable proxy for finding optical depths for specific paths through the Sun.  However, this approximation does neglect any differences in the radial distributions of hydrogen and heavier elements in the Sun, as well as any kinematic effects due to scattering off hydrogen rather than heavier atoms.  

\begin{figure}
	\begin{center}
		\includegraphics[width=2.9in,angle=270]{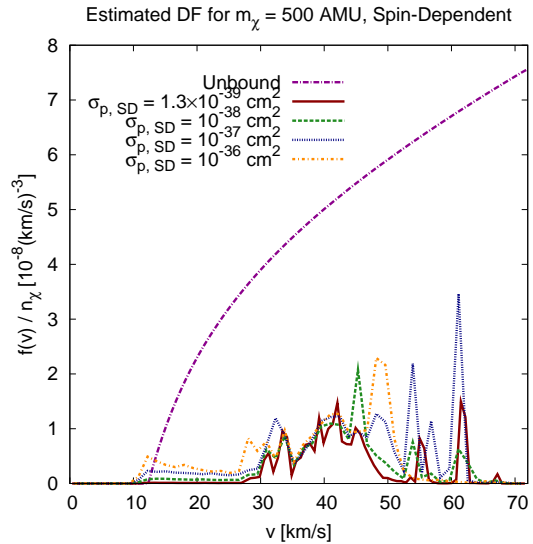}
	\end{center}
	\caption{Estimated geocentric DFs for $\sigma_p^{SD} = 1.3\times 10^{-39}$, $10^{-38}$, $10^{-37}$,$\mathrm{\mbox{ }and\mbox{ }}10^{-36}$ cm$^2$. The estimated DF for $\sigma_p^{SD} = 1.3\times 10^{-39}$ cm$^2$ is based on the DAMA simulation result, since the optical depth of the Sun for $\sigma_{p}^{SI} = 10^{-41}$ cm$^2$ is approximately the same as $\sigma_{p}^{SD} = 1.3\times 10^{-39}$ cm$^2$.}\label{fig:fixedall}
\end{figure}

We estimated DFs for $m_\chi = 60$ AMU at $\sigma_p^{SD} = 1.3\times 10^{-39}, 10^{-38}, 10^{-37}$, and $10^{-36}$ cm$^2$, and then extrapolate the results to other WIMP masses by rescaling the DFs by $\dot{N}^H_{\oplus}(m_\chi)$, the rate of scattering halo WIMPs on hydrogen to reach bound, Earth-crossing orbits.  The cross section $\sigma_p^{SD} = 1.3\times 10^{-39}$ cm$^2$ yields similar same optical depths in the Sun as $\sigma_p^{SI} = 10^{-41}$ cm$^2$.  We used 50 bootstrap resamplings for each spin-dependent cross section to estimate the DFs. The results are shown in Fig. \ref{fig:fixedall}, displaying $f(v)/n_\chi$ for each cross section against the geocentric unbound DF.

There are several key points this figure.  The central part of the DF for each cross section ($v= 30 - 45$ km s$^{-1}$) is approximately independent of cross section, which is what one would expect if Kozai cycles dominate this region and particles have lifetimes of at least one Kozai cycle.  This region is relatively unaffected by multiple scatters before the WIMPs exit the Sun for the first time because the particles on Kozai cycles originate in a part of the Sun that still has very low optical depth, even for the highest cross section considered.  The peak near $50$ km s$^{-1}$ is due to nearly radial Jupiter-crossing orbits.  The spikes in the low plateau grow for a while and then disappear, a consequence of rescattering in the Sun before WIMPs can stick to resonances. 

The most striking result of Fig. \ref{fig:fixedall} is that the low plateau is quite a bit lower than the naive prediction in Fig. \ref{fig:predict}.  It appears that, while the low plateau does rise for large WIMP-proton cross sections, rescattering in the Sun plays an integral role in severely reducing Jupiter-crossing particle lifetimes.  We find that the low plateau reaches approximately its maximum value if $\sigma_p^{SD} = 10^{-36} \hbox{ cm}^2$.  Even though the low plateau is still very slowly increasing between $\sigma_p^{SD} = 10^{-37}\hbox{ cm}^2$ for $v < 50 \hbox{ km s}^{-1}$, the plateau actually decreases between going from $\sigma_p^{SD} = 10^{-37}\hbox{ cm}^2$ to $\sigma_p^{SD} = 10^{-36}\hbox{ cm}^2$.  This is because WIMPs with geocentric speeds $v > 50 \hbox{ cm s}^{-1}$ are retrograde with respect to the planets in the solar system, and the torques from the planets are less effective for retrograde than prograde WIMPs.  Thus, the time for WIMP perihelia to exit the Sun is longer for retrograde WIMPs than prograde WIMPs, and so the probability for a retrograde WIMP to rescatter in the Sun before its perihelion exits the Sun for the first time is significantly higher than for a prograde WIMP.  Therefore, the maximum low plateau occurs for $\sigma_p^{SD} \sim 10^{-36} \hbox{ cm}^2$, or about $\sigma_p^{SI} \sim 10^{-38}\hbox{ cm}^2$.

Combining these results with the maximum DF for spin-independent solar capture in Fig. \ref{fig:max_si}, we find that particles captured to the solar system by elastic scattering in the Sun are only small population relative to the halo population at the Earth, even if the spin-dependent WIMP-proton elastic scattering cross section is quite large.  Improving on the approximations we used in this section is unlikely to change this conclusion.

\section{The Direct Detection Signal}\label{sec:dd}

Direct detection experiments look for nuclear recoil of rare WIMP-nuclear interactions in the experimental target mass.  The WIMP-nucleus scattering rate per kg of detector mass per unit recoil energy $Q$ can be expressed as \citep[cf.][]{jungman1996}
\begin{eqnarray}
	\frac{\mathrm{d}R}{\mathrm{d}Q} = \left( \frac{m_A}{\mathrm{kg}} \right)^{-1}  \int_{v_{min}}^{\infty} \mathrm{d}^3 \mathbf{v} \frac{\mathrm{d} \sigma_A}{\mathrm{d} Q} v f(\mathbf{x},\mathbf{v}), \label{eq:drdq}
\end{eqnarray}
where $\mathrm{d}\sigma_A/\mathrm{d}Q$ is the differential interaction cross section between a WIMP and a nucleus of mass $m_A$ and atomic number $A$, and $\mathbf{v}$ is the velocity of the dark matter particle with respect to the experiment.  The lower limit to the integral in Eq. (\ref{eq:drdq}) is set to
\begin{eqnarray}
v_{min} = ( m_A Q / 2 \mu^2_A)^{1/2}, \label{eq:ch5vmin}
\end{eqnarray}
the minimum WIMP speed that can yield a nuclear recoil $Q$, The dark matter DF at the Earth is $f(\mathbf{x},\mathbf{v})$.

In this section, we will determine the maximum possible contribution of the bound DF to the direct detection rate.  We focus on the maximum event rate from bound WIMPs instead of exploring how the bound WIMP event rate depends on WIMP mass and elastic scattering cross section since we expect the event rate to be small.  We are interested in both the total excess signal due to bound WIMPs for particular experiments, as well as the contribution to the differential event rate, since the latter is important for determining the WIMP mass \cite{green2007b,drees2008}.

We focus on directionally-insensitive direct detection rates for spin-independent interactions, but the results of this section can be applied qualitatively to spin-dependent interactions as well.  There is another class of direct detection experiment that is directionally sensitive \cite{alner2005b,naka2007,santos2007,nishimura2008,sciolla2008}.  In principle, the bound WIMPs should leave a unique signal in such experiments (see Fig. \ref{fig:2d}), but it would be challenging to measure this given the small bound WIMP density, current errors in directional reconstruction, and high energy thresholds.

We calculate the bound WIMP event rate for $m_\chi = 500$ AMU and $\sigma_p^{SI} = 10^{-43} \hbox{ cm}^2$ and a high spin-dependent proton cross section ($\sigma_p^{SD} = 10^{-36}$ cm$^2$, approximately the point at which the Sun becomes optically thick to WIMPs).  We choose this point in parameter space because it yields the largest DF due to WIMPs bound by solar capture.  The event rate can simply be scaled for lower (or higher) spin-independent cross sections.  The scaling for other values of $m_\chi$ and $\sigma_p^{SD}$ is different, but can be easily determined.

The geocentric bound DF is anisotropic.  Therefore, to translate the DF outside the sphere of influence of the Earth to the corresponding DF at the detector, one should use the mapping technique in Appendix \ref{sec:df_estimator}, averaged over the detector's daily motion about the Earth's rotation axis.  However, using the isotropic mapping instead of the full six-dimensional mapping in Appendix \ref{sec:df_estimator} produces errors in $\text{d}R/\text{d}Q$ of at most a few percent.  Therefore, we use this simplification for the bound WIMP DF at the surface of the Earth in calculating $\text{d}R/\text{d}Q$.

\begin{figure}
	\begin{center}
		\includegraphics[width=3.5in]{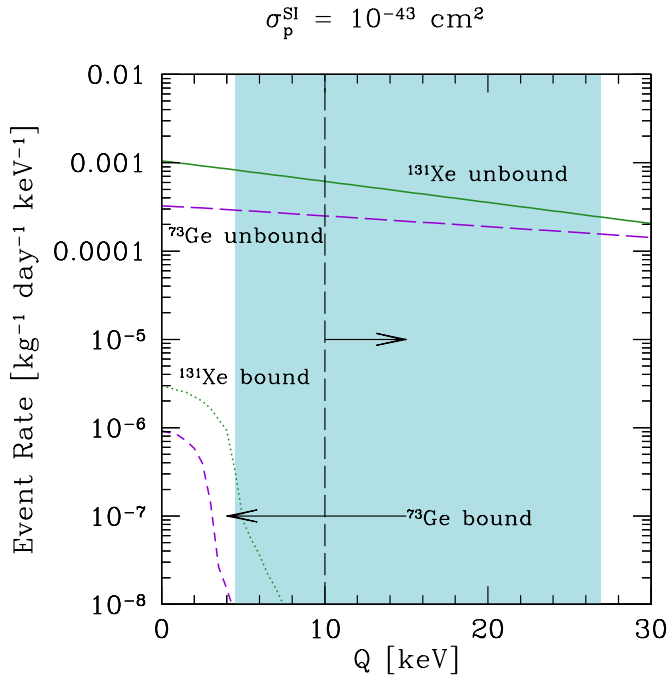}
	\end{center}
	\caption{The differential direct detection rate for $m_\chi = 500$ AMU and $\sigma_{p}^{SI} = 10^{-43}$ cm$^2$ assuming the DF is dominated by spin-dependent scattering in the Sun with $\sigma_p^{SD} = 10^{-36}\hbox{ cm}^2$.   The shaded region indicates the XENON10 analysis region \cite{angle2008}, and the vertical dashed line indicates the lower limit to the CDMS analysis window (which extends to $Q=100$ keV) \cite{cdms2008}.}\label{fig:dd_max_grav}
\end{figure}

In Fig. \ref{fig:dd_max_grav}, we show the maximal direct detection signal due to solar captured WIMPs if $m_\chi = 500$ AMU (lower two lines).  We find direct detection rates assuming $^{131}$Xe and $^{73}$Ge targets, since the current and planned experiments most sensitive to the spin-independent (and spin-dependent neutron) cross section have isotopes of either Xe or Ge as their target mass.  For comparison, we also plot the event rate expected for the halo DF in Eq. (\ref{eq:dfhaloaveraged}).  We find that bound WIMPs can only enhance the direct detection rate at very small $Q$, and that the enhancement is largest at the smallest recoil energies.  For both the germanium and xenon targets, the maximum enhancement to the total event rate is $\sim 0.5\%$ at $Q=0$.  This enhancement is actually disproportionally large compared to the enhancement in the local WIMP number density due to bound WIMPs, which is $n_{\chi,bound} \sim 10^{-4} n_{\chi,halo}$, since incoherence in the WIMP-nucleon interaction for large nuclei suppresses the elastic scattering cross section for high speed halo WIMPs.  
We also show the experimental analysis windows for the recent XENON10 and CDMS experiments in this figure \cite{angle2008,cdms2008}.  The current analysis threshold of the CDMS experiment is too high to detect bound WIMPs.  If this experiment and its successor SuperCDMS could push down their analysis thresholds, as other germanium-based rare event experiments have (e.g., CoGeNT \cite{aalseth2008}), bound WIMPs may be observed.  At $Q = 4.5$ keV, the current analysis threshold for the XENON10 experiment, the boost to the differential event rate is $\sim 0.1\%$, and the total boost in their analysis window is $\sim 10^{-3}\%$.  Thus, the bound particles only negligibly increase the total event rate (integrating $\text{d}R/\text{d}Q$ over the range of $Q$'s allowed in the analysis window), if at all.  Estimates of the WIMP mass and cross section from direct detection experiments will not be affected by solar captured particles.

\section{The Neutrino Signal from WIMP Annihilation in the Earth}\label{sec:id}
WIMPs may accumulate and annihilate in the Earth.  The signature of WIMP annihilation will be GeV to TeV neutrinos.  Neutrino observatories (e.g., Antares \cite{amram1999}, IceCube \cite{hill2006}) are sensitive to the {\v C}erenkov radiation of muons created in charged-current interactions of muon neutrinos in and around the experiment. 

The neutrino flux at a detector on the surface of the Earth is proportional to the annihilation rate $\Gamma$ of WIMPs trapped in the Earth.  Finding $\Gamma$ requires solving a differential equation for the number of WIMPs $N$ in the Earth, described by 
\begin{eqnarray}
	\dot{N} = C - 2 \Gamma, \label{eq:dotN}
\end{eqnarray}
where the capture rate of WIMPs in the Earth by elastic scattering is defined as.
\begin{multline}
	C = \int \mathrm{d}^3 \mathbf{x} \int_{v_{f} < v_{esc}(\mathbf{x})} \mathrm{d}^3 \mathbf{v} \mathrm{d} \Omega  \sum_A \frac{\mathrm{d}\sigma_A}{\mathrm{d}\Omega} n_A (\mathbf{x})  v \\
	\times \, f(\mathbf{x},\mathbf{v},t). \label{eq:earthcap}
\end{multline}
Here, $\mathrm{d}\sigma_A / \mathrm{d}\Omega$ is the WIMP-nucleus elastic scattering cross section for nuclear species $A$ and $v$ is the relative speed between the WIMP and a nucleus.  The number density of species $A$ is described by $n_A(\mathbf{x})$.  The cutoff in the velocity integral reflects the fact that the WIMP's speed after scattering $v_f$ must be less than the local escape velocity $v_{esc}(\mathbf{x})$.  

If the WIMP DF is time-independent, the annihilation rate goes as
\begin{eqnarray}
	\Gamma  = \frac{1}{2} C \tanh^2(t/t_{e}), \label{eq:gamma_analytic}
\end{eqnarray}
where 
\begin{eqnarray}
	t_{e} = ( C C_{a} )^{-1/2}\label{eq:t_e}
\end{eqnarray}
is the equilibrium timescale and $C_a$ is a constant that depends on the distribution of WIMPs in the Earth and is proportional to the annihilation cross section.

While the contribution of bound particles to the direct detection rate is expected to be minuscule, it is not unreasonable to expect that the bound particles could noticeably boost the neutrino-induced muon event rate from WIMP annihilation in the Earth.  Because the Earth's gravitational potential is shallow, it is difficult for halo WIMPs to lose enough energy during collisions with the Earth's nuclei to become bound unless the WIMP mass is nearly equal to the mass of one of the abundant nuclear species in the Earth \cite{gould1987}.  For WIMPs with mass $m_\chi > 400$ GeV, only WIMPs bound to the solar system may be captured in the Earth.  

\begin{figure}
	\begin{center}
		\includegraphics[width=3.3in]{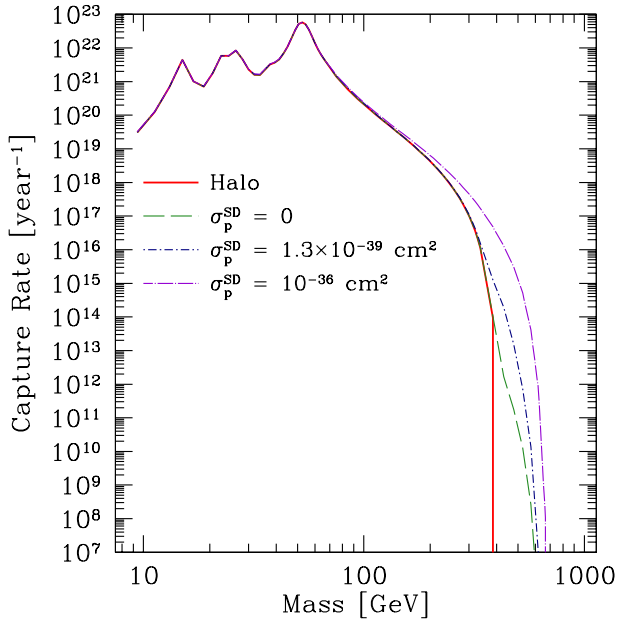}
	\end{center}
	\caption{Capture rate of WIMPs in the Earth as a function of WIMP mass for $\sigma^{SI}_{p} = 10^{-43}$ cm$^2$.  All capture rates include the capture of unbound halo WIMPs as well as the capture of bound WIMPs.}\label{fig:capearth}
\end{figure}

In Fig. \ref{fig:capearth}, we show the capture rate (Eq. \ref{eq:earthcap}) of WIMPs in the Earth as a function of mass for $\sigma_p^{SI} = 10^{-43} \hbox{ cm}^2$ for several different WIMP DFs.  We use the potential and isotope distributions in \citet{earth1} and \citet{earth2}.  The lowest line shows the capture rate of only the unbound WIMPs in the solar system.  The peaks in the capture rate correspond to points at which the WIMP mass is nearly exactly the same as a one of the common elements in the Earth, of which the iron peak ($m_{Fe}\approx 56$ AMU $= 53$ GeV) is especially prominent.  The long-dashed line represents the capture rate for both unbound particles and particles bound to the solar system by spin-independent scattering in the Sun.  We show extrapolations to the regime in which spin-dependent scattering dominates in the Sun with the short dash-dotted and long dash-dotted lines, representing $\sigma_p^{SD} = 1.3\times 10^{-39}$ cm$^2$ and $\sigma_p^{SD} = 10^{-36}$ cm$^2$ respectively.  We included unbound WIMPs in those estimates.  

From Fig. \ref{fig:fixedall}, we note that the high plateaus in the DF's if $\sigma_p^{SD} > 10^{-39} \hbox{ cm}^2$ are nearly identical; the main reason for the difference in the capture rate is the low speed DF of Jupiter-crossing WIMPs.  In fact, the capture rate is extremely sensitive to the DF of the lowest speed WIMPs.  For the relatively low capture rates in Fig. \ref{fig:capearth}, $t_e > t_\odot$, so $\Gamma \propto C^2$.  Even small variations in the low speed WIMP DF can lead to large variations in the event rate at a neutrino telescope.

To estimate a plausible range of muon event rates given the capture rates in Fig. \ref{fig:capearth}, we explore part of the MSSM parameter space.  We can in principle explore other models, but the MSSM yields, on average, somewhat larger spin-independent cross sections.  Given that iron is the most common element in the core of the Earth, and oxygen, silicon, and magnesium the most common element in the mantle, none of which has spin-dependent interactions with WIMPs, only in WIMP models with appreciable spin-independent interactions will capture in the Earth be relevant.  

We scan MSSM parameter space to estimate the neutrino-induced muon event rate for neutrino telescopes from neutralino annihilation in the Earth using routines from the publicly available DarkSUSY v.5.0.2 code \citep{gondolo2004}.  The code can also check whether a model described by a set of SUSY parameters is consistent with current collider constraints and relic density measurements.  We describe our scans in more detail in Paper III.

To estimate the muon event rate in a neutrino telescope, we set the muon energy threshold to $E^{th}_\mu = 1$ GeV.  This is somewhat optimistic for the IceCube experiment \citep{icecube2001,lundberg2004} unless muon trajectories lie near and exactly parallel to the PMT strings, but it is reasonable for the more densely packed water experiments (e.g., Super-Kamiokande).  The signal drops sharply with increasing muon energy threshold \cite{bergstrom1998}.   We assume that the material both in and surrounding the detector volume is either water or ice, since the largest current and upcoming neutrino telescopes are immersed in oceans or the Antarctic ice cap.  We include all muons oriented within a 30$^\circ$ cone relative to the direction of the center of the Earth.  

\begin{figure}
	\begin{center}
		\includegraphics[width=3.3in]{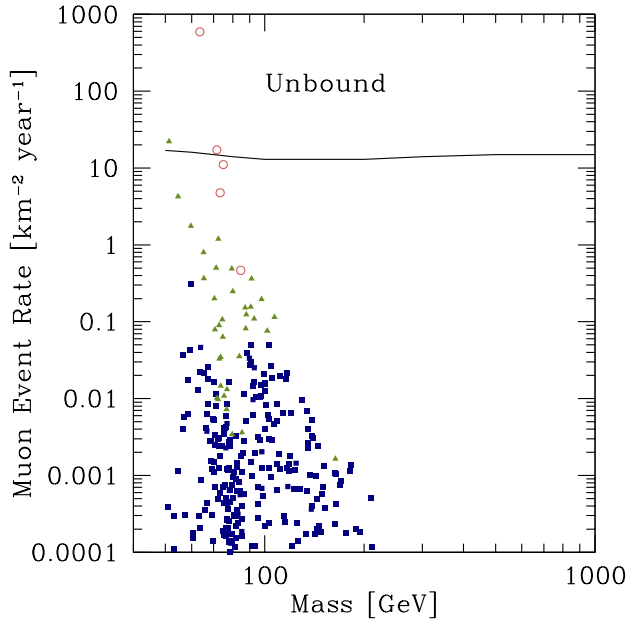}
	\end{center}
	\caption{Muon event rates from halo WIMPs unbound to the solar system.  Open circles mark MSSM models for which $\sigma_p^{SI}$ is above the 2006 CDMS limit \cite{akerib2006}, filled triangles mark those with limits between that limit and the current best limits on $\sigma_p^{SI}$ (set by XENON10 for $m_\chi < 40$ GeV \cite{angle2008} and CDMS for $m_\chi > 40$ GeV \cite{cdms2008}), and filled squares denote models consistent with the best limits on elastic scattering cross sections.  The solid line is an optimistic detection threshold for the IceCube experiment \cite[][and references therein]{lundberg2004}.}\label{fig:neutrino_unbound}
\end{figure}

In the following figures, we present muon event rates in neutrino telescopes for various DFs.  In Fig. \ref{fig:neutrino_unbound}, we show the event rates for WIMPs unbound to the solar system.  The solid black line on Fig. \ref{fig:neutrino_unbound} represents the most optimistic flux threshold for IceCube \citep[][and references therein]{lundberg2004}.  To show how the event rates depend on the SUSY models for a given spin-independent cross section, we mark the models on the figure according to which direct detection experiments bracket the cross section for a given neutralino mass.  The open circles correspond to SUSY models with $\sigma_p^{SI}$ above the  that lie above the 2006 CDMS limit \cite{akerib2006}, which is shown in Fig. \ref{fig:simpoints}.  The triangles are models for which $\sigma_p^{SI}$ lies between the 2006 CDMS limit and the current best limits on $\sigma_p^{SI}$ (a combination of XENON10 \cite{angle2008} and CDMS \cite{cdms2008} limits), and squares denote models consistent with all current direct detection experiments.  

It appears that no halo WIMPs from \emph{any} of the models found in our scan of the MSSM consistent with experiments would produce an identifiable signal in IceCube.  We cannot say that it is impossible for neutralino WIMPs from the halo to be observed by IceCube or other km$^3$-scale experiments, since we are only sampling a small part of the vast SUSY parameter space, but Fig. \ref{fig:neutrino_unbound} suggests that it is not likely.

\begin{figure}
	\begin{center}
		\includegraphics[width=3.3in]{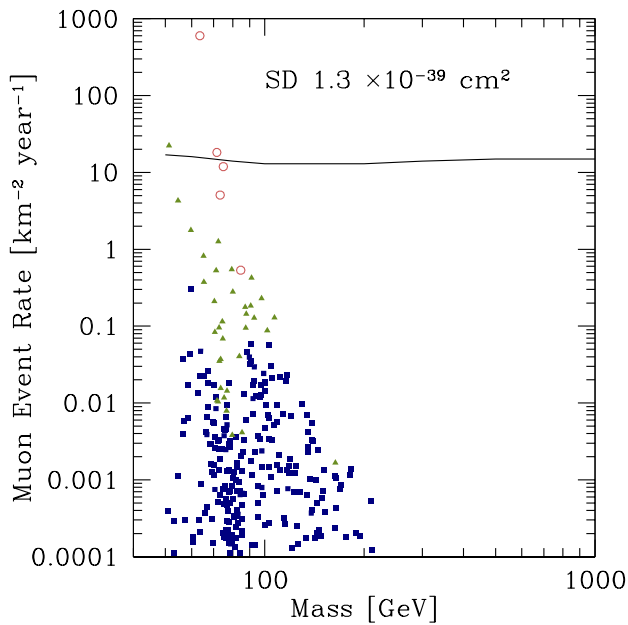}
	\end{center}
	\caption{Muon event rates including bound WIMPs.  Symbols mark the same models as in Fig. \ref{fig:neutrino_unbound}.}\label{fig:neutrino_grav}
\end{figure}

In Fig. \ref{fig:neutrino_grav}, we show the muon flux for WIMPs captured in the Earth from the halo or from the population of bound WIMPs.  We calculate the muon event rate with the bound DF for $\sigma_p^{SD} = 1.3\times 10^{-39}\hbox{ cm}^2$ no matter what the actual $\sigma_p^{SD}$ in the model is since this is near the maximum spin-dependent cross section found in the parameter scans. $\sigma_p^{SI}$ is almost always small enough that the DF due to spin-independent scattering in the Sun is subdominant to the spin-dependent-derived DF.  Therefore, the points in Fig. \ref{fig:neutrino_grav} are almost entirely \emph{upper limits} to the solar captured WIMP event rates.  This figure is almost indistinguishable from Fig. \ref{fig:neutrino_unbound}.  We find that the maximum enhancement over the halo WIMP event rate is of order 20\%.  Thus, the solar captured WIMPs produce almost no enhancement in the neutrino-induced muon event rate.

One caveat to this pessimistic result is that we estimated the event rate using only the flux of muons from outside the detector volume.  However, \citet{bergstrom1998} suggest that muons created inside the detector volume may dominate the signal for smaller WIMP masses ($m_\chi \lesssim 300$ GeV) in large (km$^3$) telescopes, although the exact enhancement has not been calculated.  But, the enhancement of the event rate due to bound WIMPs over halo WIMPs will be fixed and small.

\section{Discussion}\label{sec:discussion}
\subsection{Comparison with Damour and Krauss}\label{sec:comparison}
Here, we compare the simulation results with the semi-analytic predictions in \citet{damour1999}.  

Damour \& Krauss neglected the population of WIMPs on Jupiter-crossing orbits, arguing that it would be short-lived because of the strong perturbations from Jupiter.  This argument is plausible, but it neglects the importance of long-lived WIMPs on resonant orbits.  The presence of long-lived WIMPs on resonances suggests that Jupiter-crossing WIMPs may be important for $\sigma_p^{SI} \sim 10^{-41} \hbox{ cm}^2$ ($\sigma_p^{SD} \sim 10^{-39}\hbox{ cm}^2$).  However, such WIMPs are unlikely to contribute significantly for much larger or much smaller cross sections.  For much larger cross sections, long-lived WIMPs should be exceedingly rare; they are likely to rescatter and thermalize in the Sun before Jupiter can pull the perihelia out of the Sun.  For smaller cross sections, the rate of scattering of WIMPs onto Jupiter-crossing orbits is negligible.

Before we describe where our results diverge from Damour \& Krauss for $a < a_{\text{\jupiter}} / 2 \approx 2.6$ AU, we reemphasize the main points of their work.  They found that the main enhancement to bound WIMP DF came from a small fraction, $\sim 0.1\%$, of WIMPs scattered onto orbits with $0.5\hbox{ AU} < a < 2.6\hbox{ AU}$ on Kozai cycles.  They assumed that these WIMPs, which originated in the outskirts of the Sun, had lifetimes at least as long as the age of the solar system $t_\odot \approx 4.5 \hbox{ Gyr}$.  For this range of semi-major axes, we found two major differences between their work and ours.

First, we find that WIMPs with $1.5 \hbox{ AU}  < a < 2.6 \hbox{ AU}$ are not well described by pure Kozai cycles due to significant interactions with mean-motion resonances.  Unless the WIMP-nucleon cross section is large ($\sigma_p^{SI} \gg 10^{-41} \hbox{ cm}^2$, allowed if $m_\chi \lesssim 5\hbox{ GeV}$ or $m_\chi \gtrsim 10\hbox{ TeV}$ \cite{angle2008,cdms2008}; $\sigma_p^{SD} \gg 10^{-39}\hbox{ cm}^2$), most of the WIMPs in this semi-major axis range have lifetimes $\sim 100$ times longer than if the Sun were an isolated body.  However, this still does not increase the DF at the Earth as much as if the $1\%$ of WIMPs in this semi-major axis band (the fraction of WIMPs initially scattered onto $1.5\hbox{ AU} < a < 2.6 \hbox{ AU}$ which were on Kozai cycles in \cite{damour1999}; a higher fraction of large semi-major axis WIMPs are on Kozai cycles than WIMPs with lower semi-major axis) had lifetimes extending to $t_\odot$.  

To show why, we use the following argument.  The increase in the number density of WIMPs at the Earth $n^\prime$ over the number density without tidal torques $n$ is roughly described by
\begin{eqnarray}\label{eq:n_increase}
	\frac{n^\prime}{n} \approx \epsilon_f E_t.
\end{eqnarray}
where $\epsilon_f$ is the fraction of WIMPs disturbed enough from their orbits to have significantly longer lifetimes in the solar system.  The factor $E_t$ describes the increase in the WIMP lifetime.  Typically, $E_t \approx \min(t^\prime_{med}, t_\odot) / \min(t_{med},t_\odot)$, where $t_{med}$ is the median lifetime of the  WIMPs in the absence of gravitational torques and $t^\prime_{med}$ is the median lifetime with gravitational torques.  For our simulations, $\epsilon_f \sim 1$ since most WIMPs with $1.5\hbox{ AU} < a < 2.6 \hbox{ AU}$ were on quasi-Kozai orbits and $E_t \sim 100$, implying that $n^\prime/n \sim 100$.  However, the \citet{damour1999} prediction would be $\epsilon_f \sim 0.01$, and $E_t \sim (t_\odot / (10^3 \hbox{ yr}) = 4.5\times 10^6$, implying that $n^\prime / n \sim 4.5\times 10^4$, a factor of $\sim 500$ larger than what we found in our simulations.

However, the main reason that the density of bound WIMPs is much smaller than estimated by \citeauthor{damour1999} is that particles with $ a < 1.5\hbox{ AU}$ on Kozai cycles have lifetimes that are much less than the age of the solar system.  This is due to the fact that the typical integrated optical depth per Kozai cycle is non-negligible, so a WIMP undergoes only a finite number of Kozai cycles before rescattering in the Sun.  There are two important timescales relevant to estimating the lifetimes of WIMPs on Kozai cycles for a given WIMP-nucleon scattering cross section.  

First, in the point mass three-body problem, the period of Kozai cycles are of order \citep[cf.][]{kiseleva1998} 
\begin{eqnarray}
	T \propto \frac{P_{\jupiter}^2}{P}\frac{M_\odot}{M_{\jupiter}}.
\end{eqnarray}
Here, $P$ denotes the orbital period of a particle and $P_{\jupiter}$ represents the orbital period of Jupiter.  For typical particles, $T\lesssim 10^5$ yr. 

The other important timescale is the timescale on which the orbital perihelion is moved out of the Sun.  Although the optical depth in the outskirts of the Sun is extremely low ($\tau \sim 10^{-5}$ for an orbit with $r_p \approx 0.7 R_\odot$, $\tau \sim 10^{-6}$ for $r_p \approx 0.9R_\odot$ in the DAMA simulation, and even lower in the other simulations), it takes many orbital periods for Jupiter to pull the perihelia out of the Sun, hence making the optical depth per Kozai cycle much larger than the optical depth for a single passage through the Sun.

The rate of change in the angular momentum of a WIMP is
\begin{eqnarray}
	\frac{\text{d}J}{\text{d}t} = K_{\text{\jupiter}},
\end{eqnarray}
where $K_{\text{\jupiter}}$ is the torque on the particle orbit by Jupiter.  The torque is larger at aphelion $r_a$ for particles with $a < a_{\text{\jupiter}}/2$ than at any other point in the orbit, so the average torque can be approximated by its value at aphelion
\begin{eqnarray}
	K_{\text{\jupiter}} &\sim& r\nabla\Phi_{\text{\jupiter}} \big|_{r = r_a} \\
	  &\sim&\frac{GM_{\text{\jupiter}} a^2}{a_{\text{\jupiter}}^3} \label{eq:torque}
\end{eqnarray}
applied at aphelion, where we have expanded the potential to the $l = 2$ term in the spherical harmonic expansion (Eq. \ref{eq:legendre}).  The angular momentum must change by of order
\begin{eqnarray}
	\Delta J \sim \sqrt{GM_\odot R_\odot} \label{eq:deltaj}
\end{eqnarray}
for perihelia to be external to the Sun.  In reality, since WIMPs on Kozai cycles originate at distances from the center of the Sun $r > 0.5\hbox{R}_\odot$, Eq. (\ref{eq:deltaj}) should have a small ($\sim 0.1 -1$) coefficient in front.  Therefore, if the torques are coherent, the total time it takes for a WIMP to have its first perihelion outside the Sun is
\begin{eqnarray}
	\Delta t \sim \frac{ \Delta J}{K_{\text{\jupiter}}}.
\end{eqnarray} 
Using the expressions for $K_{\text{\jupiter}}$ and $\Delta J$ in Eqs. (\ref{eq:torque}) and (\ref{eq:deltaj}), we find
\begin{eqnarray}
	\frac{\Delta t}{P} & \sim&  \frac{M_\odot}{M_{\text{\jupiter}}} \left( \frac{a}{R_\odot} \right)^{-1/2} \left( \frac{ a }{a_{\text{\jupiter}}} \right)^{-3} \\
		&\sim& 10^4, \text{ for }a = 1\text{ AU}. 
\end{eqnarray}
Thus, a particle passes through the Sun many times during each Kozai cycle.  In the simulations, we find that the total optical depth per Kozai cycle is $\sim 10^2 - 10^3$ times the optical depth at maximum eccentricity.  Even if the optical depth at maximum eccentricity is only $10^{-6}-10^{-5}$ per orbital period (typical of the DAMA simulation), the total optical depth per Kozai cycle is $\sim 10^{-3}$.  It only takes about $1000$ Kozai cycles for such a particle to rescatter in the Sun.   The result is that the lifetimes of particles are typically less than the age of the solar system ($\sim 100$ Myr), and as such cross the Earth's orbit a factor of $\sim 50$ times than predicted by \citeauthor{damour1999}.

To compare our results to Damour \& Krauss, we use Eq. (\ref{eq:n_increase}).  They find that $\epsilon_f \sim 10^{-3}$ of WIMPs with $0.5\hbox{ AU} < a < 1.5 \hbox{ AU}$ initially captured in the Sun will be on a Kozai cycle.  For their typical WIMP-proton cross section $\sigma_p^{SI} \sim 10^{-41} \hbox{ cm}^2$, $\tau \sim 10^{-3}$, so $t_{med} \sim 10^3$ yr, and $E_t \sim 4.5\times 10^6$.  Thus, \citeauthor{damour1999} expect $n^\prime /n \sim 10^3 - 10^4$.  

However, for the same cross section, we find $t^\prime_{med} \sim 10^8$ yr, such that $E_t \sim 10^5$.  Thus, $n^\prime / n \sim 100$, which is approximately the upper limit of what is found in the simulations.  In general, we find $n^\prime / n$ somewhat smaller than $\sim 100$, both because $E_t$ decreases as $\sigma_p$ moves farther below the equilibrium value ($\sigma_P^{SI} \sim 10^{-42}\hbox{ cm}^2$), and because the median lifetime of WIMPs not on Kozai cycles but drawn from the same $a$ and $r_p$ as the Kozai cycle WIMPs is a bit higher than the population of WIMPs with $a < a_{\jupiter}$ as a whole.

We find that we can recover the \citet{damour1999} estimates of the maximum increase in direct detection experiments if the Kozai WIMPs in our simulations had never scattered.  For the DAMA simulation, the median Kozai WIMP lifetime is just short of $100$ Myr (Fig. \ref{fig:weak_tdist}).  If these WIMPs had instead rescattered on timescales longer than the age of the solar system, then we would expect the DF to have been larger by a factor of $\sim 50-100$.  We found that the maximum increase to the differential direct detection rate $\text{d}R/\hbox{d}Q$ (Eq. \ref{eq:drdq}) was $\sim 0.5\%$ of the halo event rate.  If the DF were larger by this factor of $50-100$, then the bound WIMPs would add an additional $25-50\%$ of the halo event rate at small $Q$, consistent with what is found by Damour \& Krauss.  

We can also recover the large neutrino event rate from WIMP annihilation in the Earth found by \citet{bergstrom1999} using the Damour \& Krauss results.  We found that for MSSM models consistent with limits on the WIMP-nucleon elastic scattering cross section, the capture rate of solar-captured WIMPs in the Earth was a maximum of about $10\%$ that of the halo, for $m_\chi \approx 100$ GeV.  If the DF were a factor of $50-100$ higher, the solar-bound WIMP capture rate would be $5-10$ times higher than the halo capture rate.  Since $t_e > t_\odot$ (Eq. \ref{eq:t_e}) for such capture rates, the annihilation rate of WIMPs in the Earth would scale as $\Gamma \propto C^2$, leading to an increase in the neutrino flux in neutrino detectors of $25-100$ times the halo event rate, consistent with what was found by Bergstr{\" o}m et al.  However, we note that even if the enhancement were that high, Fig. \ref{fig:neutrino_grav} shows that this signal would fall below the IceCube flux threshold for WIMP models consistent with experimental constraints.

\subsection{Planets}\label{sec:planetpredict}

Of course, all of the conclusions in this work are based on simulations in a toy solar system, consisting of Jupiter on a circular orbit about the Sun.  Dynamics in the solar system are much more complex, both because Jupiter has non-zero eccentricity and inclination and because other planets are present.  Bodies may have close encounters with any planet within its aphelion, and may be influenced by additional mean-motion and secular resonances \cite[e.g.,][]{farinella1994,duncan1997,gladman1997,dones1999,gladman2000}.  The combination of these effects yields far greater diversity of orbits in the real solar system than what we found in the toy solar system.

There are two qualitatively different ways in which a more realistic treatment of the solar system could change the WIMP distribution at the Earth.  First, additional parts of phase space become accessible.  While it is a triumph of our numerical methods that the Jacobi constant is conserved to high accuracy in our simulations, the conservation of the Jacobi constant restricts the range of motion for WIMPs.   For example, WIMPs with $a < a_{\text{\jupiter}} / 2$ experienced only minor fluctuations in the semi-major axis because they never encountered Jupiter closely enough to experience large energy changes.  Thus, according to the definition of the Jacobi constant, Eq. \ref{eq:ch3_cj}, even WIMPs on quasi-Kozai cycles only experienced relatively minor perturbations to $J_z$.  This meant that WIMPs not crossing Jupiter's orbit had heliocentric velocities perpendicular to the Earth's motion, restricting the geocentric speeds $v \gtrsim v_\oplus \approx 30 \hbox{ km s}^{-1}$.  Jupiter-crossing WIMPs were restricted to geocentric speeds $v\gtrsim 10 \hbox{ km s}^{-1}$ in the toy solar system, which we show in more detail in Chapter 5 of \cite{peter2008}.  However, encounters with other planets can push geocentric WIMP speeds below $v = 10\hbox{ km s}^{-1}$ by increasing $J_z$.  While the presence of a tail in the DF at low geocentric speeds is not significant for direct detection event rates, it can have a disproportionate effect on the capture rate of WIMPs in the Earth.

Secondly, the overall number density of bound WIMPs may change, depending largely on how efficient the planets are at increasing (or decreasing) the lifetimes of WIMPs in the solar system (Eq. \ref{eq:n_increase}).  We will argue below that the true number density of WIMPs at the Earth is unlikely to be much larger or smaller (within factors of a few) than that estimated from simulations using a toy solar system.  

We divide the discussion into three parts: (i) WIMPs with initial $a_i < 1.5$ AU, (ii) WIMPs with $1.5\hbox{ AU} < a_i < 2.6 \hbox{ AU}$ (quasi-Kozai WIMPs in the toy solar system), and (iii) Jupiter-crossing WIMPs.  Without further simulations, though, it is not possible to tell exactly by how much the DF will change.  Hence, we also discuss the challenges involved in simulating WIMPs in a more realistic solar system.

\subsubsection{$a<1.5\hbox{ AU}$}
The DF of solar-captured WIMPs could be greatly increased if the planets other than Jupiter were to either (i) pull a larger percentage of particles out of the rescattering peak and onto orbits that only occasionally enter the Sun or (ii) extend the lifetimes of particles that already did exit the Sun in the toy solar system simulations.  Here, we discuss three mechanisms for pulling additional WIMPs out of the Sun: (i) close encounters with inner planets, (ii) changes to the Kozai structure by other planets, and (iii) additional secular resonances.  Then, we will estimate the lifetimes of such WIMPs.

\emph{Close encounters:} Here, we describe how random-walk encounters with planets can pull WIMPs that were in the rescattering peak in our simulations onto long lifetime orbits in the solar system.  Close encounters with the inner planets can alter the WIMP angular momentum with respect to the Sun.  Ignoring resonant phenomena in the solar system, the close encounters can be treated as a diffusion problem.  We use the rms change in angular momentum as a function of time to estimate the timescales on which WIMP perihelia are pulled out of the Sun.  Modeling WIMP-planet interactions as two-body encounters, each time a WIMP of heliocentric speed $v$ crosses a planet's orbit, the WIMP's planet-centric speed $u$ changes in the direction perpendicular to $u$ by
\begin{eqnarray}
	\delta u \sim \frac{GM_P}{bu}, \label{eq:delta_u}
\end{eqnarray}
where $b$ is the impact parameter.  Since WIMPs with $a < 1.5\hbox{ AU}$ are on extremely eccentric orbits, to good approximation, $u = \sqrt{ v^2 + v_P^2}$, where $v^2_P = GM_\odot / a_P$.  The change in planet-centric speed can be related to the change in heliocentric speed by
\begin{eqnarray}
	\delta v &\sim& \frac{u \delta u}{v} \\
		&=& \frac{GM_P}{bv}. \label{eq:delta_v}
\end{eqnarray}
As a rough approximation, the change in angular momentum per encounter is thus
\begin{eqnarray}
	\delta J \sim a_P \delta v.
\end{eqnarray}
\indent
We use the approximation that the angular momentum undergoes a random walk to estimate the timescale on which a particle's angular momentum changes by of order $\Delta J_{\odot} \sim \sqrt{GM_\odot R_\odot}$ (Eq. \ref{eq:deltaj}) in order for the orbital perihelion to lie outside the Sun.  The rms change in angular momentum will go as
\begin{eqnarray}
	\langle (\Delta J)^2 \rangle \sim N (\delta J)^2, \label{eq:rms_j}
\end{eqnarray}
where the particle encounters planet $P$ with an impact parameter $b$ or less a total of $N$ times in a time span $t$.  In general, 
\begin{eqnarray}
	N \sim \frac{t}{(a_P/b)^2 P_\chi}, \label{eq:N_b}
\end{eqnarray}
where $P_\chi$ is the orbital period of the WIMP.  The factor $(b/a_P)^2$ is the probability per WIMP period that the WIMP comes within a distance $b$ of the planet.  Thus, with some rearranging, we find
\begin{multline}\label{eq:Delta_Jsq}
	\frac{\langle (\Delta J)^2 \rangle}{(\Delta J_{\odot})^2} \sim 10 \left( \frac{M_P}{M_\odot} \right)^2 \left( \frac{a_P}{R_\odot} \right) \left( \frac{a}{a_\oplus} \right)^{-3/2} \\ \times \left( 2 - \frac{a_P}{a} \right)^{-1} \left( \frac{t}{\text{yr}} \right),
\end{multline}
where the factor of $10$ comes from the heretofore ignored Coulomb logarithm (see \cite{binney2008}).  The singularity at $a = a_P / 2$ is artificial and would vanish in a more careful treatment of WIMP-planet encounters.  Thus, the timescale for WIMPs to diffuse out of the Sun due to the action of planet $P$ is 
\begin{multline}\label{eq:angmom_diffusion}
	t_d/ \hbox{yr} \sim 0.1 \left( \frac{M_\odot}{M_P} \right)^{2} \left( \frac{R_\odot}{a_P} \right) \left( \frac{a}{a_\oplus} \right)^{3/2} \\ \times \left( 2 - \frac{a_P}{a} \right).
\end{multline}

For both the Earth and Venus, $M_P/M_\odot \sim 3 \times 10^{5}$ and $R_\oplus / a_P \sim 10^{-2}$, yielding a diffusion time $t_d \sim 10^8$ yr for $a\sim 1$ AU.  The timescales for Mercury and Mars are $t_d \sim 10^{11}$ and $\sim 10^{10}$ years respectively.  Thus, angular momentum diffusion is dominated by the Earth and Venus.  To estimate the impact on the number density, we must find $\epsilon_f$, the fraction of WIMPs with $a < 1.5\hbox{ AU}$ that may be perturbed out of the Sun.  For the DAMA simulation, $t_{med} \sim 10^{3}$ yr, implying that $\epsilon_f \sim t_{med}/t_d \sim 10^{-5}$.  For $\sigma_p^{SI} = 10^{-43} \hbox{ cm}^2$, $\epsilon_f \sim 10^{-3}$.

To estimate the impact of this population on the number density of bound WIMPs, we must also estimate $E_t$, the ratio of the median lifetime including the gravitational effects of the planets to the lifetime if the Sun were isolated. Still ignoring resonances, we estimate the rms timescale for WIMPs to be ejected from the solar system once the perihelia are outside the Sun, 
\begin{eqnarray}
	\langle (\Delta E)^2 \rangle /E^2 \sim \langle (\Delta a)^2 \rangle / a^2 \sim 1.
\end{eqnarray}
Since
\begin{eqnarray}
	\delta a = \frac{a^2}{GM_\odot} v \delta v,
\end{eqnarray}
we use the expression for $\delta v$ in Eq. (\ref{eq:delta_v}) to find
\begin{eqnarray}
	\delta a = \left( \frac{M_P}{M_\odot} \right) \frac{a^2}{b}.
\end{eqnarray}
Using the expression for $N$ in Eq. (\ref{eq:N_b}), we find that
\begin{eqnarray}\label{eq:ejection}
	\frac{\langle (\Delta a)^2 \rangle}{a^2} \sim 10 \left( \frac{M_P}{M_\odot} \right)^2 \left( \frac{a}{a_P} \right)^2 \left( \frac{a}{a_\oplus} \right)^{-3/2} \left( \frac{t}{\text{yr}} \right),
\end{eqnarray}
where again we have included a factor of $10$ for the Coulomb logarithm.  The inner planets which will perturb the orbits the most are Venus and the Earth, yielding ejection timescales of $t_{ej} \sim 10^{10}$ yr, longer than the age of the solar system.  This yields $E_t \sim \hbox{ a few} \times 10^6$ for the DAMA simulation and $E_t \sim \hbox{ a few} \times 10^4$ if $\sigma_p^{SI} = 10^{-43}\hbox{ cm}^2$.  Combined, this would yield $n^\prime / n \sim \hbox{ a few} \times 10$, where $n$ is the number density of the rescattering peak WIMPs in the toy solar system simulations.  This is of the same order as the increase in the bound WIMP DF due to Kozai cycles in our simulations.

However, there are reasons to believe that $E_t$ is in fact significantly smaller than these estimates suggest. First, if a WIMP can diffuse out of the Sun, it can also diffuse back in.  Secondly, once a WIMP becomes Jupiter-crossing, it will be ejected from the solar system on timescales of $\sim$ Myr, which is essentially instantaneous.  

Thirdly, studies of Near Earth Object (NEO) orbits show that once small bodies reach $a \gtrsim 2$ AU, they are driven into the Sun on rather short timescales, $\sim 1-10$ Myr, mostly by secular resonances but also by mean-motion and Kozai resonances \cite{farinella1994,gladman2000}.  Given that WIMPs have significantly higher eccentricity that the typical NEO, the timescale to drive a WIMP back into the Sun via resonances may be shorter.  On the other hand, such WIMP orbits have high speeds relative to the planets, while the low eccentricity, prograde, low inclination NEO orbits have relatively low speeds.  Hence, NEOs will be more efficiently gravitationally scattered onto the mean-motion and secular resonances that drive up the eccentricity.  In spite of this latter effect, it is likely that WIMPs will be scattered back into the Sun on timescales shorter than the age of the solar system.  If the integrated optical depth in each instance that the WIMP perihelion is driven into the Sun (i.e., that the WIMP experiences many Sun-penetrating orbits each time a resonance initially drives the WIMP into the Sun), the lifetime of the WIMPs will be less than the age of the solar system, hence reducing $E_t$.

Fourthly, \citet{gladman2000} have identified additional resonances that drive some NEOs of $a < 1.9\hbox{ AU}$ into the Sun without first boosting the semi-major axis above $a = 2$ AU.  This will reduce $E_t$ for WIMPs with $a < 1.9 \hbox{ AU}$.

However, WIMPs can survive many passes through the Sun before scattering with solar nuclei onto uninteresting orbits.  The timescale for rescattering in the Sun depends crucially on how many passages WIMPs can make through the Sun before gravitational torques from the planets pull the perihelia out again.

In general, it appears that the lifetimes of WIMPs with $ a \lesssim 1.5\hbox{ AU}$ initially pulled out of the Sun by angular momentum diffusion will be shorter than those predicted by arguments based on energy diffusion, although quantifying this is difficult without a full solar system Monte Carlo simulation.  Even if WIMP lifetimes were dominated by diffusion instead of the effects listed above, the boost to the DF would only just be comparable to that due to Kozai cycles in the toy solar system.

\emph{Changes to the Kozai structure:} Next, we consider changes to the Kozai structure caused by planets other than Jupiter.  Both the inner and outer planets can affect the structure of Kozai cycles.  However, torques from the outer planets other than Jupiter are unlikely to change the number of particles whose perihelia exit the Sun.  As demonstrated in Eq. (\ref{eq:torque}), the torque on a particle by a faraway planet goes as $K \propto M_P a^2 a_P^{-3}$, where $M_P$ and $a_P$ are the mass and semi-major axis of the planet, and $a$ is the semi-major axis of the particle.  A planet will provide a torque
\begin{eqnarray}
	K_P = \frac{M_P}{M_{\text{\jupiter}}} \left( \frac{a_{\text{\jupiter}}}{a_P} \right)^3 K_{\text{\jupiter}}
\end{eqnarray}
relative to the torque from Jupiter.  Even Saturn, the next largest planet in the solar system, and the second nearest gas giant to the Earth, will only produce a torque about 5\% that from Jupiter.  Jupiter dominates the tidal field for particles that do not cross the orbits of the outer planets, and so it dominates the structure of the Kozai cycles.

Among the inner planets, \citet{michel1996} find that the Earth and Venus can dominate the structure of the Kozai cycles if the semi-major axis of the particle is near the semi-major axis of either planet, the initial eccentricity of the particle orbits is small, and the maximum inclination of the orbit is low.  However, WIMPs tend to have power-law distributed semi-major axes, high eccentricities, and are scattered isotropically in the Sun.  Therefore, we expect that the extra planets will not increase the number of particles on Kozai cycles in the inner solar system. 

\emph{Secular resonances:} There are additional secular resonances in the full solar system that do not appear in the circular restricted three-body problem considered in this work.  These occur when the rate of change of either the longitude of perihelion ($\dot{\varpi}$) or of the longitude of the ascending node ($\dot{\Omega}$) of the WIMP is almost equal to that of one of the planets.  The evolution of NEOs is greatly affected by the secular resonances with Jupiter and Saturn, although several authors show that other resonances are also important \cite{farinella1994,froeschle1995,michel1997,michel1997b,gladman2000}.  There are complications in interpreting and extending results from NEO simulations. For example, most analytic and numerical effort has focused on the regimes of prograde orbits with small $e$ and $I$ relative to typical WIMPs since most observed NEOs have such properties \cite{williams1969,knezevic1991,lemaitre1994}.

However, just like Kozai cycles, secular resonances should be able to pull WIMP perihelia outside of the Sun if the WIMP orbits originate in the outer layers of the Sun, where the orbital precession due to the Sun's potential is small.  Although there are neither analytic nor numerical investigations of secular resonances for $e>0.995$ relevant for bound WIMP orbits, extrapolating from \citet{williams1981}, it appears that for fixed $a$, the prograde resonances lie at higher inclination for higher $e$, so secular resonances will be relevant at high inclination, as for Kozai cycles \cite{williams1981}.  It is not clear how strong these resonances are, although it is unlikely that they are much stronger than Kozai resonances.

\emph{Lifetimes:}  Since Kozai WIMPs dominate the solar-captured WIMP DF at the Earth in the simulations, it is important to understand the stability of these orbits in the true solar system.  There are two important questions: (i) How long, on average, does it take for a WIMP to be perturbed off a Kozai cycle?  (ii) How does the integrated optical depth per Kozai cycle change?

Since the diffusion approximation has nothing to say about the stability of resonant orbits, we look to simulations of NEOs again for insight.  Unfortunately, NEO simulations are either fundamentally short ($ < 100$ Myr) or end when NEOs hit the Sun, making it difficult to extract estimates of the long-term stability of Kozai cycles.  There are a few hints from even those short simulations with initial conditions significantly different from those of WIMPs.  First, \citet{gladman2000} find examples of NEOs with $a < 2$ AU in Kozai cycles for tens of Myr in their 60 Myr integrations.  The lifetimes of those NEOs is limited only by the termination of the simulations at either 60 Myr or when the body hits the Sun.  Thus, it seems probable that WIMPs born on Kozai cycles will typically stay there for a least of order tens of millions of years, and maybe significantly longer.  If the timescale to perturb a WIMP off a Kozai cycle occurs on timescales similar to the ejection timescale (Eq. \ref{eq:ejection}), then WIMPs can exist on Kozai cycles of order the age of the solar system.  In this case, the DF for WIMPs with $a < 1.5\hbox{ AU}$ should be relatively unchanged.  On the other hand, if the typical timescale for the removal of a WIMP from a Kozai cycle is shorter (such that $\epsilon_f$ becomes larger), the impact on the DF depends crucially on what timescales those WIMPs are then either ejected from the solar system or rescattered in the Sun. 

The structural changes to the Kozai cycles in a more complex solar system ($a$ is no longer constant, frequent switches between librating and circulating modes, $e_{max}$ and $I_{max}$ vary) mean that the integrated optical depth per Kozai cycles may vary with time (see Figs. 7 \& 8 in \citet{gladman2000}).  In principle, this could go up or down; in the case of the quasi-Kozai cycles in our toy solar system, the mean optical depth per Kozai cycle went up due to occasional periods of very high eccentricity.  However, given the accessible phase space for WIMPs in a more realistic solar system, it is quite possible that the mean integrated optical depth per Kozai cycle will go down.  In this case, the WIMP lifetimes will be lengthened, although it is not clear by what amount.

In summary, we predict that the number density of WIMPs with $a_i < 1.5\hbox{ AU}$ will be within factors of a few of the number densities found in the toy solar system, but there are significant error bars in this prediction.  We find that the additional mechanisms to pull WIMPs out of the Sun, angular momentum diffusion or extra secular resonances, will at best yield the same number density as the WIMPs on Kozai cycles in the toy solar system.  

The total number density will likely depend largely on the behavior of WIMPs that were confined to Kozai cycles in the toy solar system.  The DF will depend on the timescales on which WIMPs are removed from Kozai cycles, and the timescales for removal from Earth-crossing orbits after they have been moved from Kozai cycles.  If the WIMP-nucleon cross section lies below the equilibrium cross section ($\sigma_p^{SI} \sim 10^{-42} \hbox{ cm}^2$ or $\sigma_p^{SD} \sim 10^{-40} \hbox{ cm}^2$) for the high plateau, perturbations by the inner planets will \emph{reduce} the Kozai WIMP DF unless both timescales are of order the age of the universe and the mean integrated optical depth per Kozai cycle is much smaller than in the toy solar system.  

However, for cross sections above the equilibrium cross sections, the Kozai WIMP number density will tend to increase. If both timescales are similar to the ejection timescale found in Eq. (\ref{eq:ejection}), and if the mean optical depth per Kozai cycle is similar to what was found in the toy solar system, the number density should be largely unchanged from what we found in this work.  If the timescale for removal of WIMPs from Kozai cycles is significantly less than the ejection timescale, or if gravitational perturbations from the planets systematically decrease the integrated optical depth per Kozai cycle, the DF could be considerably larger.

\subsubsection{$1.5\hbox{ AU} < a < 2.6\hbox{ AU}$}

Given that quasi-Kozai WIMPs have high eccentricity and/or high inclination, they also will generically have high speed encounters with planets.  Thus, we expect the timescale for WIMPs to be removed from quasi-Kozai orbits, $t_q$, to be similar to that of WIMPs on Kozai cycles.  After being removed from a quasi-Kozai orbit, the WIMP should hit the Sun again in $\sim 1-10$ Myr, according to NEO simulations, or get perturbed onto a Jupiter-crossing orbit and get ejected.

We find that $\epsilon_f \sim t_{med}/t_q$, and $E_t \sim t_{max}/ t_{med}$, where $t_{max}$ is $t_\odot$ if the perturbed WIMPs have lifetimes $t_l > t_\odot$, and is equal to the median perturbed lifetime otherwise.  If the other factors in Eq. (\ref{eq:n_increase}) are $\sim 1$, this implies that any boost or deficit in the number density of WIMPs of $1.5\hbox{ AU} < a < 2.6 \hbox{ AU}$ goes as $n^\prime / n \sim t_{max} / t_q$, where $n$ is in this case the number density of quasi-Kozai WIMPs in the toy solar system simulations.  We expect that $t_{max} \leq t_\odot$, and $t_q \gtrsim 100$ Myr (suggested by the relatively short NEO simulations of \citet{gladman2000}).  Thus, $n^\prime / n \lesssim 50$.  If the timescale for the removal of WIMPs from quasi-Kozai orbits is greater or the timescale for rescattering is less than $t_\odot$ (quite possible given the frequency with which NEOs hit the Sun), then $n^\prime / n$ will be correspondingly less.

\subsubsection{Jupiter-Crossing WIMPs}
Before discussing the effects of other planets on Jupiter-crossing particles, we summarize the main features of the Jupiter-crossing DF.  The plateau in the DF was set by $t\sim 10^7$ yr, with growth in the spikes occuring at later times due to long-lived Kozai and resonance-sticking particles.  The vast majority of Jupiter-crossing particles are lost by ejection from the solar system rather than rescattering in the Sun for $\sigma_p^{SI} \lesssim 10^{-41} \hbox{ cm}^2$, although rescattering becomes more important for larger cross sections.  

The outer planets are unlikely to affect the low plateau of the Jupiter-crossing DF.  Jupiter dominates the timescale for Jupiter-crossing WIMPs to be pulled out of the Sun; according to Eq. (\ref{eq:angmom_diffusion}), $t_d\sim 10^3$ yr, while the timescale for any of the outer planets to remove the perihelion of a passing WIMP is at least an order of magnitude longer.  Jupiter also has the shortest WIMP ejection timescale (Eq. \ref{eq:ejection}) by more than a factor of ten.  It dominates the Kozai structure of the types of orbits on which Jupiter-crossing WIMPs originate, and its mean-motion resonances are also by far the strongest in the solar system (unless the orbit of the test particle is exterior to the orbit of Neptune) \cite{thomas1996,duncan1997}.  

However, the outer planets may affect the spikes in the Jupiter-crossing WIMP DF because WIMPs that are long-lived in the toy solar system may not be long-lived in the real solar system.  If the orbital node crossings occur near one of the outer planets, the WIMP may be quickly perturbed from resonant motion and ejected.  Thus, it is possible that the long-lifetime resonance features in the WIMP DF will be less prominent than shown in this work, although even in our simulations, the Jupiter-crossing WIMPs are a subdominant contributor to the number density.  However, shortening the WIMP lifetimes in this way only strengthens our conclusion that the signal from bound WIMPs in neutrino telescopes is unobservably small compared to the signal from unbound WIMPs.

The inner planets may affect the low plateau of the bound WIMP DF for the following reason.  There is a small probability that Jupiter-crossing WIMPs will be gravitationally scattered by an inner planet onto an orbit that no longer crosses Jupiter's.  If this effect were described by diffusion, the net change to the bound WIMP number density would be of order unity; the increase in lifetime $E_t$ would be canceled by the decrease in $\epsilon_t$, since the timescale for both scattering in or out of a Jupiter-crossing orbit is the same if one inner planet dominates the gravitational interaction cross section.  However, the WIMPs could spread into the low geocentric regions of phase space inaccessible in the toy solar system, which is important for capture in the Earth.  The effect of secular or mean-motion resonances on the size of the bound WIMP population and the low speed phase space density is unclear; resonances could drive WIMPs into the Sun, as suggested by NEO and asteroid belt simulations \cite{farinella1994,gladman2000}.  In this case, the importance of resonances depends on how the typical time for rescattering in the Sun relates to the other gravity-dominated times in the solar system.

In summary, we suggest that the height of the low plateau of the Jupiter-crossing WIMP DF and the Jupiter-crossing WIMP number density will be mostly unaffected by the presence of additional planets in the solar system, although the inner planets may extend the plateau in phase space.  We expect that the long-lifetime peaks in the Jupiter-crossing WIMP DF will be lower in a more realistic solar system due to interactions with the outer planets.

\subsubsection{Future Simulations}
In order to test the our arguments above and to definitively determine the bound WIMP DF as a function of WIMP parameters, especially at the low geocentric speeds inaccessible in the three-body problem but which are so important for WIMP capture in the Earth, we would like to perform simulations of WIMP orbits using a more realistic model of the solar system.  The numerical methods presented in Section \ref{sec:num} and Appendix \ref{sec:df_estimator} should be applicable to a more complex solar system with only minor tweaking, so we are eager to use our methods for future simulations.  However, our experience with simulations in a toy solar system, as well as phenomena highlighted in earlier portions of this section, suggest specific challenges to this program.  

The main challenge will be to sample enough orbits to have a statistically significant determination of the DF, and to do this with finite computational resources.  From our simulations in the toy solar system, we have learned that it is important to determine the long-lifetime tail of the WIMP distribution, even if the overall fraction of WIMPs in this population is small.  The DFs were dominated by the small number of particles on Kozai cycles (either $a < 1.5 \hbox{ AU}$ or on Jupiter-crossing orbits) and Jupiter-crossing WIMPs on long-lifetime resonance-sticking orbits, about $\sim 0.1\%$ of all particles simulated.  These rare but long-lived WIMPs, especially the Jupiter-crossing population, also dominated the uncertainties in the DF.  However, even getting to this level of uncertainty required $\sim 10^5$ CPU-hours per simulation.  If we had simulated, say, an order of magnitude fewer WIMPs, we may not have even identified the long-lived Jupiter-crossing population.

A number of effects we identified earlier in this section for orbits in a more realistic solar system will likely affect small WIMP populations.  For example, the fraction of WIMPs with $a < 1.5\hbox{ AU}$ leaving the rescattering peak due to angular momentum diffusion will be small: $\sim 10^{-5}$ for $\sigma_p^{SI} = 10^{-41}\hbox{ cm}^2$ and $\sim 10^{-3}$ for $\sigma_p^{SI} = 10^{-43}\hbox{ cm}^2$.  It will be necessary to simulate vast numbers of WIMPs with $a < 1.5\hbox{ AU}$ to get good statistics on this population and to make sure we do not miss any important effects.

We propose the following techniques to maximize the statistics on the full solar system bound WIMP DF given finite computing time.  First, we propose a series of intermediate simulations before simulating WIMPs in complete solar system to highlight the importance of different types of behavior.  An initial step may be to simulate orbits in a solar system containing Jupiter, the Earth, and Venus (the planets that will likely dominate the behavior of WIMPs with orbits interior to Jupiter's) on circular, coplanar orbits, with the masses of the Earth and Venus scaled up by one or two orders of magnitude.  We choose a low number of planets for simplicity in understanding the simulations, and high masses for the inner planets in order to highlight the diffusion processes described in previous sections, for which the gravitational cross section scales as $M_P^2$ (e.g., Eq. \ref{eq:Delta_Jsq}).  The high planet masses should shorten the diffusion timescales by a factor of $M_P^{-2}$, which would shorten the total integration time.  One might then simulate WIMP orbits in a solar system with massive inner planets but with more realistic planet orbits (highlighting secular resonances), or to simulate WIMP orbits in a solar system with the same three planets on circular orbits, but for which the masses of the Earth and Venus are closer to their true values.  One could then add the outer planets to the simulation.  It may be possible to learn how the WIMP DF scales with the masses of the inner planets in the simulations with higher inner planet masses so the DF could be extrapolated to small planet masses without needing to simulate orbits in a solar system with the true planet masses.  Even if the latter is not possible, we would learn enough from each intermediate simulation to more efficiently run the next set of simulations.

Secondly, we propose weighting the initial conditions to achieve the best statistics with the least amount of computational time.  The optimal weighting for each intermediate simulation will be guided by the results from the previous.  For example, say that we learn from the first intermediate stage we propose, a three-planet solar system with large inner planet masses, that the population of WIMPs with $a < 1.5\hbox{ AU}$ with perihelia perturbed out of the Sun by angular momentum diffusion is significant for neutrino telescope event rates.  Perhaps the effect is dominated by WIMPs initially scattered into a very narrow range of semi-major axes.  If we then wish to simulate this population in a solar system in which the Earth and Venus have their true masses, it makes sense to focus the computational resources on this narrow semi-major axis window.  Furthermore, in order to gain good statistics for this window, we would need at least of order $10^3$ long-lived angular momentum-diffused WIMPs.  For $\sigma_p^{SD} = 10^{-41}$, $\epsilon_f \sim 10^{-5}$.  If we want a sample of at least $10^3$ WIMPs in this population, we would need to simulate $\sim 10^8$ WIMPs.  However, $\sim 0.1\%$ of these WIMPs, or $10^5$ total, should initially be on Kozai cycles.  In order to focus on the angular diffusion population instead of the Kozai population, we would simulate all $\sim 10^8$ WIMPs for a short time, $\sim 10^5-10^6$ years, which would be sufficient to identify the Kozai population.  At that point, we would only continue simulations of the WIMPs not identified as Kozai cycling.  Thus, we would have good statistics on one WIMP population without burning resources on less important populations.

Therefore, while we believe that getting good statistics for estimating the event rates in neutrino telescopes will be difficult, it will be possible given (i) a clever and adaptive simulation strategy, and (ii) patience to acquire a sufficient number of CPU cycles.

\subsection{The Halo Distribution Function}\label{sec:halodfpredict}
Throughout the simulations, we assumed that the halo WIMP DF was smooth, non-rotating in an inertial Galactocentric frame (lagging the Sun by a speed $v_\odot = 220 \hbox{ km s}^{-1}$), and had a velocity dispersion of $\sigma = v_\odot / \sqrt{2}$.  These choices are motivated by N-body simulations of Milky Way-mass dark matter halos \cite{moore2001,helmi2002}.  However, there are a few severe limitations to these N-body simulations.  First, while we hope that the simulations are a good representation of the real Milky Way, there is no way we can directly measure the dark matter phase space density.  Secondly, these simulations do not include baryons, although we know baryons dominate the gravitational potential within the solar circle.  Simulations that include a treatment of baryonic disks and the accretion of dwarf galaxies suggest that the local phase space structure of dark matter depends sensitively on the accretion history of the Milky Way \cite{read2008}. Thirdly, dark matter is fundamentally clumpy, with the smallest halos corresponding to the size of the free-streaming scale \cite{hofmann2001,green2004,diemand2005}, which for a SUSY WIMP corresponds to about $M\sim M_\oplus$ or length scales of $\sim 10^{-2}$ pc.  While high-resolution simulations show that very little ($\sim 0.1-0.5\%$ \cite{diemand2008,springel2008}) dark matter within the solar circle is in resolved subhalos, these simulations can only probe subhalo masses down to $M\sim10^5-10^6M_\odot$.  There is thus an uncertainty in the degree of clumpiness spanning more than 10 orders of magnitude in mass \cite{kamionkowski2008}.  Here, we describe how the DF will change if any of the assumptions of our fiducial halo model are challenged.  

We note that the primary change to the DF will be in normalization, not shape.  The only way to change the shape of the bound WIMP DF relative to that calculated for our fiducial model for fixed $m_\chi$ and elastic scattering cross section is to change the distribution of semi-major axes or locations of initial scatter in the Sun onto Earth-crossing orbits.  The former is robust over several orders of magnitude in WIMP mass.  The latter may be significant for large ($m_\chi \gtrsim 1$ TeV) WIMP masses if the velocity phase space is radically different from the fiducial model, but will not be significant as long as there is non-trivial phase space density of WIMPs at low heliocentric speeds.  

However, the height of the DF is proportional to $\dot{N}_\oplus$, which is increasingly sensitive to the low speed end of the halo DF for increasingly massive WIMPs.  This is because the halo WIMP energy is $E = m_\chi v_\infty^2 / 2$ (where $v_\infty$ is the heliocentric speed in the absence of the Sun's gravity) but the maximum energy a WIMP can lose in a collision with a solar nucleus is $Q_{max} = 2\mu_A^2 v^2(r) / m_A$, so it becomes hard to scatter high mass WIMPs, high energy WIMPs onto bound orbits.  If the low speed phase space density were increased, $\dot{N}_\oplus$ would increase, and the bound WIMP density would increase relative to the halo density.  This could be achieved, for example, if the WIMP halo were rotating in the same sense as the stellar disk, reducing the speed relative speed between the halo and the Sun.  Conversely, if the low speed halo WIMP density were decreased, the bound WIMP population would be even more insignificant with respect to the halo.

While clumpiness in the halo may affect the halo DF at the Earth (although it is unlikely that a subhalo is currently passing through the solar system \cite{kamionkowski2008}), it will have surprisingly little effect on the DF of WIMPs bound to the solar system if the rate at which clumps pass through the solar system is either much higher or lower than the equilibrium timescale for the bound WIMP DF.  In the former case, as long as the velocity distribution of the ensemble of subhalos is similar to that of the smooth DM component (if the rate at which clumps enter the solar system is high), the bound WIMP DF is proportional to the time-averaged capture rate in the Sun, $f(v)\propto \langle \dot{N}_\oplus \rangle$.  This is unlikely to be significantly different from $\dot{N}_\oplus$ calculated for a purely smooth halo unless the solar system is deeply embedded in a dense subhalo.  In the latter case, passages of a subhalo through the solar system are so infrequent that the DF is dominated by the smooth component in the halo.  

\citet{diemand2005} estimate that if all Earth-mass subhalos survive intact to the present, the rate at which subhalos pass through the solar system is $\sim 10^{-4} \hbox{ yr}^{-1}$, with each passage lasting $\sim 50$ yr.  \citet{diemand2004} and \citet{faltenbacher2006} find that the velocity distribution of subhalos is only slightly biased with respect to the smooth component, with the major discrepancy being a decrement of subhalos with low Galactocentric speeds due to merging.  The escape velocity from a subhalo is much smaller than either any characteristic speed in the solar system or characteristic speeds in the solar neighborhood, making it unlikely that the Sun is bound to a subhalo.  Thus, even if dark matter in the solar neighborhood were highly clumpy, the bound WIMP DF would resemble that estimated in this work.

\section{Conclusion}\label{sec:conclusion}
In conclusion, we highlight the key points of this paper:
\begin{enumerate}
	\item We have developed numerical methods to efficiently track the highly eccentric solar-captured orbits from their initial scatter in the Sun to up to 4.5 Gyr without secularly increasing errors in the Jacobi constant and without numerical precession.  These methods will be employed in future simulations of WIMPs in a more realistic solar system, and may be used to simulate eccentric orbits in other hierarchical systems in which one central body dominates the gravitational potential.  
	\item We have characterized the bound WIMP DF at the Earth as a function of WIMP mass $m_\chi$ and spin-independent $\sigma_p^{SI}$ and spin-dependent $\sigma_p^{SD}$ elastic scattering cross sections.  For the range of masses $m_\chi = 60 \hbox{ AMU} - 500 \hbox{ AMU}$, we find very little variation in the WIMP DFs aside from the mass-dependent rate at which WIMPs scatter onto Earth-crossing orbits.  In contrast to \citet{damour1999}, we find that the optical depth in the Sun to WIMPs imposes a ceiling to the size of the WIMP DF.  For WIMPs that do not intersect Jupiter's orbit, the equilibrium DF is reached for $\sigma_p^{SI} \sim 10^{-42}\hbox{ cm}^2$ and $\sigma_p^{SD} \sim 10^{-40}\hbox{ cm}^2$.  For WIMPs that intersect Jupiter's orbit, equilibrium is reached for $\sigma_p^{SI} \sim 10^{-38} \hbox{ cm}^2$ or $\sigma_p^{SD} \sim 10^{-36}\hbox{ cm}^2$.
	\item The maximum phase space density of WIMPs at the Earth consistent with current constraints on the elastic scattering cross section is significantly less than that of WIMPs unbound to the solar system.  Even though bound WIMPs occupy the low velocity phase space that disproportionally contributes to the event rates in both direct detection experiments and neutrino telescopes, the total enhancement to those event rates is negligible.  For direct detection experiments, we find that the maximum enhancement to $\hbox{d}R/\hbox{d}Q$ occurs at $Q=0$ and is $\lesssim 0.5\%$ of the halo event rate.  For the XENON10 experiment, we predict the maximum enhancement integrated over their analysis window is of order $10^{-3}\%$.  In the MSSM, we find less than order unity enhancements to the neutrino-induced muon event rate in neutrino telescopes from the annihilation of solar-captured WIMPs in the Earth.
	\item Although we only include one planet (Jupiter) in our toy solar system, we do not expect that our conclusions would be significantly different than if we had included more planets in our simulations.  If the other planets are efficient at putting solar-captured WIMPs at geocentric speeds $v < 30 \hbox{ km s}^{-1}$, there may be large increase in the event rate at neutrino detectors due to WIMP annihilation in the Earth.  However, it is unlikely that the boost will be sufficient to move the event rate above the detection threshold for the IceCube neutrino telescope unless the halo WIMP DF is significantly different from the fiducial model.
\end{enumerate}
In two other papers in this series, we examine the impact of the finite optical depth in the Sun and gravitational interactions between WIMPs and Jupiter on the rate of WIMP annihilation in the Sun (Paper II); and we characterize the population of WIMPs bound to the solar system by gravitational interactions with Jupiter (Paper III).

\begin{acknowledgments}
We thank Scott Tremaine for advising this project, and Aldo Serenelli and Carlos Penya-Garay for providing tables of isotope abundances in the Sun. We acknowledge financial support from NASA grants NNG04GL47G and NNX08AH24G and from the Gordon and Betty Moore Foundation. The simulations were performed using computing resources at Princeton University supported by the Department of Astrophysical Sciences (NSF AST-0216105), the Department of Physics, and the TIGRESS High Performance Computing Center.  
\end{acknowledgments}

\appendix
\section{WIMP Elastic Scattering}\label{app:elastic}

\subsection{Spin-Independent Scattering}\label{app:si}
For particle physics models of dark matter, the general spin-independent (``SI''; scalar) scattering cross section has the form \citep{jungman1996,hooper2007}:
\begin{eqnarray}\label{dsdQ}
	\frac{d\sigma^{SI}}{dQ} = \frac{2m_A}{\pi g_A^2} \left[Z f_p + (A-Z)f_n\right]^2 F_{SI}^2(Q),
\end{eqnarray}
where $Q$ is the energy transferred from the WIMP to a nucleus of mass $m_A$ (with atomic mass $A$ and charge $Z$) during the scatter, $g_A$ is the relative velocity between the particles, $f_p$ and $f_n$ are the proton and neutron effective couplings to the WIMP, and $F_{SI}(Q)$ is a nuclear form factor.  The nuclear form factor used in this set of calculations is of the standard exponential form,
\begin{eqnarray}\label{ff}
	F_{SI}(Q) = e^{ - Q/2Q_A},
\end{eqnarray}
where the coherence energy is
\begin{eqnarray}
	Q_A = \frac{1.5 \hbar^2}{m_A R_A^2},
\end{eqnarray}
and the coherence length (the radius of the nucleus A) is set to
\begin{eqnarray}
	R_A = 1 \text{ fm} [0.3 + 0.91 (m_A / (\text{GeV}/c^2) )^{1/3} ].
\end{eqnarray}
The nuclear form factor quantifies the extent to which the WIMP interacts coherently with the nucleus as a whole (if the de Broglie wavelength of the nucleus is small), or incoherently with the nucleons individually.  
\newline\indent
It is often more convenient to use the center-of-mass differential cross section.  Using the functional form of the energy transfer 
\begin{eqnarray}
	Q = 2 \frac{ \mu_A^2}{m_A} g_A^2 \left( \frac{1 - \cos \theta}{2} \right), \label{eq:Q}
\end{eqnarray}
where 
\begin{eqnarray}
\mu_A = \frac{m_A m_{\chi}}{m_A + m_{\chi}}, 
\end{eqnarray}
the differential cross section is
\begin{eqnarray}\label{dsig_dOmega}
	\frac{d\sigma^{SI}}{d\Omega} &=& \frac{1}{2\pi} \frac{dQ}{d(\cos\theta)} \frac{d \sigma}{d Q}  \\
		&=& \frac{1}{2\pi} \frac{\mu_A^2}{m_A} g_A^2 \left( \frac{d\sigma}{dQ} \right) \\
		&=& \frac{1}{4\pi} \frac{4}{\pi} \mu_A^2 \left[ Z f_p + (A-Z) f_n \right]^2 F^2(Q) \\
		&=& \frac{ \sigma^{SI}_A F^2(Q(\cos \theta))}{4\pi}. \label{eq:dsdosi}
\end{eqnarray}
We have parameterized the strength of the interaction by $\sigma_A$.  If $f_p = f_n$, which is often a good approximation for both supersymmetric and UED models, 
\begin{eqnarray}
	\sigma^{SI}_A = \frac{4}{\pi}\mu_A^2 A^2 f_n^2,
\end{eqnarray}
so that the strength of the coupling between a nucleus and the WIMP depends only on the atomic number of the nucleus.  This coupling can also be parameterized in terms of the strength of the WIMP-proton (or -neutron) cross section:
\begin{eqnarray}
	\sigma^{SI}_A = \frac{\mu_A^2}{\mu_p^2} A^2 \sigma^{SI}_p,
\end{eqnarray}
which is useful since experimental constraints on the spin-independent cross section are reported in terms of the WIMP-nucleon cross section.  In the limit of high WIMP mass,
\begin{eqnarray}
	\mu_A &\rightarrow & m_A \\
	\mu_p &\rightarrow & m_p \\
	\sigma^{SI}_A &\rightarrow & \frac{m_A^2}{m_p^2} A^2 \sigma^{SI}_p \\
		&\approx& A^4 \sigma^{SI}_p,
\end{eqnarray}
where the last approximation can be made since $m_A \approx A m_p$.

\subsection{Spin-Dependent Scattering}\label{app:sd}
The likely WIMP candidates for both the MSSM (neutrino $\chi$) and UED (Kaluza-Klein photon $B^{(1)}$) theories can have elastic axial-vector interactions with quarks, via squarks in the MSSM or the lightest  Kaluza-Klein excitation of quarks $q^{(1)}$ in UED models.  In both cases, the spin-dependent (SD) WIMP interaction with a nucleus of atomic number $A$ can be parameterized as \citep{jungman1996,servant2002}
\begin{eqnarray}\label{eq:dsdqsd}
	\frac{\mathrm{d}\sigma^{SD}}{\mathrm{d}Q} = \alpha \times \frac{2 m_A}{\pi g_A^2} \Lambda^2 J(J+1) F_{SD}^2(|\mathbf{q}|),
\end{eqnarray}
where
\begin{eqnarray}
	\alpha = \begin{cases}
	8 G_F^2 &\text{MSSM} \\
	\frac{\displaystyle 1}{\displaystyle 6}\frac{\displaystyle g^{\prime 4}}{ \displaystyle (m^2_{B^{(1)}} - m^2_{q^{(1)}})^2 } &\text{UED}
	\end{cases}
\end{eqnarray}
parameterizes the coupling in each theory.  Here, $g^\prime$ is the coupling constant for the $B$ boson in electroweak theory, and $m_{B^{(1)}}$ and $m_{q^{(1)}}$ are the masses of the $B^{(1)}$ and $q^{(1)}$ particles respectively.  The other quantities in Eq. (\ref{eq:dsdqsd}) depend on nuclear properties.  Here $J$ is the total angular momentum of the nucleus, and
\begin{eqnarray}
	\Lambda = \frac{1}{J} \left[ a_p \langle S_p \rangle + a_n \langle S_n \rangle \right],
\end{eqnarray}
where $a_n$ and $a_p$ describe the WIMP couplings to the neutron and proton, and $\langle S_n \rangle$ and $\langle S_p \rangle$ are the spin expectation values for the neutrons and protons within the nucleus.  The couplings $a_n$ and $a_p$ are derived from specific WIMP models, while the spin expectation values must be calculated using detailed nuclear physics models \citep[e.g.,][]{dimitrov1995,jungman1996,ressell1993,ressell1997}, and calculations using different techniques often yield different results.  The function $F_{SD}(|\mathbf{q}|)$ is the spin-dependent nuclear form factor as a function of the momentum transfer $|\mathbf{q}|$.  Its form must be carefully calculated for each nucleus of interest \citep[][and references therein]{gondolo1996}.
\newline\indent
There are several important differences between the form of the spin-dependent and spin-independent cross sections that have major implications for detection experiment design.  The first point is that nuclei with even numbers of protons and neutrons will have \emph{zero} spin-dependent interactions with WIMPs.  Secondly, the spin-dependent cross section has a much weaker dependence on the atomic mass than the spin-independent cross section.  This is apparent if Eq. (\ref{eq:dsdqsd}) is written in the same form as Eq. (\ref{eq:dsdosi}),
\begin{eqnarray}
	\frac{\mathrm{d}\sigma^{SD}}{\mathrm{d}\Omega} &=& \frac{1}{2\pi} \frac{\mathrm{d}Q}{\mathrm{d} \cos \theta} \frac{\mathrm{d}\sigma^{SD}}{\mathrm{d}Q} \\
	& = & \frac{1}{2\pi} \frac{ \mu^2_A g_A^2}{m_A} \frac{2 m_A}{\pi g_A^2} J (J+1 ) \alpha \Lambda^2 \nonumber \\
	& & \; \; \; \; \; \; \; \; \; \times \, F_{SD}^2(|\mathbf{q}|) \\
	& = & \frac{1}{4\pi} \sigma_A^{SD} F^2(|\mathbf{q}|), 
\end{eqnarray}
where
\begin{eqnarray}
	\sigma_A^{SD} = \frac{4}{\pi} \mu_A^2 J (J+1) \alpha \Lambda^2.
\end{eqnarray}
In the limit that $m_{WIMP} \gg m_A$,
\begin{eqnarray}
	\sigma_A^{SD} \propto A^2,
\end{eqnarray}
unlike
\begin{eqnarray}
	\sigma_A^{SI} \propto A^4
\end{eqnarray}
for the spin-independent case.  Therefore, even if $\sigma_p^{SD} > \sigma_p^{SI}$ or $\sigma_n^{SD} > \sigma_n^{SI}$, the spin-independent cross section may dominate for heavy nuclei.  The spin-dependent cross section could be large if $J$ scaled with $A$ (since $\sigma_A \propto J^2$), but this is not the case for heavy nuclei.  Note that, in contrast to predictions for spin-independent scattering, the spin-dependent WIMP-proton and WIMP-neutron cross sections are generally \emph{not} the same to within a few percent.

\section{Subsequent Scattering in the Sun}\label{sec:tau}

Each time a particle passes through the Sun, there is a probability
\begin{eqnarray}
	P_{\mathrm{scatt}} = 1 - e^{-\tau}
\end{eqnarray}
that it will be scattered at least once, given the optical depth $\tau$ for one jaunt through the Sun.  Since the WIMP-nucleon cross sections relevant to this paper imply low opacity in the Sun ($\tau \lesssim 10^{-3}$), the scattering probability per solar passage is well approximated by
\begin{eqnarray}
	P_{scatt} &=& 1 - \left( 1 - \tau + \mathcal{O} (\tau^2) \right) \\
		  &\approx & \tau.
\end{eqnarray}

Instead of calculating the scattering probability $\tau$ on the fly, we create a table for optical depth indexed by the semi-major axis and Kepler perihelion of the orbit, and then interpolate for a particular orbit through the Sun.
\newline\indent
The optical depth in differential form is given by
\begin{eqnarray}
	\frac{\mathrm{d}\tau}{\mathrm{d} \mathit{l} \mathrm{d} Q}  &=& \sum_A \frac{\mathrm{d}\tau_A}{\mathrm{d} \mathit{l} \mathrm{d} Q} \\ 
	&=& \sum_A n_A(\mathit{l}) \frac{\mathrm{d} \sigma_A}{\mathrm{d} Q},
\end{eqnarray}
where $\mathit{l}$ denotes the particle trajectory, $n_A(\mathit{l})$ is the number density of element $A$ in the Sun at position $\mathit{l}$ along the path, and $\mathrm{d}\sigma_A / \mathrm{d} Q$ is the differential elastic scattering cross section with respect to the energy transfer $Q$ between element $A$ and the WIMP.  Since we assume that spin-independent scattering dominates in the Sun, the integral over energy transfer can be computed using the form of the differential cross section in Eq. (\ref{dsdQ}) and the form factor in Eq. (\ref{ff}):
\begin{eqnarray}
	\frac{\mathrm{d} \tau}{\mathrm{d} \mathit{l} } &=& \sum_A n_A( \mathit{l} ) \int_0^{Q_{max}} \frac{\mathrm{d} \sigma_A}{\mathrm{d} Q} \\
		&= &\sum_A n_A( \mathit{l} ) \frac{2m_A}{\pi v(\mathit{l})^2} [ Zf_p + (A-Z) f_n]^2 Q_A \\
		& & \hskip2.6cm \times \left( 1 - e^{- Q_{\max, A} / Q_A} \right), \nonumber \label{eq:dtau_dl}
\end{eqnarray}
where we have used the approximation of a zero-tem\-per\-a\-ture Sun to set $v_{rel} = v(\mathit{l})$.  Using Eq. (\ref{Q}), we find the maximum energy transfer
\begin{eqnarray}
	Q_{\max, A} = 2 \frac{\mu_A^2}{m_A} v(\mathit{l})^2.
\end{eqnarray}
\indent
The integration of Eq. (\ref{eq:dtau_dl}) is greatly simplified because the torque on the particle by Jupiter is negligible in the Sun compared to the rest of the orbit.  Therefore,
\begin{eqnarray}
	\mathrm{d} \mathit{l} &=& v(t) \mathrm{d}t \\
			      &=& v(t(r)) \left| \frac{ \mathrm{d} t}{ \mathrm{d} r } \right| \mathrm{d} r \\
			      &=& \frac{v(r(t))}{\left|v_r(r(t))\right|} \mathrm{d} r,
\end{eqnarray}
where 
\begin{eqnarray}\label{magvel}
	v(E, r) = \sqrt{ 2[E - \Phi_\odot(r)] }
\end{eqnarray}
is the particle's speed and 
\begin{eqnarray}
	|v_r(E, J, r)| = \sqrt{ 2[E - \Phi_\odot(r)] -J^2/r^2 }
\end{eqnarray} 
is the radial velocity of the particle. Thus, 
\begin{eqnarray}
	\frac{\mathrm{d} \tau(E,J)}{\mathrm{d} r } = \frac{ v(r) }{|v_r(r)|} \frac{\mathrm{d} \tau}{\mathrm{d} \mathit{l}}.
\end{eqnarray}
and the total optical depth along the path is
\begin{multline}
	\tau( E, J ) = \frac{4}{\pi} \sum_A m_A Q_{A} [Z f_p + (A-Z) f_n]^2  \label{eq:tau} \\
		\times \int_{r_p}^{R_\odot} \text{d}r \frac{n_A( r )\left( 1 - e^{- 2 \mu_A^2 v^2(E,r)/ m_A Q_A} \right)}{v(E, r) | v_r( E, J, r) |} .
\end{multline}

In order to express the optical depth $\tau$ as a function of the semi-major axis and eccentricity, We use the relations
\begin{eqnarray}
	E &=& \pm \frac{GM_c}{2a} \\
	J^2 &=& \pm GM_c a (e^2 - 1),
\end{eqnarray}
where $M_c = \mu M_\odot$ is the central mass, as determined by Eq. (\ref{mu}), and the upper (lower) sign is used for hyperbolic (elliptical) orbits.  Therefore,
\begin{eqnarray}
	\tilde{v}(a,r) &= &v(a,r)/\sqrt{GM_c} \nonumber \\
		       &= &\sqrt{2\left[\pm \frac{1}{2a} - \tilde{\Phi}_\odot (r)\right] } 
\end{eqnarray}
\begin{eqnarray}
	|\tilde{v}(a,e,r)| &= &|v_r(a,e,r)|/\sqrt{GM_c}\nonumber \\
			   &= &\sqrt{ 2\left[\pm \frac{1}{2a} - \tilde{\Phi}_\odot (r)\right] \mp \frac{a(e^2 - 1)}{r^2}},
\end{eqnarray}
where $\tilde{\Phi}_\odot = \Phi_\odot / GM_c$.  If we insert these expression into Eq. (\ref{eq:tau}),
\begin{eqnarray}
\tau( a, e ) &=& \frac{4}{\pi} \frac{1}{GM_c}\sum_A m_A Q_{A} [Z f_p + (A-Z) f_n]^2 \\
	     & & \times \int_{r_p}^{R_\odot} \mathrm{d}r \frac{n_A( r ) \left( 1 - e^{- 2 \mu_A^2 GM_c \tilde{v}^2(a, r)/ m_A Q_A} \right)}{\tilde{v}(a, e, r) | \tilde{v}_r( a, e, r) |}. \nonumber
\end{eqnarray}
We make a look-up table for $\tau$ using for the choice $\mu = 1$, and then scale $\tau$ by a factor of $\mu^{-1}$.  There is also a factor of $\mu$ in the exponent.  However, its impact on $\tau$ is negligible since $| \mu -1 | \lesssim 10^{-6} - 10^{-5}$.
\newline\indent
If the particle scatters in the Sun, its new phase space coordinates can be determined by sampling the scattering distribution 
\begin{eqnarray}\label{tau_dist}
	\frac{\mathrm{d}\tau( E, J)}{\mathrm{d} r \mathrm{d} \Omega} &=& \sum_A n_A(r) \frac{ v(E, r)}{|v_r(E, J, r)|} \frac{\mathrm{d}\sigma_A}{\mathrm{d} \Omega},
\end{eqnarray}
where $\Omega$ is the center-of-mass scattering solid angle.

\section{Distribution Function Estimators}\label{sec:df_estimator}
In this section, we describe the outputs of the simulations, and how to estimate the bound distribution function from these data.

Our method is to find the average DF along Earth's path. We record the phase space coordinates of particles passing near the Earth's orbit.  Since we treat the Earth's orbit as circular and coplanar with Jupiter's orbit, this means that we focus on particles passing through the wall of a cylinder of height $2z_c$ centered on the reference plane and radius $a_\oplus$ from the Sun.  Thus, the raw data product is the flux of dark matter particles through the Earth's orbit as a function of time.  

To convert the flux at position $\mathbf{x}$ and time $t$, $F(\mathbf{x},t)$, into a DF $f(\mathbf{x},\mathbf{v},t)$, we assume that the timescale of variation in the distribution function is much larger than the typical dynamical timescale of particles in the solar system ($\sim$ year). We adopt the usual argument \citep[cf.][]{reif1965} to relate the flux as a function of velocity $\hbox{d}F/\hbox{d}\mathbf{v}$ to the distribution function.  Consider particles passing \emph{outward} through a wall of area $\delta A$ with a unit vector normal to the surface $\hat{\mathbf{n}}$.  For particles with velocity between $\mathbf{v}$ and $\mathbf{v} + \delta \mathbf{v}$, the particles that pass through the wall in time $\delta t$ inhabit a prism volume of base $\delta A$, long side $v\delta t$, and height $\delta t \mathbf{v} \cdot \hat{\mathbf{n}}$.  The total number of particles with velocity between $\mathbf{v}$ and $\mathbf{v} + \delta \mathbf{v}$ passing \emph{out} through the surface $\delta A$ per unit time $\delta t$ is
\begin{eqnarray}
	\frac{ \mathrm{d} F(\mathbf{x}, t)}{\mathrm{d}\mathbf{v}} \mathrm{d} \mathbf{v} \delta A \delta t  &= f(\mathbf{x}, \mathbf{v}, t) (\mathbf{v}  \delta t) \cdot (\delta A \hat{\mathbf{n}}) \mathrm{d} \mathbf{v} \\
		&= f (\mathbf{x},\mathbf{v},t) v \cos \gamma \mathrm{d} \mathbf{v} \delta A \delta t,
\end{eqnarray}
where $\cos \gamma = \mathbf{v} \cdot \hat{\mathbf{n}} / v$.  In the simulations, we do not care if the particles pass inward or outward through the wall of the cylinder, so we estimate the distribution function from the simulations using
\begin{eqnarray}\label{fluxarea}
	\left| \frac{ \text{d} F(\mathbf{x},t)}{\text{d}\mathbf{v}} \right| \mathrm{d} \mathbf{v} \delta A \delta t = f(\mathbf{x}, \mathbf{v}, t) v |\cos \gamma| \mathrm{d} \mathbf{v} \delta A \delta t, 
\end{eqnarray}
or
\begin{eqnarray}
	f (\mathbf{x}, \mathbf{v}, t) &=& \left| \frac{ \text{d}F(\mathbf{x}, t) / \text{d}\mathbf{v} }{v \cos \gamma }\right| \\
				&=& \left| \frac{\text{d}F(\mathbf{x},t) / \text{d} \mathbf{v} }{| v_r|} \right| \label{eq:f_from_F},
\end{eqnarray}
since $v_r = v \cos \gamma$ is the velocity component normal to the wall of the cylinder (i.e., the radial component of the velocity).

We now describe in detail how to estimate the distribution function from the data obtained in the simulations.  For each simulation, we start integrating the orbits of $N_p$ particles (Table \ref{tab:weakic}) at time $t_i$ since the birth of the solar system.  Particles scatter onto bound, Earth-crossing orbits at a rate $\dot{N}_\oplus(t_i)$, where $t_i$ is the time at which the particle first scatters onto a bound orbit.  In principle, $\dot{N}_\oplus$ can vary with time if the halo dark matter distribution function varies on timescales shorter than the age of the solar system, but we assume that the halo distribution function is static, so that $\dot{N}_\oplus(t_i) = \dot{N}_\oplus$.  

Each time a particle $\alpha$ crosses through the cylinder wall, we record the time of passage $t_{\alpha\beta}$ (here, $\beta$ labels the particular passage of the particle $\alpha$ through the Earth's orbit) since the start of the simulation at $t_i$, position $\mathbf{x}_{\alpha\beta}$, and velocity $\mathbf{v}_{\alpha\beta}$.  The height $z_c$ is chosen to be larger than the radius of the Earth $R_\oplus$ in order to improve statistics, but is small enough ($z_c \ll 1$ AU) so that the estimate should be unaffected by gradients in flux as a function of height above the reference plane.

Each particle crossing can be characterized as one point in a six-dimensional phase space: $\mathbf{n}_{\alpha\beta}$, the vector describing the orientation $(\phi, z)$ of the particle when it crosses the cylinder of radius $a_\oplus$; the three components of the velocity $\mathbf{v}_{\alpha\beta}$; and $t_{\alpha\beta}$.  The vector $\mathbf{n}_{\alpha\beta}$ only has two independent coordinates since the radial component of $\mathbf{x}_{\alpha\beta}$ is fixed.  We estimate the flux of particles passing passing through a patch of the cylinder at position $\mathbf{n}$ in the cylinder at time $t$ since the birth of the solar system, for which the particles had initial scattering time in the Sun at time $t_i$, with velocity between $\mathbf{v}$ and $\mathbf{v} + \mathrm{d}\mathbf{v}$,  as
\begin{multline}
	\frac{\mathrm{d}\hat{F}}{\mathrm{d}\mathbf{v} \mathrm{d}t_i} = \left( 1/ \int \text{d} \mathbf{\lambda} \sum^{N_p}_{\alpha = 1} w(\mathbf{\lambda})\delta( \mathbf{\lambda} - \mathbf{\lambda}_\alpha) \right)  \label{eq:largeflux} \\
	\times \sum^{N_p}_{\alpha = 1} \sum^{N_\alpha}_{\beta = 1} \dot{N}_\oplus w(\mathbf{\lambda}_\alpha) \delta^{(6)}( \mathbf{n} - \mathbf{n}_{\alpha\beta}, \\ 
	\mathbf{v} - \mathbf{v}_{\alpha\beta}, t - (t_i + t_{\alpha\beta})) 
\end{multline}
for each experiment.  Here, $\hat{F}$ denotes that this is an \emph{estimator} for the true flux $F$.  The total flux can be estimated by integrating Eq. (\ref{eq:largeflux}) over $t_i$ and $\mathbf{v}$.  $N_\alpha$ is the total number of times particle $\alpha$ crosses the Earth's orbit. The weight function $w(\mathbf{\lambda})$ describes how we sample the initial conditions $\mathbf{\lambda}$ relative to the initial particle distribution at the first scatter.  The denominator of Eq. (\ref{eq:largeflux}) normalizes the flux.    
\newline\indent
Since we sample the bound, Earth-crossing WIMPs to the same density as they scatter onto such orbits in the solar system, $w = 1$ for each particle $\alpha$.  Thus,
\begin{eqnarray}
	\sum_{\alpha = 1}^{N_p} w(\mathbf{\lambda}) \delta(\mathbf{\lambda} - \mathbf{\lambda}_\alpha)&=& \sum_{\alpha = 1}^{N_p} \delta (\lambda - \lambda_\alpha)
\end{eqnarray}
so that
\begin{eqnarray}
	\int \text{d}\mathbf{\lambda} \sum_{\alpha =1}^{N_p} w(\mathbf{\lambda}) \delta(\mathbf{\lambda} - {\lambda}_\alpha) = N_p,
\end{eqnarray}
where the integral over $\mathbf{\lambda}$ spans the entire range of $\mathbf{\lambda}_\alpha$.  The flux at position $\mathbf{n}$ as a function of velocity, observation time, and initial time $t_i$ is
\begin{multline}
	\frac{\mathrm{d}\hat{F}}{\mathrm{d}\mathbf{v}\mathrm{d}t_i}  = \frac{\dot{N}_\oplus}{N_p} \sum^{N_p}_{\alpha = 1} \sum^{N_\alpha}_{\beta = 1} \delta^{(6)}( \mathbf{n} - \mathbf{n}_{\alpha\beta}, \mathbf{v} - \mathbf{v}_{\alpha\beta}, \\
	t - (t_i + t_{\alpha\beta})).
\end{multline}

We are interested in the flux arising from particles entering the solar system at all times prior to the present, not just at a particular time $t_i$.  Therefore, to estimate the total flux in a unit volume of velocity-space, one must integrate Eq. (\ref{eq:largeflux}) over $t_i$, in the range between the time of the formation of the solar system and the time at which the flux is measured, 
\begin{eqnarray}\label{dFdt}
	\frac{\text{d}\hat{F}}{\text{d}\mathbf{v}} &=& \int^{t}_0 \mathrm{d}t_i \frac{\mathrm{d}\hat{F}}{\text{d}\mathbf{v}\mathrm{d}t_i} \\ 
					    &=& 
\frac{\displaystyle \dot{N}_\oplus}{\displaystyle N_p} \displaystyle \sum^{N_p}_{\alpha = 1} \sum^{N_\alpha}_{\beta = 1} \delta^{(5)}( \mathbf{n} - \mathbf{n}_{\alpha\beta}, \mathbf{v} - \mathbf{v}_{\alpha\beta}) \nonumber \\
	& & \; \times \,\Theta( t - t_{\alpha\beta}) \nonumber
\end{eqnarray}
\indent
In order to get better statistics for the flux through the Earth, we average the flux in Eq. (\ref{dFdt}) over all positions $\mathbf{n}$ on the cylinder wall.  In this case, 
\begin{eqnarray}
	\int_{cylinder} \text{d}^2 \mathbf{n} = \delta A = 2 \times 2\pi a_\oplus z_c,
\end{eqnarray} 
the whole area through which we count particle crossings.  This implies that the averaged flux is 
\begin{eqnarray}
	\frac{\text{d}\hat{\overline{F}}( \mathbf{n}, t)}{\text{d}\mathbf{v}} &=& \frac{1}{\delta A}\int_{cylinder} \mathrm{d}^2 \mathbf{n} \frac{\text{d} \hat{F}}{\text{d}\mathbf{v}} \\
				&=& 
\displaystyle \frac{\dot{N}_\oplus}{N_p} \frac{1}{\delta A} \sum^{N_p}_{\alpha = 1} \sum^{N_\alpha}_{\beta = 1} \delta^{(3)}(\mathbf{v} - \mathbf{v}_{\alpha\beta}) \nonumber \\
	& & \times \, \Theta( t -  t_{\alpha\beta}). \label{eq:final_flux} \nonumber \\
\end{eqnarray}
In effect, we are averaging the flux over the Earth's orbit.  We find the local estimate of the distribution function by inserting Eq. (\ref{eq:final_flux}) into Eq. (\ref{eq:f_from_F}).
\newline\indent
To find the distribution function in the frame of the Earth, we make a Galilean transformation $\mathbf{u} = \mathbf{v} - \mathbf{v}_\oplus$, where $\mathbf{v}_\oplus$ is the circular velocity of the Earth about the Sun, to find
\begin{eqnarray}
	\hat{f}_\oplus ( \mathbf{x}, \mathbf{u}, t) = \hat{f}(\mathbf{x}, \mathbf{u} + \mathbf{v}_\oplus, t).
\end{eqnarray}

\subsection{Estimating Distribution Functions in Practice}
In practice, there are $10^8-10^9$ Earth-orbit crossings in each simulation.  In order to present and use the DFs in a manageable form, we use a small $z_c$ and bin the distribution function in velocity space.  We set $z_c = 10 R_\oplus$, but using different $z_c$ up to $z_c = 10^{-3}$ AU (the largest value we tried) yields consistent DFs, demonstrating the desired result that the estimate for the DF does not depend on the choice of $z_c$.

The most straightforward way of estimating uncertainty in the distribution function and any calculations derived from it is to use bootstrap resampling.  Bootstrap resampling yields accurate parameter and error estimation if the data sample the underlying distribution well.  In each resampling, we select $N_p$ initial conditions with replacement from the $N_p$ WIMPs.  We then calculate all distribution functions and event rates using the trajectories and crossings of the new sample as described in the previous section.

\subsection{The Distribution Function in the Earth}\label{sec:dfearth}
In the previous section, we found DFs in the absence of the Earth's gravity.  However, since both direct detection experiments and neutrino telescopes are sensitive to particles within the potential well of the Earth, it is necessary to find the mapping between the velocity coordinates at distances $\ll 1$ AU from the Earth but well outside the Earth's gravitational field and those at distances at which the Earth's gravity is significant.  Let $\mathbf{v} = (v, \theta, \phi)$ denote the velocity outside the Earth's gravitational field in an inertial frame centered on and moving with the Earth, with the polar axis along the Earth's direction of motion, and the velocity $\mathbf{v}_{loc} = (v_{loc}, \theta_{loc}, \phi_{loc})$ be in the Earth's gravitational field at a position $\mathbf{R} = (R, \zeta, \psi)$ from the Earth's center, where $\mathbf{v}_{loc}$ is also in an inertial frame centered on and moving with the Earth.  In these coordinates, the angles $\theta,\mbox{ }\theta_{loc}$, and $\zeta$ are measured relative to the direction of motion of the Earth with respect to the Sun, and the $\phi,\mbox{ }\phi_{loc}$, and $\psi$ angles are azimuthal.
\newline\indent
Since the particle energy $E$ and angular momentum $J$ with respect to the Earth are approximately conserved near the Earth, the local DF $f_{\text{loc}}$ of dark matter in the gravitational field of the Earth can be written as
\begin{eqnarray}
	f_{loc}(\mathbf{R}, \mathbf{v}_{loc}) = f(\mathbf{v}(\mathbf{v}_{loc}, \mathbf{R} ) ). \label{eq:floc}
\end{eqnarray}
Here, $f(\mathbf{v})$ is the dark matter DF in the frame of the Earth but far from the Earth's center.  Eq. (\ref{eq:floc}) is a restatement of Liouville's theorem.  The number of particles in an interval between $(\mathbf{R}, \mathbf{v}_{loc})$ and $(\mathbf{R} + \mathrm{d} \mathbf{R}, \mathbf{v}_{loc} + \mathbf{d} \mathbf{v}_{loc})$ is
\begin{eqnarray}
	\mathrm{d} N = f_{loc}(\mathbf{R}, \mathbf{v}_{loc}) \mathrm{d}^3 \mathbf{R} \mathrm{d}^3 \mathbf{v}_{loc}.
\end{eqnarray}
\indent
If the DF $f(\mathbf{v})$ were isotropic, then the mapping between velocity coordinates would be greatly simplified.  In such a situation, the speeds $v$ and $v_{loc}$ are related through conservation of energy, 
\begin{eqnarray}
	E = \frac{1}{2} v^2 = \frac{1}{2} v_{\text{loc}}^2(R) + \Phi_\oplus ( R ) \label{eq:ch4e},
\end{eqnarray}
assuming that the Earth's potential $\Phi_\oplus$ is spherical.  Therefore, the number of dark matter particles with positions between $\mathbf{R}$ and $\mathbf{R} + \mathrm{d}\mathbf{R}$ and speeds between $\mathbf{v}_{loc}$ and $\mathbf{v}_{loc} + \text{d}\mathbf{v}_{loc}$ would be represented as
\begin{eqnarray}
	\text{d} N_{iso} = 4 \pi v_{loc}^2 f(v(R,v_{loc})) \text{d}^3 \mathbf{R} \mathrm{d} v_{loc}.
\end{eqnarray} 
\indent
However, the DFs are not isotropic in the frame of the Earth.  Thus, it is necessary to find $\mathbf{v}$ in terms of the velocity $\mathbf{v}_{loc}$ at position $\mathbf{R}$.  The speeds are still related by Eq. (\ref{eq:ch4e}), so that $v$ is a function of only two variables, $v_{loc}$ and $R$.  The angular coordinates $(\theta, \phi)$, however, will now be a complicated function of all six local phase space coordinates, so that the number of particles at $(\mathbf{R},\mathbf{v}_{loc}$) is described as
\begin{multline}
	\text{d}N = f(v(R,v_{loc}), \theta (\mathbf{R},\mathbf{v}_{loc}), \phi(\mathbf{R}, \mathbf{v}_{loc}) ) \\
	\times \, R^2 v_{loc}^2 \text{d} R \text{d} \cos \zeta \text{d}\psi  \text{d} v_{loc} \text{d} \cos \theta_{loc} \text{d} \phi_{loc}.
\end{multline} 
\indent 
To relate the angular coordinates, we make use of angular momentum conservation as well as energy conservation, and the fact that the problem reduces to a spherically symmetric two-body problem.  Since orbits are confined to a plane, $\mathbf{R}$ and $\mathbf{v}_{loc}$ are a set of basis vectors for the orbital plane if the vectors are not parallel.  Then, in general, the position $\mathbf{R}_{far}$ and velocity $\mathbf{v}$ far from the Earth can be described by
\begin{eqnarray}
	\mathbf{R}_{far} &=& \alpha \mathbf{R} + \beta \mathbf{v}_{loc}, \label{eq:rgauss}\\
	\mathbf{v} &=& \gamma \mathbf{R} + \delta \mathbf{v}_{loc}, \label{eq:vgauss}
\end{eqnarray}
where the coefficients $\alpha$, $\beta$, $\gamma$, and $\delta$ only depend on the local coordinates $\mathbf{R}$ and $\mathbf{v}_{loc}$, $E$, and $J$.  If the Earth's potential were purely Keplerian, $\alpha$ and $\beta$ would be the Gauss $f$ and $g$ functions \citep[see Section 2.5 in][]{murray2000}, with $\gamma = \dot{\alpha}$ and $\delta = \dot{\beta}$.  The functional form of the coefficients is different in the case of non-Keplerian spherically symmetric potentials, but the general framework of Eqs. (\ref{eq:rgauss}) and (\ref{eq:vgauss}) holds.  Therefore, Eqs. (\ref{eq:rgauss}) and (\ref{eq:vgauss}) describe the mapping between coordinates in the gravitational field of the Earth to those outside the Earth's sphere of influence.


\begin{thebibliography}{113}
\expandafter\ifx\csname natexlab\endcsname\relax\def\natexlab#1{#1}\fi
\expandafter\ifx\csname bibnamefont\endcsname\relax
  \def\bibnamefont#1{#1}\fi
\expandafter\ifx\csname bibfnamefont\endcsname\relax
  \def\bibfnamefont#1{#1}\fi
\expandafter\ifx\csname citenamefont\endcsname\relax
  \def\citenamefont#1{#1}\fi
\expandafter\ifx\csname url\endcsname\relax
  \def\url#1{\texttt{#1}}\fi
\expandafter\ifx\csname urlprefix\endcsname\relax\def\urlprefix{URL }\fi
\providecommand{\bibinfo}[2]{#2}
\providecommand{\eprint}[2][]{\url{#2}}

\bibitem[{\citenamefont{{Jungman} et~al.}(1996)\citenamefont{{Jungman},
  {Kamionkowski}, and {Griest}}}]{jungman1996}
\bibinfo{author}{\bibfnamefont{G.}~\bibnamefont{{Jungman}}},
  \bibinfo{author}{\bibfnamefont{M.}~\bibnamefont{{Kamionkowski}}},
  \bibnamefont{and} \bibinfo{author}{\bibfnamefont{K.}~\bibnamefont{{Griest}}},
  \bibinfo{journal}{Phys. Rep.} \textbf{\bibinfo{volume}{267}},
  \bibinfo{pages}{195} (\bibinfo{year}{1996}), \eprint{arXiv:hep-ph/9506380}.

\bibitem[{\citenamefont{{Hooper} and {Profumo}}(2007)}]{hooper2007}
\bibinfo{author}{\bibfnamefont{D.}~\bibnamefont{{Hooper}}} \bibnamefont{and}
  \bibinfo{author}{\bibfnamefont{S.}~\bibnamefont{{Profumo}}},
  \bibinfo{journal}{Phys. Rep.} \textbf{\bibinfo{volume}{453}},
  \bibinfo{pages}{29} (\bibinfo{year}{2007}), \eprint{arXiv:hep-ph/0701197}.

\bibitem[{\citenamefont{{Servant} and {Tait}}(2002)}]{servant2002}
\bibinfo{author}{\bibfnamefont{G.}~\bibnamefont{{Servant}}} \bibnamefont{and}
  \bibinfo{author}{\bibfnamefont{T.~M.~P.} \bibnamefont{{Tait}}},
  \bibinfo{journal}{New J. Phys.} \textbf{\bibinfo{volume}{4}},
  \bibinfo{pages}{99} (\bibinfo{year}{2002}), \eprint{arXiv:hep-ph/0209262}.

\bibitem[{\citenamefont{{Cheng} et~al.}(2002)\citenamefont{{Cheng}, {Feng}, and
  {Matchev}}}]{cheng2002}
\bibinfo{author}{\bibfnamefont{H.-C.} \bibnamefont{{Cheng}}},
  \bibinfo{author}{\bibfnamefont{J.~L.} \bibnamefont{{Feng}}},
  \bibnamefont{and} \bibinfo{author}{\bibfnamefont{K.~T.}
  \bibnamefont{{Matchev}}}, \bibinfo{journal}{Phys. Rev. Lett.}
  \textbf{\bibinfo{volume}{89}}, \bibinfo{pages}{211301}
  (\bibinfo{year}{2002}), \eprint{arXiv:hep-ph/0207125}.

\bibitem[{\citenamefont{{Hubisz} and {Meade}}(2005)}]{hubisz2005}
\bibinfo{author}{\bibfnamefont{J.}~\bibnamefont{{Hubisz}}} \bibnamefont{and}
  \bibinfo{author}{\bibfnamefont{P.}~\bibnamefont{{Meade}}},
  \bibinfo{journal}{\prd} \textbf{\bibinfo{volume}{71}},
  \bibinfo{pages}{035016} (\bibinfo{year}{2005}),
  \eprint{arXiv:hep-ph/0411264}.

\bibitem[{\citenamefont{{Birkedal-Hansen} and {Wacker}}(2004)}]{birkedal2004}
\bibinfo{author}{\bibfnamefont{A.}~\bibnamefont{{Birkedal-Hansen}}}
  \bibnamefont{and} \bibinfo{author}{\bibfnamefont{J.~G.}
  \bibnamefont{{Wacker}}}, \bibinfo{journal}{\prd}
  \textbf{\bibinfo{volume}{69}}, \bibinfo{pages}{065022}
  (\bibinfo{year}{2004}), \eprint{arXiv:hep-ph/0306161}.

\bibitem[{\citenamefont{{Birkedal} et~al.}(2006)\citenamefont{{Birkedal},
  {Noble}, {Perelstein}, and {Spray}}}]{birkedal2006}
\bibinfo{author}{\bibfnamefont{A.}~\bibnamefont{{Birkedal}}},
  \bibinfo{author}{\bibfnamefont{A.}~\bibnamefont{{Noble}}},
  \bibinfo{author}{\bibfnamefont{M.}~\bibnamefont{{Perelstein}}},
  \bibnamefont{and} \bibinfo{author}{\bibfnamefont{A.}~\bibnamefont{{Spray}}},
  \bibinfo{journal}{\prd} \textbf{\bibinfo{volume}{74}},
  \bibinfo{pages}{035002} (\bibinfo{year}{2006}),
  \eprint{arXiv:hep-ph/0603077}.

\bibitem[{\citenamefont{{Komatsu} et~al.}(2008)}]{komatsu2008}
\bibinfo{author}{\bibfnamefont{E.}~\bibnamefont{{Komatsu}}}
  \bibnamefont{et~al.} (\bibinfo{year}{2008}), \eprint{arXiv:0803.0547}.

\bibitem[{\citenamefont{{Wai} et~al.}(2007)}]{wai2007}
\bibinfo{author}{\bibfnamefont{L.}~\bibnamefont{{Wai}}} \bibnamefont{et~al.},
  in \emph{\bibinfo{booktitle}{SUSY06}}, edited by
  \bibinfo{editor}{\bibfnamefont{J.~L.} \bibnamefont{{Feng}}}
  (\bibinfo{publisher}{American Institute of Physics, Melville, NY},
  \bibinfo{year}{2007}), vol. \bibinfo{volume}{903} of
  \emph{\bibinfo{series}{AIP Conference Series}}, pp.
  \bibinfo{pages}{599--602}.

\bibitem[{\citenamefont{{Kuhlen} et~al.}(2008)\citenamefont{{Kuhlen},
  {Diemand}, and {Madau}}}]{kuhlen2008}
\bibinfo{author}{\bibfnamefont{M.}~\bibnamefont{{Kuhlen}}},
  \bibinfo{author}{\bibfnamefont{J.}~\bibnamefont{{Diemand}}},
  \bibnamefont{and} \bibinfo{author}{\bibfnamefont{P.}~\bibnamefont{{Madau}}},
  \bibinfo{journal}{\apj} \textbf{\bibinfo{volume}{686}}, \bibinfo{pages}{262}
  (\bibinfo{year}{2008}), \eprint{arXiv:0805.4416}.

\bibitem[{\citenamefont{{Chang} et~al.}(2008)}]{chang2008}
\bibinfo{author}{\bibfnamefont{J.}~\bibnamefont{{Chang}}} \bibnamefont{et~al.},
  \bibinfo{journal}{\nat} \textbf{\bibinfo{volume}{456}}, \bibinfo{pages}{362}
  (\bibinfo{year}{2008}).

\bibitem[{\citenamefont{{Adriani} et~al.}(2008)}]{pamela2008}
\bibinfo{author}{\bibfnamefont{O.}~\bibnamefont{{Adriani}}}
  \bibnamefont{et~al.} (\bibinfo{year}{2008}), \eprint{arXiv:0810.4995}.

\bibitem[{\citenamefont{{Boliev} et~al.}(1996)\citenamefont{{Boliev}, {Bugaev},
  {Butkevich}, {Chudakov}, {Mikheyev}, {Suvorova}, and
  {Zakidyshev}}}]{boliev1996}
\bibinfo{author}{\bibfnamefont{M.~M.} \bibnamefont{{Boliev}}},
  \bibinfo{author}{\bibfnamefont{E.~V.} \bibnamefont{{Bugaev}}},
  \bibinfo{author}{\bibfnamefont{A.~V.} \bibnamefont{{Butkevich}}},
  \bibinfo{author}{\bibfnamefont{A.~E.} \bibnamefont{{Chudakov}}},
  \bibinfo{author}{\bibfnamefont{S.~P.} \bibnamefont{{Mikheyev}}},
  \bibinfo{author}{\bibfnamefont{O.~V.} \bibnamefont{{Suvorova}}},
  \bibnamefont{and} \bibinfo{author}{\bibfnamefont{V.~N.}
  \bibnamefont{{Zakidyshev}}}, \bibinfo{journal}{Nucl. Phys. B Proc. Suppl.}
  \textbf{\bibinfo{volume}{48}}, \bibinfo{pages}{83} (\bibinfo{year}{1996}).

\bibitem[{\citenamefont{{Ambrosio} et~al.}(1999)}]{ambrosio1999}
\bibinfo{author}{\bibfnamefont{M.}~\bibnamefont{{Ambrosio}}}
  \bibnamefont{et~al.}, \bibinfo{journal}{Phys. Rev. D}
  \textbf{\bibinfo{volume}{60}}, \bibinfo{pages}{082002}
  (\bibinfo{year}{1999}).

\bibitem[{\citenamefont{{Desai} et~al.}(2004)}]{desai2004}
\bibinfo{author}{\bibfnamefont{S.}~\bibnamefont{{Desai}}} \bibnamefont{et~al.},
  \bibinfo{journal}{\prd} \textbf{\bibinfo{volume}{70}},
  \bibinfo{pages}{083523} (\bibinfo{year}{2004}),
  \eprint{arXiv:hep-ex/0404025}.

\bibitem[{\citenamefont{{Achterberg} et~al.}(2006)}]{achterberg2006}
\bibinfo{author}{\bibfnamefont{A.}~\bibnamefont{{Achterberg}}}
  \bibnamefont{et~al.}, \bibinfo{journal}{Astropart. Phys.}
  \textbf{\bibinfo{volume}{26}}, \bibinfo{pages}{129} (\bibinfo{year}{2006}).

\bibitem[{\citenamefont{{Ackermann} et~al.}(2006)}]{ackermann2006}
\bibinfo{author}{\bibfnamefont{M.}~\bibnamefont{{Ackermann}}}
  \bibnamefont{et~al.}, \bibinfo{journal}{Astropart. Phys.}
  \textbf{\bibinfo{volume}{24}}, \bibinfo{pages}{459} (\bibinfo{year}{2006}).

\bibitem[{\citenamefont{{Lim} and {for the ANTARES
  Collaboration}}(2007)}]{lim2007}
\bibinfo{author}{\bibfnamefont{G.}~\bibnamefont{{Lim}}} \bibnamefont{and}
  \bibinfo{author}{\bibnamefont{{for the ANTARES Collaboration}}}
  (\bibinfo{year}{2007}), \eprint{arXiv:0710.3685}.

\bibitem[{\citenamefont{{de los Heros} et~al.}(2008)}]{delosheros2008}
\bibinfo{author}{\bibfnamefont{C.}~\bibnamefont{{de los Heros}}}
  \bibnamefont{et~al.} (\bibinfo{year}{2008}), \eprint{arXiv:0802.0147}.

\bibitem[{\citenamefont{{de Wolf}}(2008)}]{dewolf2008}
\bibinfo{author}{\bibfnamefont{E.}~\bibnamefont{{de Wolf}}},
  \bibinfo{journal}{Nucl. Instrum. Meth. Phys. Res. A}
  \textbf{\bibinfo{volume}{588}}, \bibinfo{pages}{86} (\bibinfo{year}{2008}).

\bibitem[{\citenamefont{{Hime}}(2007)}]{hime2007}
\bibinfo{author}{\bibfnamefont{A.}~\bibnamefont{{Hime}}}, \bibinfo{journal}{APS
  Meeting Abstracts} pp. \bibinfo{pages}{14002--+} (\bibinfo{year}{2007}).

\bibitem[{\citenamefont{{Gaitskell}}(2007)}]{gaitskell2007}
\bibinfo{author}{\bibfnamefont{R.}~\bibnamefont{{Gaitskell}}},
  \bibinfo{journal}{APS Meeting Abstracts} pp. \bibinfo{pages}{H3002+}
  (\bibinfo{year}{2007}), \bibinfo{note}{slides available at
  http://xenon.astro.columbia.edu/talks/\newline
  APS2007/070415\_DM\_Noble\_Gaitskell\_v08.pdf}.

\bibitem[{\citenamefont{{Brink} et~al.}(2005)}]{brink2005}
\bibinfo{author}{\bibfnamefont{P.~L.} \bibnamefont{{Brink}}}
  \bibnamefont{et~al.} (\bibinfo{year}{2005}), \eprint{arXiv:astro-ph/0503583}.

\bibitem[{\citenamefont{{SuperCDMS Collaboration}}(2005)}]{schnee2005}
\bibinfo{author}{\bibnamefont{{SuperCDMS Collaboration}}}
  (\bibinfo{year}{2005}), \eprint{arXiv:astro-ph/0502435}.

\bibitem[{\citenamefont{{Akerib} et~al.}(2006{\natexlab{a}})}]{akerib2006c}
\bibinfo{author}{\bibfnamefont{D.~S.} \bibnamefont{{Akerib}}}
  \bibnamefont{et~al.}, \bibinfo{journal}{Nucl. Instr. Meth. A}
  \textbf{\bibinfo{volume}{559}}, \bibinfo{pages}{411}
  (\bibinfo{year}{2006}{\natexlab{a}}).

\bibitem[{\citenamefont{{Brunetti} et~al.}(2005)}]{brunetti2005}
\bibinfo{author}{\bibfnamefont{R.}~\bibnamefont{{Brunetti}}}
  \bibnamefont{et~al.}, \bibinfo{journal}{New Astron. Rev.}
  \textbf{\bibinfo{volume}{49}}, \bibinfo{pages}{265} (\bibinfo{year}{2005}),
  \eprint{arXiv:astro-ph/0405342}.

\bibitem[{\citenamefont{{Aprile} et~al.}(2002)}]{aprile2002}
\bibinfo{author}{\bibfnamefont{E.}~\bibnamefont{{Aprile}}} \bibnamefont{et~al.}
  (\bibinfo{year}{2002}), \bibinfo{note}{astro-ph/0207670},
  \eprint{arXiv:astro-ph/0207670}.

\bibitem[{\citenamefont{{The Xmass Collaboration}}(2005)}]{xmass2005}
\bibinfo{author}{\bibnamefont{{The Xmass Collaboration}}},
  \bibinfo{journal}{Nucl. Phys. B Proc. Suppl.} \textbf{\bibinfo{volume}{143}},
  \bibinfo{pages}{506} (\bibinfo{year}{2005}).

\bibitem[{\citenamefont{{Zacek}}(2007)}]{zacek2007}
\bibinfo{author}{\bibfnamefont{V.}~\bibnamefont{{Zacek}}}
  (\bibinfo{year}{2007}), \eprint{arXiv:0707.0472}.

\bibitem[{\citenamefont{{Baudis}}(2008)}]{baudis2008}
\bibinfo{author}{\bibfnamefont{L.}~\bibnamefont{{Baudis}}}
  (\bibinfo{year}{2008}),
  \bibinfo{note}{http://xenon.astro.columbia.edu/\newline
  presentations/baudis\_idm08.pdf}.

\bibitem[{\citenamefont{{Green}}(2007)}]{green2007b}
\bibinfo{author}{\bibfnamefont{A.~M.} \bibnamefont{{Green}}},
  \bibinfo{journal}{JCAP} \textbf{\bibinfo{volume}{8}}, \bibinfo{pages}{22}
  (\bibinfo{year}{2007}), \eprint{arXiv:hep-ph/0703217}.

\bibitem[{\citenamefont{{Angle} et~al.}(2008)}]{angle2008}
\bibinfo{author}{\bibfnamefont{J.}~\bibnamefont{{Angle}}} \bibnamefont{et~al.},
  \bibinfo{journal}{Phys. Rev. Lett.} \textbf{\bibinfo{volume}{100}},
  \bibinfo{pages}{021303} (\bibinfo{year}{2008}),
  \eprint{arXiv:astro-ph/0706.0039}.

\bibitem[{\citenamefont{{Aalseth} et~al.}(2008)}]{aalseth2008}
\bibinfo{author}{\bibfnamefont{C.~E.} \bibnamefont{{Aalseth}}}
  \bibnamefont{et~al.} (\bibinfo{year}{2008}), \eprint{arXiv:0807.0879}.

\bibitem[{\citenamefont{{Gould}}(1988)}]{gould1988}
\bibinfo{author}{\bibfnamefont{A.}~\bibnamefont{{Gould}}},
  \bibinfo{journal}{\apj} \textbf{\bibinfo{volume}{328}}, \bibinfo{pages}{919}
  (\bibinfo{year}{1988}).

\bibitem[{\citenamefont{{Gould}}(1991)}]{gould1991}
\bibinfo{author}{\bibfnamefont{A.}~\bibnamefont{{Gould}}},
  \bibinfo{journal}{\apj} \textbf{\bibinfo{volume}{368}}, \bibinfo{pages}{610}
  (\bibinfo{year}{1991}).

\bibitem[{\citenamefont{{Lundberg} and {Edsj{\"o}}}(2004)}]{lundberg2004}
\bibinfo{author}{\bibfnamefont{J.}~\bibnamefont{{Lundberg}}} \bibnamefont{and}
  \bibinfo{author}{\bibfnamefont{J.}~\bibnamefont{{Edsj{\"o}}}},
  \bibinfo{journal}{\prd} \textbf{\bibinfo{volume}{69}},
  \bibinfo{pages}{123505} (\bibinfo{year}{2004}),
  \eprint{arXiv:astro-ph/0401113}.

\bibitem[{\citenamefont{{Damour} and {Krauss}}(1999)}]{damour1999}
\bibinfo{author}{\bibfnamefont{T.}~\bibnamefont{{Damour}}} \bibnamefont{and}
  \bibinfo{author}{\bibfnamefont{L.~M.} \bibnamefont{{Krauss}}},
  \bibinfo{journal}{\prd} \textbf{\bibinfo{volume}{59}},
  \bibinfo{pages}{063509} (\bibinfo{year}{1999}),
  \eprint{arXiv:astro-ph/9807099}.

\bibitem[{\citenamefont{{Bergstr{\"o}m} et~al.}(1999)}]{bergstrom1999}
\bibinfo{author}{\bibfnamefont{L.}~\bibnamefont{{Bergstr{\"o}m}}}
  \bibnamefont{et~al.}, \bibinfo{journal}{JHEP} \textbf{\bibinfo{volume}{8}},
  \bibinfo{pages}{10} (\bibinfo{year}{1999}), \eprint{arXiv:hep-ph/9905446}.

\bibitem[{\citenamefont{{Peter}}(2009{\natexlab{a}})}]{peter2009b}
\bibinfo{author}{\bibfnamefont{A.~H.~G.} \bibnamefont{{Peter}}}
  (\bibinfo{year}{2009}{\natexlab{a}}), \bibinfo{note}{arXiv:0902.1347}.

\bibitem[{\citenamefont{{Peter}}(2009{\natexlab{b}})}]{peter2009c}
\bibinfo{author}{\bibfnamefont{A.~H.~G.} \bibnamefont{{Peter}}}
  (\bibinfo{year}{2009}{\natexlab{b}}), \bibinfo{note}{arXiv:0902.1348}.

\bibitem[{\citenamefont{{Kozai}}(1962)}]{kozai1962}
\bibinfo{author}{\bibfnamefont{Y.}~\bibnamefont{{Kozai}}},
  \bibinfo{journal}{Astron. J.} \textbf{\bibinfo{volume}{67}},
  \bibinfo{pages}{591} (\bibinfo{year}{1962}).

\bibitem[{\citenamefont{{Thomas} and {Morbidelli}}(1996)}]{thomas1996}
\bibinfo{author}{\bibfnamefont{F.}~\bibnamefont{{Thomas}}} \bibnamefont{and}
  \bibinfo{author}{\bibfnamefont{A.}~\bibnamefont{{Morbidelli}}},
  \bibinfo{journal}{Cel. Mech. Dyn. Astron.} \textbf{\bibinfo{volume}{64}},
  \bibinfo{pages}{209} (\bibinfo{year}{1996}).

\bibitem[{\citenamefont{{Kozai}}(1979)}]{kozai1979}
\bibinfo{author}{\bibfnamefont{Y.}~\bibnamefont{{Kozai}}}, in
  \emph{\bibinfo{booktitle}{Dynamics of the Solar System}}, edited by
  \bibinfo{editor}{\bibfnamefont{R.~L.} \bibnamefont{{Duncombe}}}
  (\bibinfo{year}{1979}), vol.~\bibinfo{volume}{81} of
  \emph{\bibinfo{series}{IAU Symposium}}, pp. \bibinfo{pages}{231--236}.

\bibitem[{\citenamefont{{Michel} and {Thomas}}(1996)}]{michel1996}
\bibinfo{author}{\bibfnamefont{P.}~\bibnamefont{{Michel}}} \bibnamefont{and}
  \bibinfo{author}{\bibfnamefont{F.}~\bibnamefont{{Thomas}}},
  \bibinfo{journal}{Astron. Astrophys.} \textbf{\bibinfo{volume}{307}},
  \bibinfo{pages}{310} (\bibinfo{year}{1996}).

\bibitem[{\citenamefont{{Ford} et~al.}(2000)\citenamefont{{Ford}, {Kozinsky},
  and {Rasio}}}]{ford2000}
\bibinfo{author}{\bibfnamefont{E.~B.} \bibnamefont{{Ford}}},
  \bibinfo{author}{\bibfnamefont{B.}~\bibnamefont{{Kozinsky}}},
  \bibnamefont{and} \bibinfo{author}{\bibfnamefont{F.~A.}
  \bibnamefont{{Rasio}}}, \bibinfo{journal}{\apj}
  \textbf{\bibinfo{volume}{535}}, \bibinfo{pages}{385} (\bibinfo{year}{2000}).

\bibitem[{\citenamefont{{Fabrycky} and {Tremaine}}(2007)}]{fabrycky2007}
\bibinfo{author}{\bibfnamefont{D.}~\bibnamefont{{Fabrycky}}} \bibnamefont{and}
  \bibinfo{author}{\bibfnamefont{S.}~\bibnamefont{{Tremaine}}},
  \bibinfo{journal}{Astrophys. J.} \textbf{\bibinfo{volume}{669}},
  \bibinfo{pages}{1298} (\bibinfo{year}{2007}), \eprint{arXiv:0705.4285}.

\bibitem[{\citenamefont{{Wisdom} and {Holman}}(1991)}]{wisdom1991}
\bibinfo{author}{\bibfnamefont{J.}~\bibnamefont{{Wisdom}}} \bibnamefont{and}
  \bibinfo{author}{\bibfnamefont{M.}~\bibnamefont{{Holman}}},
  \bibinfo{journal}{Astron. J.} \textbf{\bibinfo{volume}{102}},
  \bibinfo{pages}{1528} (\bibinfo{year}{1991}).

\bibitem[{\citenamefont{{Saha} and {Tremaine}}(1994)}]{saha1994}
\bibinfo{author}{\bibfnamefont{P.}~\bibnamefont{{Saha}}} \bibnamefont{and}
  \bibinfo{author}{\bibfnamefont{S.}~\bibnamefont{{Tremaine}}},
  \bibinfo{journal}{Astron. J.} \textbf{\bibinfo{volume}{108}},
  \bibinfo{pages}{1962} (\bibinfo{year}{1994}),
  \eprint{arXiv:astro-ph/9403057}.

\bibitem[{\citenamefont{{Chambers}}(1999)}]{chambers1999}
\bibinfo{author}{\bibfnamefont{J.~E.} \bibnamefont{{Chambers}}},
  \bibinfo{journal}{Mon. Not. Roy. Astron. Soc.}
  \textbf{\bibinfo{volume}{304}}, \bibinfo{pages}{793} (\bibinfo{year}{1999}).

\bibitem[{\citenamefont{{Preto} and {Tremaine}}(1999)}]{preto1999}
\bibinfo{author}{\bibfnamefont{M.}~\bibnamefont{{Preto}}} \bibnamefont{and}
  \bibinfo{author}{\bibfnamefont{S.}~\bibnamefont{{Tremaine}}},
  \bibinfo{journal}{Astron. J.} \textbf{\bibinfo{volume}{118}},
  \bibinfo{pages}{2532} (\bibinfo{year}{1999}),
  \eprint{arXiv:astro-ph/9906322}.

\bibitem[{\citenamefont{{Mikkola} and {Tanikawa}}(1999)}]{mikkola1999}
\bibinfo{author}{\bibfnamefont{S.}~\bibnamefont{{Mikkola}}} \bibnamefont{and}
  \bibinfo{author}{\bibfnamefont{K.}~\bibnamefont{{Tanikawa}}},
  \bibinfo{journal}{Cel. Mech. Dyn. Astron.} \textbf{\bibinfo{volume}{74}},
  \bibinfo{pages}{287} (\bibinfo{year}{1999}).

\bibitem[{\citenamefont{{Peter}}(2008{\natexlab{c}})}]{peter2008}
\bibinfo{author}{\bibfnamefont{A.~H.~G.} \bibnamefont{{Peter}}}, Ph.D. thesis,
  \bibinfo{school}{Princeton University} (\bibinfo{year}{2008}{\natexlab{c}}).

\bibitem[{\citenamefont{{Gunn} et~al.}(1979)\citenamefont{{Gunn}, {Knapp}, and
  {Tremaine}}}]{gunn1979}
\bibinfo{author}{\bibfnamefont{J.~E.} \bibnamefont{{Gunn}}},
  \bibinfo{author}{\bibfnamefont{G.~R.} \bibnamefont{{Knapp}}},
  \bibnamefont{and} \bibinfo{author}{\bibfnamefont{S.~D.}
  \bibnamefont{{Tremaine}}}, \bibinfo{journal}{Astron. J.}
  \textbf{\bibinfo{volume}{84}}, \bibinfo{pages}{1181} (\bibinfo{year}{1979}).

\bibitem[{\citenamefont{{Kerr} and {Lynden-Bell}}(1986)}]{kerr1986}
\bibinfo{author}{\bibfnamefont{F.~J.} \bibnamefont{{Kerr}}} \bibnamefont{and}
  \bibinfo{author}{\bibfnamefont{D.}~\bibnamefont{{Lynden-Bell}}},
  \bibinfo{journal}{Mon. Not. Roy. Astron. Soc.}
  \textbf{\bibinfo{volume}{221}}, \bibinfo{pages}{1023} (\bibinfo{year}{1986}).

\bibitem[{\citenamefont{{Kamionkowski} and
  {Koushiappas}}(2008)}]{kamionkowski2008}
\bibinfo{author}{\bibfnamefont{M.}~\bibnamefont{{Kamionkowski}}}
  \bibnamefont{and} \bibinfo{author}{\bibfnamefont{S.~M.}
  \bibnamefont{{Koushiappas}}} (\bibinfo{year}{2008}),
  \eprint{arXiv:0801.3269}.

\bibitem[{\citenamefont{{Bahcall} et~al.}(2005)\citenamefont{{Bahcall},
  {Serenelli}, and {Basu}}}]{bahcall2005}
\bibinfo{author}{\bibfnamefont{J.~N.} \bibnamefont{{Bahcall}}},
  \bibinfo{author}{\bibfnamefont{A.~M.} \bibnamefont{{Serenelli}}},
  \bibnamefont{and} \bibinfo{author}{\bibfnamefont{S.}~\bibnamefont{{Basu}}},
  \bibinfo{journal}{Astrophys. J.} \textbf{\bibinfo{volume}{621}},
  \bibinfo{pages}{L85} (\bibinfo{year}{2005}), \eprint{arXiv:astro-ph/0412440}.

\bibitem[{\citenamefont{{Murray} and {Dermott}}(2000)}]{murray2000}
\bibinfo{author}{\bibfnamefont{C.~D.} \bibnamefont{{Murray}}} \bibnamefont{and}
  \bibinfo{author}{\bibfnamefont{S.~F.} \bibnamefont{{Dermott}}},
  \emph{\bibinfo{title}{{Solar System Dynamics}}}
  (\bibinfo{publisher}{Cambridge, UK, Cambridge University Press},
  \bibinfo{year}{2000}).

\bibitem[{\citenamefont{{Ito} and {Tanikawa}}(2002)}]{ito2002}
\bibinfo{author}{\bibfnamefont{T.}~\bibnamefont{{Ito}}} \bibnamefont{and}
  \bibinfo{author}{\bibfnamefont{K.}~\bibnamefont{{Tanikawa}}},
  \bibinfo{journal}{Mon. Not. Roy. Astron. Soc.}
  \textbf{\bibinfo{volume}{336}}, \bibinfo{pages}{483} (\bibinfo{year}{2002}).

\bibitem[{\citenamefont{{Belli} et~al.}(2000)\citenamefont{{Belli}, {Bernabei},
  {Bottino}, {Donato}, {Fornengo}, {Prosperi}, and {Scopel}}}]{belli2000}
\bibinfo{author}{\bibfnamefont{P.}~\bibnamefont{{Belli}}},
  \bibinfo{author}{\bibfnamefont{R.}~\bibnamefont{{Bernabei}}},
  \bibinfo{author}{\bibfnamefont{A.}~\bibnamefont{{Bottino}}},
  \bibinfo{author}{\bibfnamefont{F.}~\bibnamefont{{Donato}}},
  \bibinfo{author}{\bibfnamefont{N.}~\bibnamefont{{Fornengo}}},
  \bibinfo{author}{\bibfnamefont{D.}~\bibnamefont{{Prosperi}}},
  \bibnamefont{and} \bibinfo{author}{\bibfnamefont{S.}~\bibnamefont{{Scopel}}},
  \bibinfo{journal}{\prd} \textbf{\bibinfo{volume}{61}},
  \bibinfo{pages}{023512} (\bibinfo{year}{2000}),
  \eprint{arXiv:hep-ph/9903501}.

\bibitem[{\citenamefont{{Bernabei} et~al.}(2000)}]{bernabei2000a}
\bibinfo{author}{\bibfnamefont{R.}~\bibnamefont{{Bernabei}}}
  \bibnamefont{et~al.}, \bibinfo{journal}{Phys. Lett. B}
  \textbf{\bibinfo{volume}{480}}, \bibinfo{pages}{23} (\bibinfo{year}{2000}).

\bibitem[{\citenamefont{{Malyshkin} and {Tremaine}}(1999)}]{malyshkin1999}
\bibinfo{author}{\bibfnamefont{L.}~\bibnamefont{{Malyshkin}}} \bibnamefont{and}
  \bibinfo{author}{\bibfnamefont{S.}~\bibnamefont{{Tremaine}}},
  \bibinfo{journal}{Icarus} \textbf{\bibinfo{volume}{141}},
  \bibinfo{pages}{341} (\bibinfo{year}{1999}), \eprint{arXiv:astro-ph/9808172}.

\bibitem[{\citenamefont{{Duncan} and {Levison}}(1997)}]{duncan1997}
\bibinfo{author}{\bibfnamefont{M.~J.} \bibnamefont{{Duncan}}} \bibnamefont{and}
  \bibinfo{author}{\bibfnamefont{H.~F.} \bibnamefont{{Levison}}},
  \bibinfo{journal}{Science} \textbf{\bibinfo{volume}{276}},
  \bibinfo{pages}{1670} (\bibinfo{year}{1997}).

\bibitem[{\citenamefont{{Gould}}(1992)}]{gould1992}
\bibinfo{author}{\bibfnamefont{A.}~\bibnamefont{{Gould}}},
  \bibinfo{journal}{Astrophys. J.} \textbf{\bibinfo{volume}{388}},
  \bibinfo{pages}{338} (\bibinfo{year}{1992}).

\bibitem[{\citenamefont{{CDMS Collaboration}}(2008)}]{cdms2008}
\bibinfo{author}{\bibnamefont{{CDMS Collaboration}}} (\bibinfo{year}{2008}),
  \eprint{arXiv:astro-ph/0802.3530}.

\bibitem[{\citenamefont{{Dine} and {Nelson}}(1993)}]{dine1993}
\bibinfo{author}{\bibfnamefont{M.}~\bibnamefont{{Dine}}} \bibnamefont{and}
  \bibinfo{author}{\bibfnamefont{A.~E.} \bibnamefont{{Nelson}}},
  \bibinfo{journal}{\prd} \textbf{\bibinfo{volume}{48}}, \bibinfo{pages}{1277}
  (\bibinfo{year}{1993}), \eprint{arXiv:hep-ph/9303230}.

\bibitem[{\citenamefont{{Dine} et~al.}(1995)\citenamefont{{Dine}, {Nelson}, and
  {Shirman}}}]{dine1995}
\bibinfo{author}{\bibfnamefont{M.}~\bibnamefont{{Dine}}},
  \bibinfo{author}{\bibfnamefont{A.~E.} \bibnamefont{{Nelson}}},
  \bibnamefont{and}
  \bibinfo{author}{\bibfnamefont{Y.}~\bibnamefont{{Shirman}}},
  \bibinfo{journal}{\prd} \textbf{\bibinfo{volume}{51}}, \bibinfo{pages}{1362}
  (\bibinfo{year}{1995}), \eprint{arXiv:hep-ph/9408384}.

\bibitem[{\citenamefont{{Dine} et~al.}(1996)\citenamefont{{Dine}, {Nelson},
  {Nir}, and {Shirman}}}]{dine1996}
\bibinfo{author}{\bibfnamefont{M.}~\bibnamefont{{Dine}}},
  \bibinfo{author}{\bibfnamefont{A.~E.} \bibnamefont{{Nelson}}},
  \bibinfo{author}{\bibfnamefont{Y.}~\bibnamefont{{Nir}}}, \bibnamefont{and}
  \bibinfo{author}{\bibfnamefont{Y.}~\bibnamefont{{Shirman}}},
  \bibinfo{journal}{\prd} \textbf{\bibinfo{volume}{53}}, \bibinfo{pages}{2658}
  (\bibinfo{year}{1996}), \eprint{arXiv:hep-ph/9507378}.

\bibitem[{\citenamefont{{Baltz} and {Murayama}}(2003)}]{baltz2003}
\bibinfo{author}{\bibfnamefont{E.~A.} \bibnamefont{{Baltz}}} \bibnamefont{and}
  \bibinfo{author}{\bibfnamefont{H.}~\bibnamefont{{Murayama}}},
  \bibinfo{journal}{JHEP} \textbf{\bibinfo{volume}{5}}, \bibinfo{pages}{67}
  (\bibinfo{year}{2003}), \eprint{arXiv:astro-ph/0108172}.

\bibitem[{\citenamefont{{Lee} et~al.}(2007)}]{lee2007}
\bibinfo{author}{\bibfnamefont{H.~S.} \bibnamefont{{Lee}}}
  \bibnamefont{et~al.}, \bibinfo{journal}{Phys. Rev. Lett.}
  \textbf{\bibinfo{volume}{99}}, \bibinfo{pages}{091301}
  (\bibinfo{year}{2007}), \eprint{arXiv:astro-ph/0704.0423}.

\bibitem[{\citenamefont{{Behnke} et~al.}(2008)}]{behnke2008}
\bibinfo{author}{\bibfnamefont{E.}~\bibnamefont{{Behnke}}}
  \bibnamefont{et~al.}, \bibinfo{journal}{Science}
  \textbf{\bibinfo{volume}{319}}, \bibinfo{pages}{933} (\bibinfo{year}{2008}),
  \eprint{arXiv:astro-ph/0804.2886}.

\bibitem[{\citenamefont{{Drees} and {Shan}}(2008)}]{drees2008}
\bibinfo{author}{\bibfnamefont{M.}~\bibnamefont{{Drees}}} \bibnamefont{and}
  \bibinfo{author}{\bibfnamefont{C.-L.} \bibnamefont{{Shan}}},
  \bibinfo{journal}{JCAP} \textbf{\bibinfo{volume}{6}}, \bibinfo{pages}{12}
  (\bibinfo{year}{2008}), \eprint{0803.4477}.

\bibitem[{\citenamefont{{Alner} et~al.}(2005)}]{alner2005b}
\bibinfo{author}{\bibfnamefont{G.~J.} \bibnamefont{{Alner}}}
  \bibnamefont{et~al.}, \bibinfo{journal}{Nucl. Instr. Meth. A}
  \textbf{\bibinfo{volume}{555}}, \bibinfo{pages}{173} (\bibinfo{year}{2005}).

\bibitem[{\citenamefont{{Naka} et~al.}(2007)}]{naka2007}
\bibinfo{author}{\bibfnamefont{T.}~\bibnamefont{{Naka}}} \bibnamefont{et~al.},
  \bibinfo{journal}{Nucl. Instr. Meth. A} \textbf{\bibinfo{volume}{581}},
  \bibinfo{pages}{761} (\bibinfo{year}{2007}).

\bibitem[{\citenamefont{{Santos} et~al.}(2007)}]{santos2007}
\bibinfo{author}{\bibfnamefont{D.}~\bibnamefont{{Santos}}}
  \bibnamefont{et~al.}, \bibinfo{journal}{J. Phys. Conf. Ser.}
  \textbf{\bibinfo{volume}{65}}, \bibinfo{pages}{012012}
  (\bibinfo{year}{2007}), \eprint{arXiv:astro-ph/0703310}.

\bibitem[{\citenamefont{{Nishimura} et~al.}(2008)}]{nishimura2008}
\bibinfo{author}{\bibfnamefont{H.}~\bibnamefont{{Nishimura}}}
  \bibnamefont{et~al.}, \bibinfo{journal}{J. Phys. Conf. Ser.}
  \textbf{\bibinfo{volume}{120}}, \bibinfo{pages}{042025}
  (\bibinfo{year}{2008}).

\bibitem[{\citenamefont{{Sciolla} et~al.}(2008)}]{sciolla2008}
\bibinfo{author}{\bibfnamefont{G.}~\bibnamefont{{Sciolla}}}
  \bibnamefont{et~al.} (\bibinfo{year}{2008}), \bibinfo{note}{arXiv:0805.2431}.

\bibitem[{\citenamefont{{Amram} et~al.}(1999)}]{amram1999}
\bibinfo{author}{\bibfnamefont{P.}~\bibnamefont{{Amram}}} \bibnamefont{et~al.},
  \bibinfo{journal}{Nucl. Phys. B Proc. Suppl.} \textbf{\bibinfo{volume}{75}},
  \bibinfo{pages}{415} (\bibinfo{year}{1999}).

\bibitem[{\citenamefont{{Hill} et~al.}(2006)}]{hill2006}
\bibinfo{author}{\bibfnamefont{G.~C.} \bibnamefont{{Hill}}}
  \bibnamefont{et~al.} (\bibinfo{year}{2006}), \eprint{arXiv:astro-ph/0611773}.

\bibitem[{\citenamefont{{Gould}}(1987)}]{gould1987}
\bibinfo{author}{\bibfnamefont{A.}~\bibnamefont{{Gould}}},
  \bibinfo{journal}{\apj} \textbf{\bibinfo{volume}{321}}, \bibinfo{pages}{571}
  (\bibinfo{year}{1987}).

\bibitem[{\citenamefont{{Encyclop{\ae}dia Britannica}}(1994-1999)}]{earth1}
\bibinfo{author}{\bibnamefont{{Encyclop{\ae}dia Britannica}}},
  \emph{\bibinfo{title}{{The Earth: Its Properties, Composition, and
  Structure}}} (\bibinfo{publisher}{Britannica CD, Version 99. Encyclop{\ae}dia
  Britannica, Inc.}, \bibinfo{year}{1994-1999}).

\bibitem[{\citenamefont{{McDonough}}(2003)}]{earth2}
\bibinfo{author}{\bibfnamefont{W.~F.} \bibnamefont{{McDonough}}},
  \emph{\bibinfo{title}{{Treatise on Geochemistry}}}, vol.~\bibinfo{volume}{2}
  (\bibinfo{publisher}{Amsterdam, Elsevier}, \bibinfo{year}{2003}).

\bibitem[{\citenamefont{{Gondolo} et~al.}(2004)}]{gondolo2004}
\bibinfo{author}{\bibfnamefont{P.}~\bibnamefont{{Gondolo}}}
  \bibnamefont{et~al.}, \bibinfo{journal}{JCAP} \textbf{\bibinfo{volume}{7}},
  \bibinfo{pages}{8} (\bibinfo{year}{2004}), \eprint{arXiv:astro-ph/0406204}.

\bibitem[{\citenamefont{{The IceCube Collaboration}}(2001)}]{icecube2001}
\bibinfo{author}{\bibnamefont{{The IceCube Collaboration}}}
  (\bibinfo{year}{2001}), \bibinfo{note}{http://www.icecube.wisc.edu/\newline
  science/publications/pdd/pdd.pdf}.

\bibitem[{\citenamefont{{Bergstr{\"o}m}
  et~al.}(1998)\citenamefont{{Bergstr{\"o}m}, {Edsj{\"o}}, and
  {Gondolo}}}]{bergstrom1998}
\bibinfo{author}{\bibfnamefont{L.}~\bibnamefont{{Bergstr{\"o}m}}},
  \bibinfo{author}{\bibfnamefont{J.}~\bibnamefont{{Edsj{\"o}}}},
  \bibnamefont{and}
  \bibinfo{author}{\bibfnamefont{P.}~\bibnamefont{{Gondolo}}},
  \bibinfo{journal}{Phys. Rev. D} \textbf{\bibinfo{volume}{58}},
  \bibinfo{pages}{103519} (\bibinfo{year}{1998}),
  \eprint{arXiv:hep-ph/9806293}.

\bibitem[{\citenamefont{{Akerib} et~al.}(2006{\natexlab{b}})}]{akerib2006}
\bibinfo{author}{\bibfnamefont{D.~S.} \bibnamefont{{Akerib}}}
  \bibnamefont{et~al.}, \bibinfo{journal}{Phys. Rev. Lett.}
  \textbf{\bibinfo{volume}{96}}, \bibinfo{pages}{011302}
  (\bibinfo{year}{2006}{\natexlab{b}}).

\bibitem[{\citenamefont{{Kiseleva} et~al.}(1998)\citenamefont{{Kiseleva},
  {Eggleton}, and {Mikkola}}}]{kiseleva1998}
\bibinfo{author}{\bibfnamefont{L.~G.} \bibnamefont{{Kiseleva}}},
  \bibinfo{author}{\bibfnamefont{P.~P.} \bibnamefont{{Eggleton}}},
  \bibnamefont{and}
  \bibinfo{author}{\bibfnamefont{S.}~\bibnamefont{{Mikkola}}},
  \bibinfo{journal}{Mon. Not. Roy. Astron. Soc.}
  \textbf{\bibinfo{volume}{300}}, \bibinfo{pages}{292} (\bibinfo{year}{1998}).

\bibitem[{\citenamefont{{Farinella} et~al.}(1994)\citenamefont{{Farinella},
  {Froeschle}, {Gonczi}, {Hahn}, {Morbidelli}, and
  {Valsecchi}}}]{farinella1994}
\bibinfo{author}{\bibfnamefont{P.}~\bibnamefont{{Farinella}}},
  \bibinfo{author}{\bibfnamefont{C.}~\bibnamefont{{Froeschle}}},
  \bibinfo{author}{\bibfnamefont{R.}~\bibnamefont{{Gonczi}}},
  \bibinfo{author}{\bibfnamefont{G.}~\bibnamefont{{Hahn}}},
  \bibinfo{author}{\bibfnamefont{A.}~\bibnamefont{{Morbidelli}}},
  \bibnamefont{and} \bibinfo{author}{\bibfnamefont{G.~B.}
  \bibnamefont{{Valsecchi}}}, \bibinfo{journal}{\nat}
  \textbf{\bibinfo{volume}{371}}, \bibinfo{pages}{314} (\bibinfo{year}{1994}).

\bibitem[{\citenamefont{{Dones} et~al.}(1999)\citenamefont{{Dones}, {Gladman},
  {Melosh}, {Tonks}, {Levison}, and {Duncan}}}]{dones1999}
\bibinfo{author}{\bibfnamefont{L.}~\bibnamefont{{Dones}}},
  \bibinfo{author}{\bibfnamefont{B.}~\bibnamefont{{Gladman}}},
  \bibinfo{author}{\bibfnamefont{H.~J.} \bibnamefont{{Melosh}}},
  \bibinfo{author}{\bibfnamefont{W.~B.} \bibnamefont{{Tonks}}},
  \bibinfo{author}{\bibfnamefont{H.~F.} \bibnamefont{{Levison}}},
  \bibnamefont{and} \bibinfo{author}{\bibfnamefont{M.}~\bibnamefont{{Duncan}}},
  \bibinfo{journal}{Icarus} \textbf{\bibinfo{volume}{142}},
  \bibinfo{pages}{509} (\bibinfo{year}{1999}).

\bibitem[{\citenamefont{{Gladman} et~al.}(2000)\citenamefont{{Gladman},
  {Michel}, and {Froeschl{\'e}}}}]{gladman2000}
\bibinfo{author}{\bibfnamefont{B.}~\bibnamefont{{Gladman}}},
  \bibinfo{author}{\bibfnamefont{P.}~\bibnamefont{{Michel}}}, \bibnamefont{and}
  \bibinfo{author}{\bibfnamefont{C.}~\bibnamefont{{Froeschl{\'e}}}},
  \bibinfo{journal}{Icarus} \textbf{\bibinfo{volume}{146}},
  \bibinfo{pages}{176} (\bibinfo{year}{2000}).

\bibitem[{\citenamefont{{Gladman} et~al.}(1997)\citenamefont{{Gladman},
  {Migliorini}, {Morbidelli}, {Zappala}, {Michel}, {Cellino}, {Froeschle},
  {Levison}, {Bailey}, and {Duncan}}}]{gladman1997}
\bibinfo{author}{\bibfnamefont{B.~J.} \bibnamefont{{Gladman}}},
  \bibinfo{author}{\bibfnamefont{F.}~\bibnamefont{{Migliorini}}},
  \bibinfo{author}{\bibfnamefont{A.}~\bibnamefont{{Morbidelli}}},
  \bibinfo{author}{\bibfnamefont{V.}~\bibnamefont{{Zappala}}},
  \bibinfo{author}{\bibfnamefont{P.}~\bibnamefont{{Michel}}},
  \bibinfo{author}{\bibfnamefont{A.}~\bibnamefont{{Cellino}}},
  \bibinfo{author}{\bibfnamefont{C.}~\bibnamefont{{Froeschle}}},
  \bibinfo{author}{\bibfnamefont{H.~F.} \bibnamefont{{Levison}}},
  \bibinfo{author}{\bibfnamefont{M.}~\bibnamefont{{Bailey}}}, \bibnamefont{and}
  \bibinfo{author}{\bibfnamefont{M.}~\bibnamefont{{Duncan}}},
  \bibinfo{journal}{Science} \textbf{\bibinfo{volume}{277}},
  \bibinfo{pages}{197} (\bibinfo{year}{1997}).

\bibitem[{\citenamefont{{Binney} and {Tremaine}}(2008)}]{binney2008}
\bibinfo{author}{\bibfnamefont{J.}~\bibnamefont{{Binney}}} \bibnamefont{and}
  \bibinfo{author}{\bibfnamefont{S.}~\bibnamefont{{Tremaine}}},
  \emph{\bibinfo{title}{{Galactic Dynamics}}} (\bibinfo{publisher}{Princeton,
  NJ, Princeton University Press}, \bibinfo{year}{2008}).

\bibitem[{\citenamefont{{Froeschle} et~al.}(1995)\citenamefont{{Froeschle},
  {Hahn}, {Gonczi}, {Morbidelli}, and {Farinella}}}]{froeschle1995}
\bibinfo{author}{\bibfnamefont{C.}~\bibnamefont{{Froeschle}}},
  \bibinfo{author}{\bibfnamefont{G.}~\bibnamefont{{Hahn}}},
  \bibinfo{author}{\bibfnamefont{R.}~\bibnamefont{{Gonczi}}},
  \bibinfo{author}{\bibfnamefont{A.}~\bibnamefont{{Morbidelli}}},
  \bibnamefont{and}
  \bibinfo{author}{\bibfnamefont{P.}~\bibnamefont{{Farinella}}},
  \bibinfo{journal}{Icarus} \textbf{\bibinfo{volume}{117}}, \bibinfo{pages}{45}
  (\bibinfo{year}{1995}).

\bibitem[{\citenamefont{{Michel}}(1997)}]{michel1997}
\bibinfo{author}{\bibfnamefont{P.}~\bibnamefont{{Michel}}},
  \bibinfo{journal}{Icarus} \textbf{\bibinfo{volume}{129}},
  \bibinfo{pages}{348} (\bibinfo{year}{1997}).

\bibitem[{\citenamefont{{Michel} et~al.}(1997)\citenamefont{{Michel},
  {Froeschl{\'e}}, and {Farinella}}}]{michel1997b}
\bibinfo{author}{\bibfnamefont{P.}~\bibnamefont{{Michel}}},
  \bibinfo{author}{\bibfnamefont{C.}~\bibnamefont{{Froeschl{\'e}}}},
  \bibnamefont{and}
  \bibinfo{author}{\bibfnamefont{P.}~\bibnamefont{{Farinella}}},
  \bibinfo{journal}{Cel. Mech. Dyn. Astron.} \textbf{\bibinfo{volume}{69}},
  \bibinfo{pages}{133} (\bibinfo{year}{1997}).

\bibitem[{\citenamefont{{Williams}}(1969)}]{williams1969}
\bibinfo{author}{\bibfnamefont{J.~G.} \bibnamefont{{Williams}}}, Ph.D. thesis,
  \bibinfo{school}{University of California, Los Angeles}
  (\bibinfo{year}{1969}).

\bibitem[{\citenamefont{{Knezevic} et~al.}(1991)\citenamefont{{Knezevic},
  {Milani}, {Farinella}, {Froeschle}, and {Froeschle}}}]{knezevic1991}
\bibinfo{author}{\bibfnamefont{Z.}~\bibnamefont{{Knezevic}}},
  \bibinfo{author}{\bibfnamefont{A.}~\bibnamefont{{Milani}}},
  \bibinfo{author}{\bibfnamefont{P.}~\bibnamefont{{Farinella}}},
  \bibinfo{author}{\bibfnamefont{C.}~\bibnamefont{{Froeschle}}},
  \bibnamefont{and}
  \bibinfo{author}{\bibfnamefont{C.}~\bibnamefont{{Froeschle}}},
  \bibinfo{journal}{Icarus} \textbf{\bibinfo{volume}{93}}, \bibinfo{pages}{316}
  (\bibinfo{year}{1991}).

\bibitem[{\citenamefont{{Lemaitre} and {Morbidelli}}(1994)}]{lemaitre1994}
\bibinfo{author}{\bibfnamefont{A.}~\bibnamefont{{Lemaitre}}} \bibnamefont{and}
  \bibinfo{author}{\bibfnamefont{A.}~\bibnamefont{{Morbidelli}}},
  \bibinfo{journal}{Cel. Mech. Dyn. Astron.} \textbf{\bibinfo{volume}{60}},
  \bibinfo{pages}{29} (\bibinfo{year}{1994}).

\bibitem[{\citenamefont{{Williams} and {Faulkner}}(1981)}]{williams1981}
\bibinfo{author}{\bibfnamefont{J.~G.} \bibnamefont{{Williams}}}
  \bibnamefont{and}
  \bibinfo{author}{\bibfnamefont{J.}~\bibnamefont{{Faulkner}}},
  \bibinfo{journal}{Icarus} \textbf{\bibinfo{volume}{46}}, \bibinfo{pages}{390}
  (\bibinfo{year}{1981}).

\bibitem[{\citenamefont{{Moore} et~al.}(2001)\citenamefont{{Moore},
  {Calc{\'a}neo-Rold{\'a}n}, {Stadel}, {Quinn}, {Lake}, {Ghigna}, and
  {Governato}}}]{moore2001}
\bibinfo{author}{\bibfnamefont{B.}~\bibnamefont{{Moore}}},
  \bibinfo{author}{\bibfnamefont{C.}~\bibnamefont{{Calc{\'a}neo-Rold{\'a}n}}},
  \bibinfo{author}{\bibfnamefont{J.}~\bibnamefont{{Stadel}}},
  \bibinfo{author}{\bibfnamefont{T.}~\bibnamefont{{Quinn}}},
  \bibinfo{author}{\bibfnamefont{G.}~\bibnamefont{{Lake}}},
  \bibinfo{author}{\bibfnamefont{S.}~\bibnamefont{{Ghigna}}}, \bibnamefont{and}
  \bibinfo{author}{\bibfnamefont{F.}~\bibnamefont{{Governato}}},
  \bibinfo{journal}{Phys. Rev. D} \textbf{\bibinfo{volume}{64}},
  \bibinfo{pages}{063508} (\bibinfo{year}{2001}),
  \eprint{arXiv:astro-ph/0106271}.

\bibitem[{\citenamefont{{Helmi} et~al.}(2002)\citenamefont{{Helmi}, {White},
  and {Springel}}}]{helmi2002}
\bibinfo{author}{\bibfnamefont{A.}~\bibnamefont{{Helmi}}},
  \bibinfo{author}{\bibfnamefont{S.~D.} \bibnamefont{{White}}},
  \bibnamefont{and}
  \bibinfo{author}{\bibfnamefont{V.}~\bibnamefont{{Springel}}},
  \bibinfo{journal}{\prd} \textbf{\bibinfo{volume}{66}},
  \bibinfo{pages}{063502} (\bibinfo{year}{2002}),
  \eprint{arXiv:astro-ph/0201289}.

\bibitem[{\citenamefont{{Read} et~al.}(2008)\citenamefont{{Read}, {Lake},
  {Agertz}, and {Debattista}}}]{read2008}
\bibinfo{author}{\bibfnamefont{J.~I.} \bibnamefont{{Read}}},
  \bibinfo{author}{\bibfnamefont{G.}~\bibnamefont{{Lake}}},
  \bibinfo{author}{\bibfnamefont{O.}~\bibnamefont{{Agertz}}}, \bibnamefont{and}
  \bibinfo{author}{\bibfnamefont{V.~P.} \bibnamefont{{Debattista}}},
  \bibinfo{journal}{Mon. Not. Roy. Astron. Soc.}
  \textbf{\bibinfo{volume}{389}}, \bibinfo{pages}{1041} (\bibinfo{year}{2008}),
  \eprint{arXiv:0803.2714}.

\bibitem[{\citenamefont{{Hofmann} et~al.}(2001)\citenamefont{{Hofmann},
  {Schwarz}, and {St{\"o}cker}}}]{hofmann2001}
\bibinfo{author}{\bibfnamefont{S.}~\bibnamefont{{Hofmann}}},
  \bibinfo{author}{\bibfnamefont{D.~J.} \bibnamefont{{Schwarz}}},
  \bibnamefont{and}
  \bibinfo{author}{\bibfnamefont{H.}~\bibnamefont{{St{\"o}cker}}},
  \bibinfo{journal}{Phys. Rev. D} \textbf{\bibinfo{volume}{64}},
  \bibinfo{pages}{083507} (\bibinfo{year}{2001}),
  \eprint{arXiv:astro-ph/0104173}.

\bibitem[{\citenamefont{{Green} et~al.}(2004)\citenamefont{{Green}, {Hofmann},
  and {Schwarz}}}]{green2004}
\bibinfo{author}{\bibfnamefont{A.~M.} \bibnamefont{{Green}}},
  \bibinfo{author}{\bibfnamefont{S.}~\bibnamefont{{Hofmann}}},
  \bibnamefont{and} \bibinfo{author}{\bibfnamefont{D.~J.}
  \bibnamefont{{Schwarz}}}, \bibinfo{journal}{Mon. Not. Roy. Astron. Soc.}
  \textbf{\bibinfo{volume}{353}}, \bibinfo{pages}{L23} (\bibinfo{year}{2004}),
  \eprint{arXiv:astro-ph/0309621}.

\bibitem[{\citenamefont{{Diemand} et~al.}(2005)\citenamefont{{Diemand},
  {Moore}, and {Stadel}}}]{diemand2005}
\bibinfo{author}{\bibfnamefont{J.}~\bibnamefont{{Diemand}}},
  \bibinfo{author}{\bibfnamefont{B.}~\bibnamefont{{Moore}}}, \bibnamefont{and}
  \bibinfo{author}{\bibfnamefont{J.}~\bibnamefont{{Stadel}}},
  \bibinfo{journal}{\nat} \textbf{\bibinfo{volume}{433}}, \bibinfo{pages}{389}
  (\bibinfo{year}{2005}), \eprint{arXiv:astro-ph/0501589}.

\bibitem[{\citenamefont{{Diemand} et~al.}(2008)\citenamefont{{Diemand},
  {Kuhlen}, {Madau}, {Zemp}, {Moore}, {Potter}, and {Stadel}}}]{diemand2008}
\bibinfo{author}{\bibfnamefont{J.}~\bibnamefont{{Diemand}}},
  \bibinfo{author}{\bibfnamefont{M.}~\bibnamefont{{Kuhlen}}},
  \bibinfo{author}{\bibfnamefont{P.}~\bibnamefont{{Madau}}},
  \bibinfo{author}{\bibfnamefont{M.}~\bibnamefont{{Zemp}}},
  \bibinfo{author}{\bibfnamefont{B.}~\bibnamefont{{Moore}}},
  \bibinfo{author}{\bibfnamefont{D.}~\bibnamefont{{Potter}}}, \bibnamefont{and}
  \bibinfo{author}{\bibfnamefont{J.}~\bibnamefont{{Stadel}}},
  \bibinfo{journal}{\nat} \textbf{\bibinfo{volume}{454}}, \bibinfo{pages}{735}
  (\bibinfo{year}{2008}), \eprint{arXiv:0805.1244}.

\bibitem[{\citenamefont{{Springel} et~al.}(2008)\citenamefont{{Springel},
  {Wang}, {Vogelsberger}, {Ludlow}, {Jenkins}, {Helmi}, {Navarro}, {Frenk}, and
  {White}}}]{springel2008}
\bibinfo{author}{\bibfnamefont{V.}~\bibnamefont{{Springel}}},
  \bibinfo{author}{\bibfnamefont{J.}~\bibnamefont{{Wang}}},
  \bibinfo{author}{\bibfnamefont{M.}~\bibnamefont{{Vogelsberger}}},
  \bibinfo{author}{\bibfnamefont{A.}~\bibnamefont{{Ludlow}}},
  \bibinfo{author}{\bibfnamefont{A.}~\bibnamefont{{Jenkins}}},
  \bibinfo{author}{\bibfnamefont{A.}~\bibnamefont{{Helmi}}},
  \bibinfo{author}{\bibfnamefont{J.~F.} \bibnamefont{{Navarro}}},
  \bibinfo{author}{\bibfnamefont{C.~S.} \bibnamefont{{Frenk}}},
  \bibnamefont{and} \bibinfo{author}{\bibfnamefont{S.~D.~M.}
  \bibnamefont{{White}}} (\bibinfo{year}{2008}), \eprint{arXiv:0809.0898}.

\bibitem[{\citenamefont{{Diemand} et~al.}(2004)\citenamefont{{Diemand},
  {Moore}, and {Stadel}}}]{diemand2004}
\bibinfo{author}{\bibfnamefont{J.}~\bibnamefont{{Diemand}}},
  \bibinfo{author}{\bibfnamefont{B.}~\bibnamefont{{Moore}}}, \bibnamefont{and}
  \bibinfo{author}{\bibfnamefont{J.}~\bibnamefont{{Stadel}}},
  \bibinfo{journal}{Mon. Not. Roy. Astron. Soc.}
  \textbf{\bibinfo{volume}{352}}, \bibinfo{pages}{535} (\bibinfo{year}{2004}),
  \eprint{arXiv:astro-ph/0402160}.

\bibitem[{\citenamefont{{Faltenbacher} and {Diemand}}(2006)}]{faltenbacher2006}
\bibinfo{author}{\bibfnamefont{A.}~\bibnamefont{{Faltenbacher}}}
  \bibnamefont{and}
  \bibinfo{author}{\bibfnamefont{J.}~\bibnamefont{{Diemand}}},
  \bibinfo{journal}{Mon. Not. Roy. Astron. Soc.}
  \textbf{\bibinfo{volume}{369}}, \bibinfo{pages}{1698} (\bibinfo{year}{2006}),
  \eprint{arXiv:astro-ph/0602197}.

\bibitem[{\citenamefont{{Dimitrov} et~al.}(1995)\citenamefont{{Dimitrov},
  {Engel}, and {Pittel}}}]{dimitrov1995}
\bibinfo{author}{\bibfnamefont{V.~I.} \bibnamefont{{Dimitrov}}},
  \bibinfo{author}{\bibfnamefont{J.}~\bibnamefont{{Engel}}}, \bibnamefont{and}
  \bibinfo{author}{\bibfnamefont{S.}~\bibnamefont{{Pittel}}},
  \bibinfo{journal}{Phys. Rev. D} \textbf{\bibinfo{volume}{51}},
  \bibinfo{pages}{291} (\bibinfo{year}{1995}), \eprint{arXiv:hep-ph/9408246}.

\bibitem[{\citenamefont{{Ressell} et~al.}(1993)\citenamefont{{Ressell},
  {Aufderheide}, {Bloom}, {Griest}, {Mathews}, and {Resler}}}]{ressell1993}
\bibinfo{author}{\bibfnamefont{M.~T.} \bibnamefont{{Ressell}}},
  \bibinfo{author}{\bibfnamefont{M.~B.} \bibnamefont{{Aufderheide}}},
  \bibinfo{author}{\bibfnamefont{S.~D.} \bibnamefont{{Bloom}}},
  \bibinfo{author}{\bibfnamefont{K.}~\bibnamefont{{Griest}}},
  \bibinfo{author}{\bibfnamefont{G.~J.} \bibnamefont{{Mathews}}},
  \bibnamefont{and} \bibinfo{author}{\bibfnamefont{D.~A.}
  \bibnamefont{{Resler}}}, \bibinfo{journal}{Phys. Rev. D}
  \textbf{\bibinfo{volume}{48}}, \bibinfo{pages}{5519} (\bibinfo{year}{1993}),
  \eprint{arXiv:hep-ph/9307228}.

\bibitem[{\citenamefont{{Ressell} and {Dean}}(1997)}]{ressell1997}
\bibinfo{author}{\bibfnamefont{M.~T.} \bibnamefont{{Ressell}}}
  \bibnamefont{and} \bibinfo{author}{\bibfnamefont{D.~J.}
  \bibnamefont{{Dean}}}, \bibinfo{journal}{Phys. Rev. C}
  \textbf{\bibinfo{volume}{56}}, \bibinfo{pages}{535} (\bibinfo{year}{1997}),
  \eprint{arXiv:hep-ph/9702290}.

\bibitem[{\citenamefont{{Gondolo}}(1996)}]{gondolo1996}
\bibinfo{author}{\bibfnamefont{P.}~\bibnamefont{{Gondolo}}}, in
  \emph{\bibinfo{booktitle}{Dark Matter in Cosmology Quantam Measurements
  Experimental Gravitation}}, edited by
  \bibinfo{editor}{\bibfnamefont{R.}~\bibnamefont{{Ansari}}},
  \bibinfo{editor}{\bibfnamefont{Y.}~\bibnamefont{{Giraud-Heraud}}},
  \bibnamefont{and} \bibinfo{editor}{\bibfnamefont{J.}~\bibnamefont{{Tran Thanh
  van}}} (\bibinfo{year}{1996}), pp. \bibinfo{pages}{41--+}.

\bibitem[{\citenamefont{{Reif}}(1965)}]{reif1965}
\bibinfo{author}{\bibfnamefont{F.}~\bibnamefont{{Reif}}},
  \emph{\bibinfo{title}{{Fundamentals of Statistical and Thermal Physics}}}
  (\bibinfo{publisher}{New York, McGraw-Hill}, \bibinfo{year}{1965}).

\end{thebibliography}

\end{document}